\documentclass[prb,preprint]{revtex4}

\usepackage{amsfonts,amsbsy,amssymb,amsmath,graphicx,float,subfigure,color} 
\usepackage[linktoc=all,colorlinks=true,linkcolor=magenta]{hyperref}
\pdfoutput=1

\begin{document}

\title{Tilting and Squeezing: \\ Phase space geometry of Hamiltonian saddle-node bifurcation \\ and its influence on chemical reaction dynamics}

\author{V\'ictor J. Garc\'ia-Garrido}
\email{vjose.garcia@uah.es}
\affiliation{Departamento de F\'isica y Matem\'aticas, Universidad de Alcal\'a, Alcal\'a de Henares, 28871, Spain.}

\author{Shibabrat Naik}
\email{s.naik@bristol.ac.uk}
\affiliation{School of Mathematics, University of Bristol, Bristol BS8 1TW, United Kingdom.}

\author{Stephen Wiggins}
\email{s.wiggins@bristol.ac.uk}
\affiliation{School of Mathematics, University of Bristol, Bristol BS8 1TW, United Kingdom.}

\date{\today}

\begin{abstract}

In this article we present the influence of a Hamiltonian saddle-node bifurcation on the high-dimensional phase space structures that mediate reaction dynamics. To achieve this goal, we identify the phase space invariant manifolds using Lagrangian descriptors, which is a trajectory-based diagnostic suitable for the construction of a complete ``phase space tomography'' by means of analyzing dynamics on low-dimensional slices. First, we build a Hamiltonian system with one degree-of-freedom (DoF) that models \textit{reaction}, and study the effect of adding a parameter to the potential energy function that controls the depth of the well. Then, we extend this framework to a saddle-node bifurcation for a two DoF Hamiltonian, constructed by coupling a harmonic oscillator, i.e. \textit{a bath mode}, to the other \textit{reactive} DoF in the system. For this problem, we describe the phase space structures associated with the rank-1 saddle equilibrium point in the bottleneck region, which is a Normally Hyperbolic Invariant Manifold (NHIM) and its stable and unstable manifolds. Finally, we address the qualitative changes in the reaction dynamics of the Hamiltonian system due to changes in the well depth of the potential energy surface that gives rise to the saddle-node bifurcation. 

\end{abstract}

\pacs{needs pacs number}

\maketitle


\tableofcontents

\section{Introduction}
\label{sec:intro}
\paragraph*{\bf Background.}
Molecular reaction dynamics is concerned with the breaking and forming of bonds between the atoms that make up a molecule. The description of such mechanisms requires analyzing how the bonds change in time, that is, it is a {\em dynamical} mechanism requiring a phase space description. The phase space will have coordinates that correspond to the bond configurations with a corresponding canonical momentum for each bond configuration coordinate. Thus, the dynamics of chemical bonds can be formulated in the language of Hamiltonian dynamics and its resulting phase space geometry. In recent years there has been significant developments in the mathematical description and analysis of chemical reaction dynamics in phase space~\cite{Komatsuzaki00,uzer2002geometry,Komatsuzaki02a}.

Our objective is to describe the change in a bond that signals the occurrence of a chemical reaction in terms of a unique characteristic of a trajectory of Hamilton's equations. This characteristic requires an understanding of the geometry of the phase space of the molecule in a way that enables us to divide an appropriate volume of phase space into a region corresponding to ``reactants'' and a region corresponding to ``products''. The passage from reactants to products, that is ``reaction'', occurs when a trajectory crosses the ``dividing surface'' (DS) between the reactants and products. In this framework for understanding chemical reactions, the flux through such a dividing surface would be related to the reaction rate, and therefore the construction of DS between such regions is of interest in reaction dynamics. The description of the regions of reactants and products can often be inferred from the nature of the development of the coordinates used in the mathematical model of the chemical reaction. Once the model is developed, the DS must be constructed in the context of this model. In relating the flux through the DS to a reaction rate it is desirable that trajectories crossing the DS from the reactant side proceed to the product side before possible recrossing the DS back to the reactant region. Thus, we require the DS to have the ``no-recrossing'' property\cite{MacKay90, waalkens2004direct}.  Hamiltonian dynamics conserves energy and therefore the trajectories evolve on a fixed energy surface of one less dimension (referred to as {\it codimension one}) than the phase space. The DS is required to be codimension one in the energy surface. Hence, we require a DS between reactants and products to be codimension one in the energy surface and to have the no-recrossing property. We note that Wigner had already described these properties for a DS in phase space much earlier\cite{Wigner38, Wigner39}, and a review of classical and quantum versions of DS constructed in phase space can be found in \cite{WaalkensSchubertWiggins08}. 

Traditionally, the construction of DS was initially focused on critical points of the potential energy surface (PES), that is, in the configuration space describing the molecular system. Critical points on the PES do have significance in phase space; they are the equilibrium points for zero momentum. But they continue to have influence for nonzero momentum for a range of energies above the energy of the equilibrium point. The precise manner of this dynamical influence has only been understood recently and we will describe this shortly~\cite{Komatsuzaki97,Komatsuzaki00,waalkens2004direct}. The construction of a DS separating the phase space into two parts, reactants and products, has been a focus from the dynamical systems point of view in recent years. However, the lack of a firm theoretical basis for the construction of such surfaces for molecular systems with three and more degree-of-freedom (DoF) has until recently been a major obstacle in the development of the theory. In phase space, that is for nonzero momentum, the role of the {\em saddle point} is played by an {\em invariant manifold} of saddle stability type, the normally hyperbolic invariant manifold (NHIM) (see \S:~\ref{sec:HSN_2DOF})~\cite{Wiggins88,wiggins90,wiggins2013normally}. In order to fully appreciate the NHIM and its role in reaction rate theory, it is useful to begin with a precursor concept -- the \emph{periodic orbit dividing surface} or PODS. For systems with two DoF described by a natural Hamiltonian, kinetic plus potential energy, the problem of constructing the DS in phase space was solved during the 1970s by McLafferty, Pechukas and Pollak \cite{Pechukas73,Pechukas77,Pollak78,Pechukas79}. They demonstrated that the DS at a specific energy is related to an invariant phase space structure, an unstable periodic orbit (UPO). The UPO defines (it is the boundary of) the bottleneck in phase space through which the reaction occurs and the DS which intersects trajectories evolving from reactants to products can be shown to have the geometry of a hemisphere in phase space whose boundary is the unstable PO \cite{wiggins2001impenetrable,waalkens2004direct}. The same construction can be carried out for a DS intersecting trajectories crossing from products to reactants and these two hemispheres form a sphere for which the UPO is the equator. Generalisation of this construction of DS to high dimensional systems has been a central question in reaction dynamics and has only received a satisfactory answer in recent years \cite{wiggins2001impenetrable,uzer2002geometry}. The key difficulty concerns the high dimensional analogue of the unstable PO used in the two DoF system for the construction of the DS. This difficulty is resolved by considering the NHIM, which has the appropriate dimensionality for anchoring the dividing surface in phase space.

Results from dynamical systems theory show that transport in phase space is controlled by high dimensional manifolds, NHIMs, which are the natural generalisation of the UPO of the two DoF case \cite{wiggins90}. Normal hyperbolicity of these invariant manifolds means that their stability, in a precise sense, is of saddle type in the transverse direction, which implies that they possess stable and unstable invariant manifolds that are impenetrable barriers and mediate transport in phase space. These invariant manifolds of the NHIM are structurally stable, that is, stable under perturbation \cite{wiggins2013normally}. For two DoF systems, the NHIM is an unstable PO, and for an $n > 2$ DoF system at a fixed energy, the NHIM has the topology of a $(2n-3)$-dimensional sphere and is the equator of a $(2n-2)$-dimensional sphere which constitutes the DS. The DS can be used to divide the $(2n-1)$-dimensional energy surface into two parts, reactants and products\cite{Gillilan91, Komatsuzaki96, Komatsuzaki97, Komatsuzaki00, Komatsuzaki02a}. An elementary description of the role of the NHIM in reaction dynamics is given in \cite{wiggins2016}. Fundamental theorems assure the existence of the phase space structures \textemdash NHIM and its invariant manifolds \textemdash for a range of energies above that of the saddle \cite{wiggins2013normally}. However, the precise extent of this range, as well as the nature and consequences of any bifurcations of the phase space structures that might occur as energy is increased, is not known and is a topic of continuing research\cite{Li09,Inarrea11, Allahem12, mauguiere2013bifurcations, mackay2014bifurcations, MacKay2015}.

\paragraph*{\bf Motivation.} 
In this article, we investigate the changes in geometry of the phase space structures in a normal form Hamiltonian that undergoes saddle-node bifurcation. In general, a Hamiltonian system with a saddle equilibrium point whose eigenvalues depend on the potential well depth parameter will exhibit  bifurcation when the potential energy barrier height decreases. If the potential energy surface has a single well, then the result is a collision of the stable and saddle equilibrium points as we illustrate in Figs. \ref{pes_1dof_alpha} and \ref{figpes_2dof}. This leads to drastic changes in the geometry of phase space structures \textemdash~NHIM and its invariant manifolds \textemdash~which can then be used to account for corrections to Kramers' reaction rate as barrier height decreases~\cite{hathcock2019renormalization}. Furthermore, quantifying rates of crossing low or vanishing barrier is significant for experimental study of single bond dynamics of molecules and control of micro and nano-electromechanical devices~\cite{hathcock2019renormalization,husson2009force,miller2012escape,herbert2017predictability}. We would also like to point out that the questions addressed in this article are further motivated by the work of Borondo and co-authors \cite{borondo1995,borondo1996,revuelta2019unveiling} who noted the significance of a saddle-node bifurcation in the isomerization of LiCN/LiNC molecule. 

Conservative dynamics on an open potential well has received considerable attention because the geometry of phase space structures explains the intricate fractal structure of ionization 
rates~\cite{mitchell_geometry_2003_I,mitchell_geometry_2003_II,mitchell_chaos-induced_2004}. 
Furthermore, the discrepancies in observed and predicted ionization rates in atomic systems has 
also been explained by accounting for the topology of these phase space structures. These have been connected with the breakdown of the ergodic assumption that is the basis for using ionization and dissociation rate formulae~\cite{de_leon_intramolecular_1981}. This rich literature on chaotic escape of electrons from atoms sets a precedent for delineating the changes in phase space structures due to saddle-node bifurcation that can be expected in open potential wells where the eigenvalues of the saddle equilibrium point depend on the well depth parameter~\cite{mitchell_analysis_2004,mitchell_chaos-induced_2004,mitchell_nonlinear_2009,mitchell_structure_2007,wang_photoionization_2010}.

\paragraph*{\bf Approach in this article.}
We first develop a Hamiltonian model that exhibits saddle-node bifurcation when the well depth is decreased and obtain a relationship between the parameters of the model for the bifurcation in the 2 DoF system. Then, the NHIM (an unstable PO in the 2 DoF system) is computed using differential correction and numerical continuation, and its invariant manifolds are computed using globalization. The role of these invariant manifolds in reaction dynamics and the implications of the change in their geometry by varying well depth is discussed. We compare the results with another numerical method that can be used for identifying high dimensional phase space (4 or more dimensions) structures called \textit{Lagrangian descriptors} (LDs). Lagrangian descriptors is a trajectory diagnostic method for revealing invariant manifolds that mediate phase space transport. The method was originally developed in the context of Lagrangian transport studies in fluid dynamics~\cite{madrid2009}, but the wide applicability of the method has recently been recognized as a tool to construct the dividing surface via identification of NHIM~\cite{craven2016deconstructing, craven2015lagrangian, craven2017lagrangian, revuelta2019unveiling, junginger2017chemical, feldmaier2017obtaining, patra2018detecting}. The method is straightforward to implement computationally, and it provides a high resolution method for locating high dimensional invariant manifolds using low dimensional slices of the high dimensional phase space~\cite{demian2017,Naik2019a,Naik2019b}. Therefore, this technique provides us with a first step towards realizing a complete ``phase space tomography'' of high dimensional invariant manifolds. It also applies to both Hamiltonian and non-Hamiltonian systems~\cite{lopesino2017} as well as to stochastic dynamical systems~\cite{balibrea2016lagrangian}. Moreover, Lagrangian descriptors can be applied directly to data sets obtained from the numerical solution of geophysical models and satelite observations~\cite{gg2016,ramos2018,balibrea2019}. We provide more details on our use of Lagrangian descriptors in Appendix \ref{sec:appA}.

This paper is outlined as follows. Section \ref{sec:HSN_MODELS} is devoted to the description of the Hamiltonian models with one and two DoF that we will use in our study of saddle-node bifurcation phenomena in phase space. In Section \ref{sec:RD} we explore the implications of these model problems for reaction dynamics, and discuss our results by analyzing the phase space geometrical structures responsible for the escaping dynamics from the PES and their relevance for chemical reactions. Finally, in Section \ref{sec:concl} we summarize the conclusions of this work. The reader can find all the relevant information related to the method of Lagrangian Descriptors and its implementation details in Appendix \ref{sec:appA}. Furthermore, Appendix \ref{sec:appB} describes the numerical techniques used (differential correction and manifold globalization) for the computation of the NHIM and its stable and unstable manifolds associated to the rank-1 saddle point in the 2 DoF Hamiltonian system that characterizes  the saddle-node phase space bifurcation. 

\section{Hamiltonian Saddle-Node Bifurcation Models}
\label{sec:HSN_MODELS}

In this section we introduce the normal form Hamiltonian that exhibit saddle-node bifurcation and their implications for reaction dynamics. First, we describe the normal form Hamiltonian system with one DoF, and parametrize this model by adding a parameter that controls the potential well depth. Next, we extend this model to a two DoF Hamiltonian system by coupling a bath mode (a harmonic oscillator) to the reaction DoF.

\subsection{One Degree-of-Freedom Hamiltonian}
\label{sec:HSN_1DOF}

The normal form for the one DoF Hamiltonian that undergoes a saddle-node bifurcation in phase space\cite{Wiggins2017book} can be written as
\begin{equation}
H(u,v) = \frac{1}{2} \, v^2  - \mu \, u + \frac{1}{3} \,u^3 \;,
\label{eqn:ham_1dof}
\end{equation}
where $\mu \in \mathbb{R}$ is the bifurcation parameter. Hamilton's equations are given by
\begin{equation}
\left\{
\begin{aligned}
\dot{u} &= \dfrac{\partial H}{\partial v} = v \\[.3cm]
\dot{v} & = -\dfrac{\partial H}{\partial u} = \mu - u^2
\end{aligned}
\right. \;,
\label{eq:hameq1}
\end{equation}
and the equilibria of this system are located at $(u,v) = (\pm\sqrt{\mu},0)$. We observe that the dynamical system has two different equilibrium points for $\mu  > 0$, which approach each other as the bifurcation parameter goes to zero. When $\mu = 0$, both equilibria ``collide'' into one equilibrium point at $(0,0)$. For $\mu < 0$, the resulting dynamical system has no equilibria. 

In order to study the stability of the equilibria, we linearize the vector field and calculate the Jacobian 
\begin{equation}
\mathbb{J}(u,v) = 
\begin{pmatrix}
\dfrac{\partial^2 H}{\partial u \partial v} & \dfrac{\partial^2 H}{\partial v^2} \\[.3cm]
-\dfrac{\partial^2 H}{\partial u^2} & -\dfrac{\partial^2 H}{\partial v \partial u}
\end{pmatrix} = 
\begin{pmatrix}
0 & 1 \\
-2 u & 0
\end{pmatrix}.
\end{equation}
Evaluating the Jacobian at the equilibrium point $(-\sqrt{\mu},0)$, we get
\begin{equation}
\mathbb{J}(-\sqrt{\mu},0) = 
\begin{pmatrix}
0 & 1 \\
2 \sqrt{\mu} & 0
\end{pmatrix} \;,
\end{equation}
which has eigenvalues $\pm \sqrt[4]{4\mu}$, and hence the equilibrium point $(-\sqrt{\mu},0)$ is a saddle. The Jacobian at $(\sqrt{\mu},0)$ is
\begin{equation}
\mathbb{J}(\sqrt{\mu},0) = 
\begin{pmatrix}
0 & 1 \\
-2 \sqrt{\mu} & 0
\end{pmatrix};,    
\end{equation}
which has eigenvalues $\pm \sqrt[4]{4 \, |\mu|} \, i$,yielding thst the equilibrium point $(\sqrt{\mu},0)$ is a center. For $\mu = 0$, zero is the only eigenvalue of the Jacobian matrix, and therefore the linearization does not provide enough information, and one needs to include higher order terms to determine the stability of the equilibrium point $(0,0)$.

\begin{figure}[htbp]
	\begin{center}
		A)\includegraphics[scale=0.32]{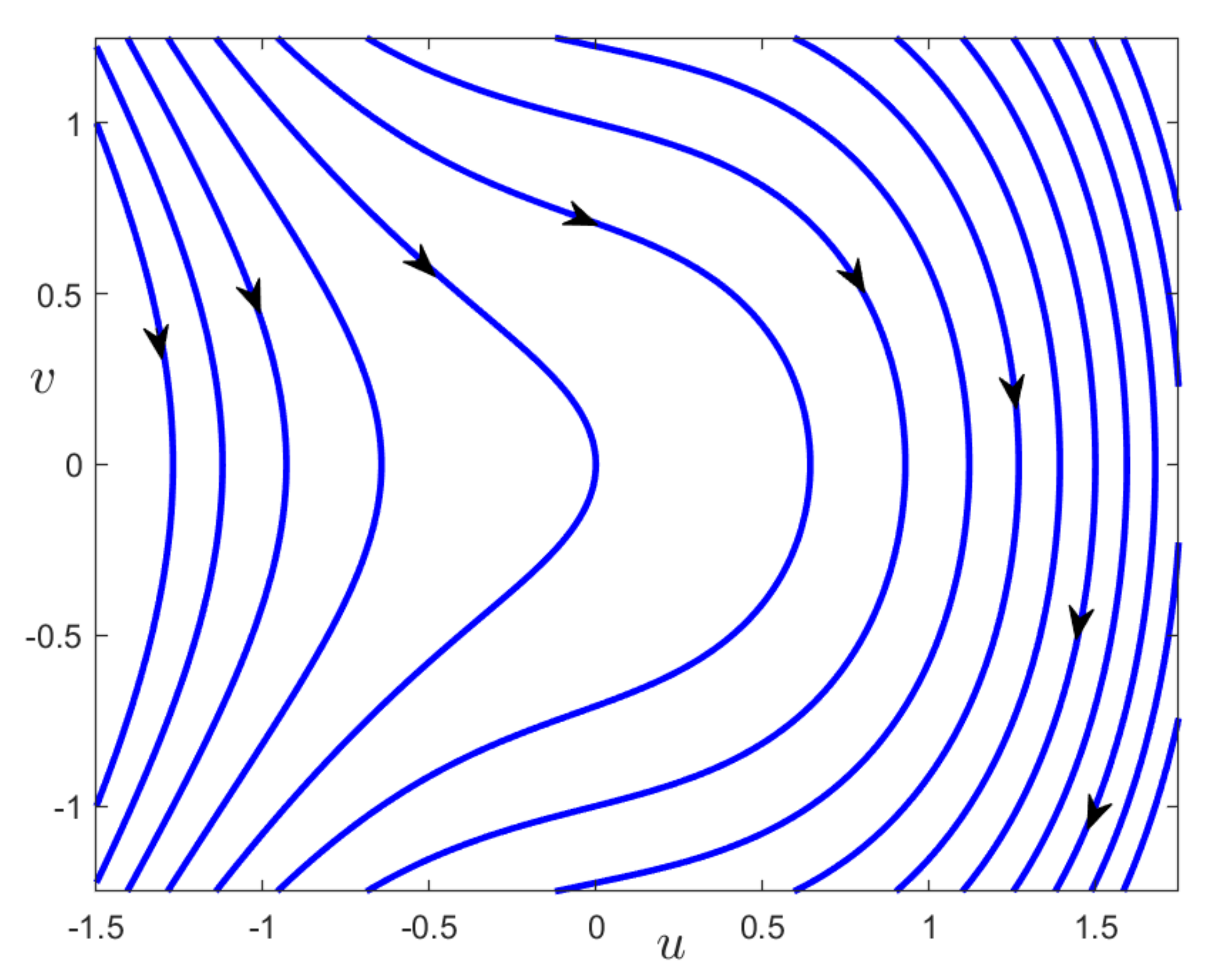}
		B)\includegraphics[scale=0.36]{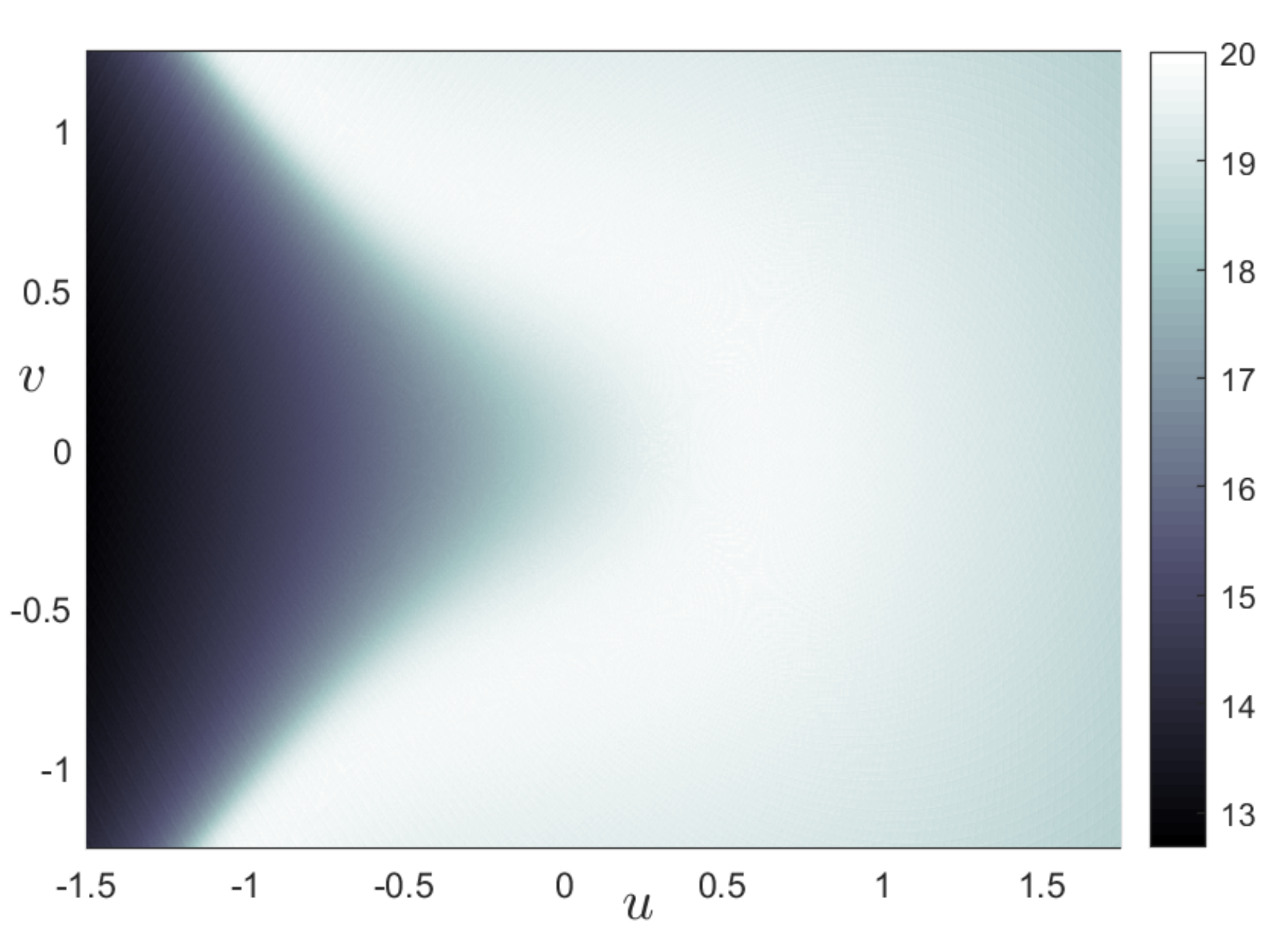}
		C)\includegraphics[scale=0.32]{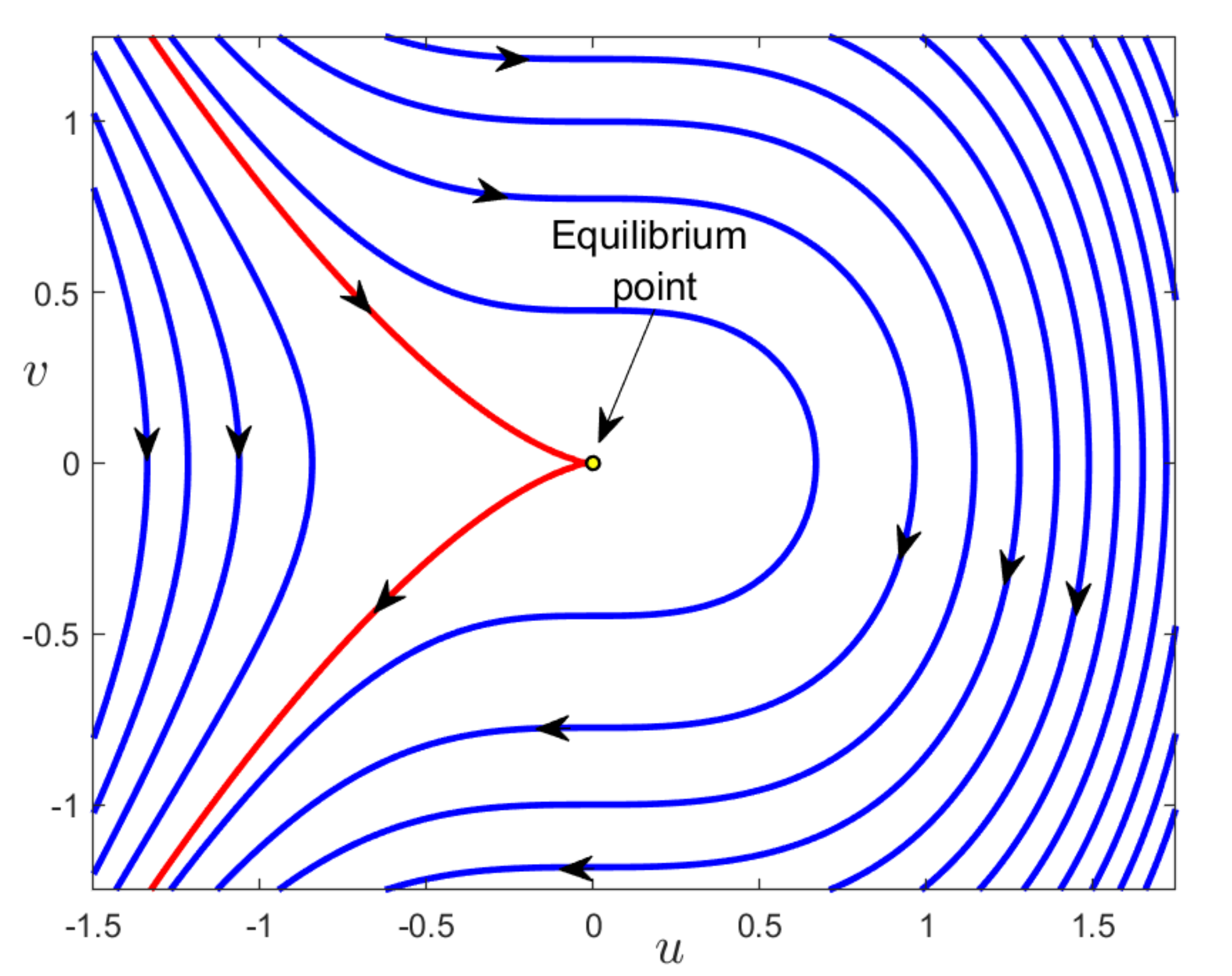}
		D)\includegraphics[scale=0.36]{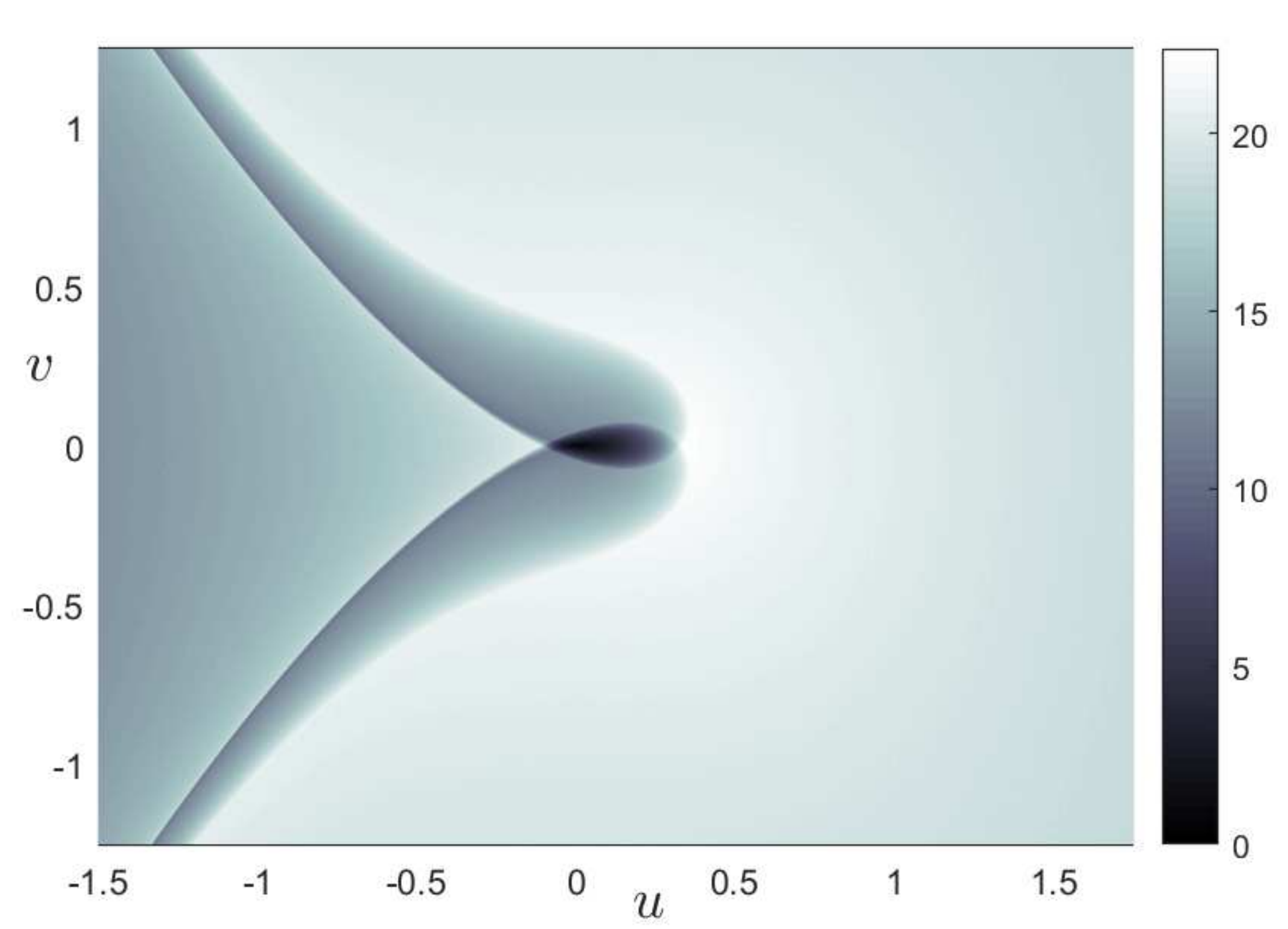}
		E)\includegraphics[scale=0.32]{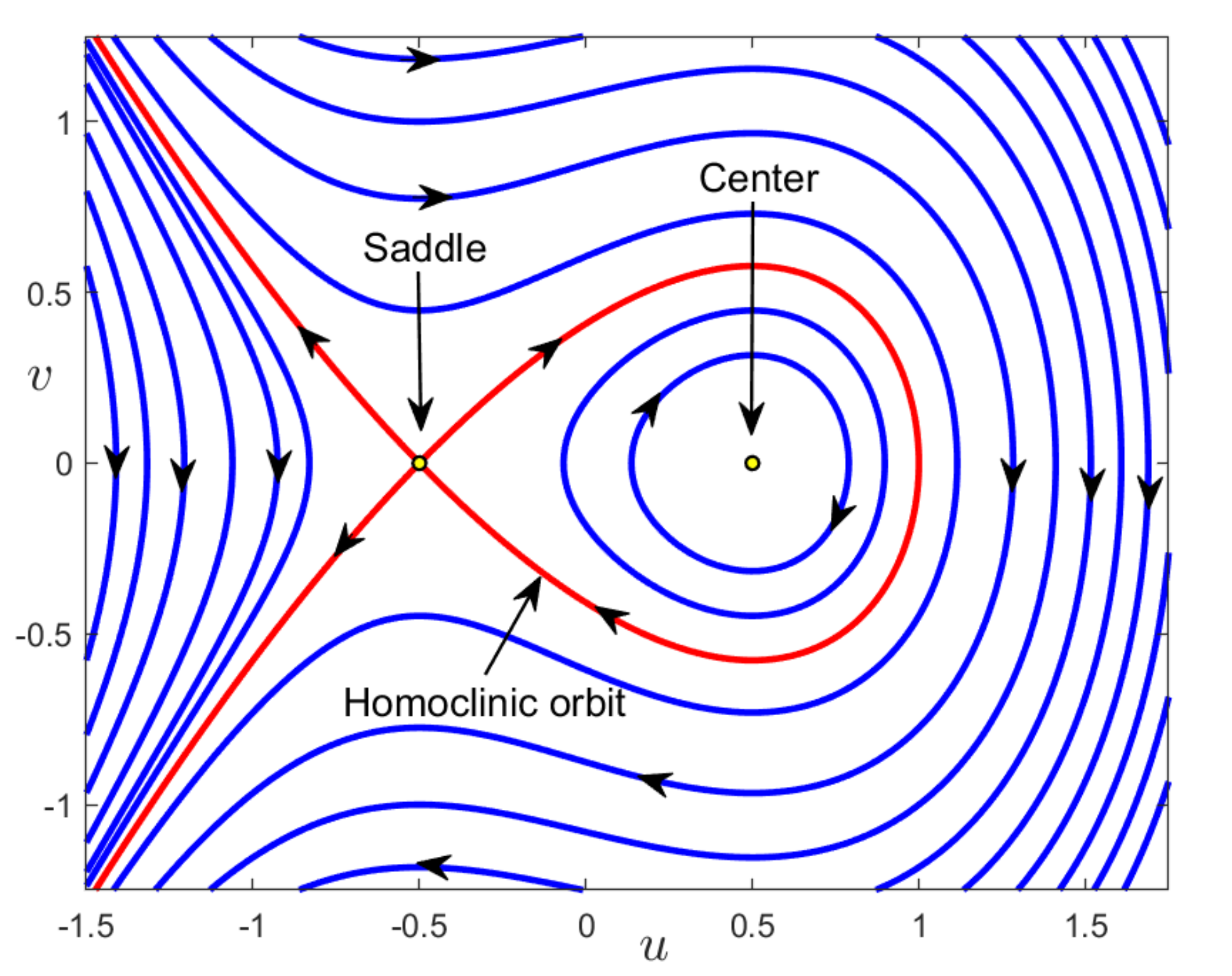}
		F)\includegraphics[scale=0.36]{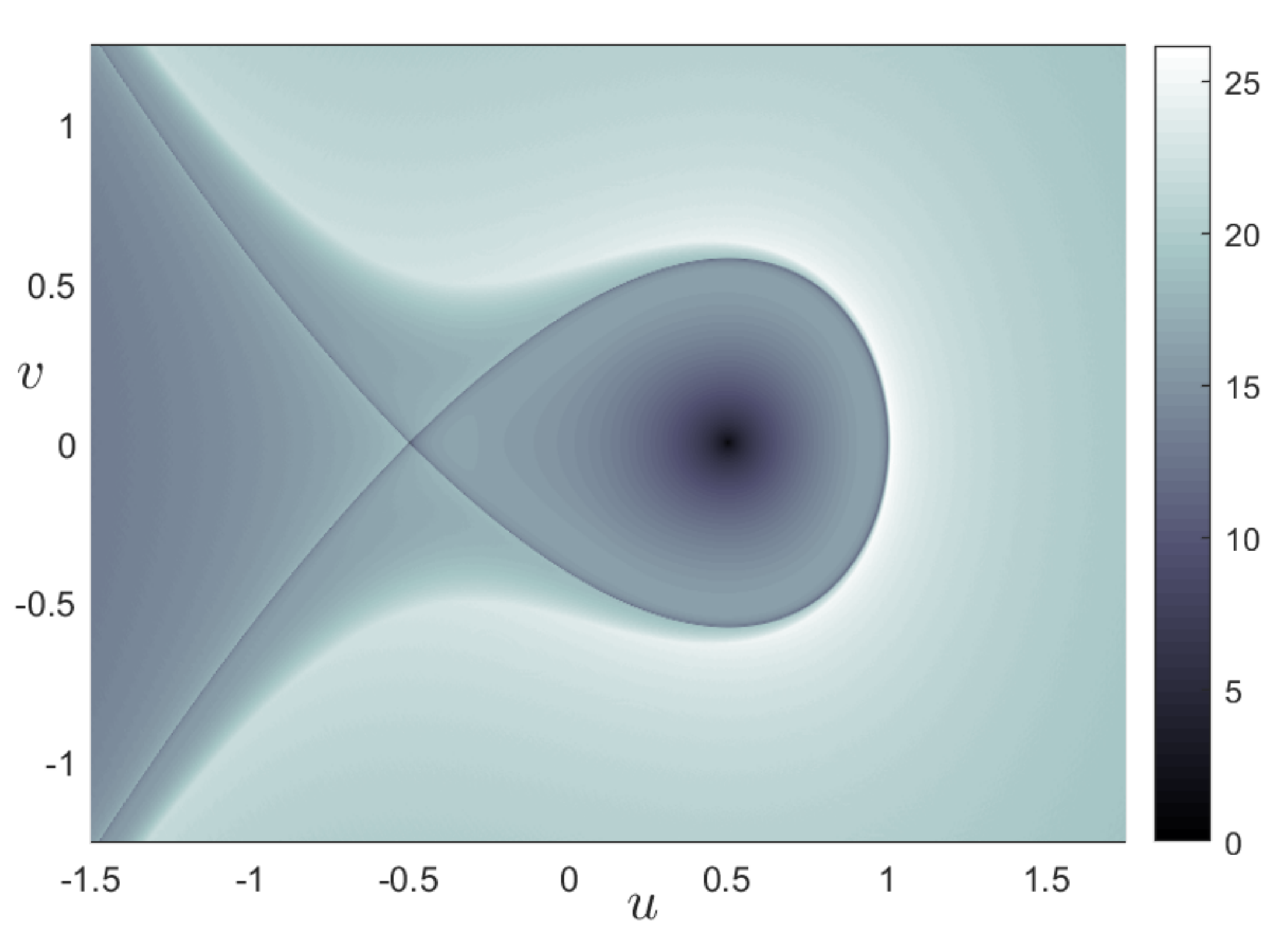}
	\end{center}
	\vspace{-3ex}
	\caption{Comparison between the phase space portrait of the dynamical system represented by Eq. \eqref{eq:hameq1} (left column) and the output of variable time LDs calculated using the $p$-norm definition with $p = 1/2$ and $\tau = 8$ (right column). A) and B) correspond to the value of the bifurcation parameter $\mu = -0.25$; C) and D) for $\mu = 0$; D) and E) for $\mu = 0.25$. We have marked  different trajectories in the phase portraits in blue, invariant manifolds in red, and equilibrium points as yellow dots.}
	\label{ld_bif_sn}
\end{figure}

We compute LDs with the goal of detecting the invariant manifolds of the dynamical system given in Eq. \eqref{eq:hameq1}. As shown by the phase portraits in Fig.~\ref{ld_bif_sn}, trajectories of~\eqref{eq:hameq1} can escape off to infinity in finite time depending on the total energy of the system and the location of the initial conditions. Therefore, this issue needs to be accounted for in the computation of LDs as explained in Appendix \ref{sec:appA}, and the approach we will follow to resolve this problem has also been adopted in the chemical reactions literature~\cite{craven2015lagrangian,craven2017lagrangian}. In particular we will use the $p$-norm in the LD definition~\eqref{Mp_function} with $p = 1/2$. Therefore, we evolve initial conditions forward and backward in time for $\tau = 8$ and whenever a trajectory leaves the domain defined by the circle of radius $15$ about the origin, we stop the numerical integration of that trajectory. In Fig.~\ref{ld_bif_sn} we show the LD contour maps for three different values of the bifurcation parameter $\mu = -0.25, \, 0, \, 0.25$ and compare the results obtained by means of LDs with the corresponding phase portraits. We can clearly see in Fig.~\ref{ld_bif_sn} how the method succesfully recovers all the relevant phase space structures. The stable and unstable manifolds are highlighted by the singularities present in the LD contour map. Furthermore, another interesting aspect to highlight from Fig. \ref{ld_bif_sn}D is that, for $\mu = 0$, the LD values in the neighborhood of the origin seem to indicate that a bifurcation is going to take place in the phase space structure, as one can observe that a `ghost' center structure is about to be created close to the cusp at the origin.

\paragraph{Fixing the saddle point at the origin.}
In the Hamiltonian saddle-node model~\eqref{eqn:ham_1dof}, the equilibria move as the bifurcation parameter $\mu$ is varied. In order to simplify the bifurcation analysis necessary to address the implications for chemical reaction dynamics, we fix the saddle point at the origin. Moreover, this form of \textit{non-moving saddle} will become useful when discussing the 2 DoF model of the saddle-node bifurcation. This transformation will also facilitate the construction of the dividing surface from the NHIM associated with the `non-moving' rank-1 saddle. 

Let us consider the linear change of coordinates
\begin{equation}
\left\{
\begin{aligned}
u & = q - \sqrt{\mu} \\
v & = p
\end{aligned}
\right. \quad , \quad \mu \in \mathbb{R}^{+} \cup \lbrace0\rbrace \, ,
\end{equation}
and substitute this transformation into Eq.~\eqref{eq:hameq1} to yield the following dynamical system in the new coordinates
\begin{equation}
\left\{
\begin{aligned}
\dot{q} & = p \\
\dot{p} & = 2  \sqrt{\mu} \, q - q^2
\end{aligned}
\right. \; ,
\label{eq:hameq2}
\end{equation}
with corresponding Hamiltonian
\begin{equation}
H(q,p) = \frac{1}{2} \, p^2 - \sqrt{\mu} \, q^2 + \frac{1}{3} \, q^3 \;.
\label{eq:ham_saddle_origin_1dof}
\end{equation}
The equilibrium points of Eq. \eqref{eq:hameq2} are $(0,0)$ and $(2 \sqrt{\mu}, 0)$. Their stability is characterized by the eigenvalues of the Jacobian matrix
\begin{equation}
\mathbb{J}(q,p) = 
\begin{pmatrix}
	\dfrac{\partial^2 H}{\partial q \partial p} & \dfrac{\partial^2 H}{\partial p^2} \\[.3cm]
	-\dfrac{\partial^2 H}{\partial q^2} & -\dfrac{\partial^2 H}{\partial p \partial q}
\end{pmatrix} = 
\begin{pmatrix}
0 & 1 \\
2 \sqrt{\mu} -2 q  & 0
\end{pmatrix}
\end{equation}
evaluated at each equilibrium point
\begin{equation}
\mathbb{J}(0,0) = 
\begin{pmatrix}
0 & 1 \\
2 \sqrt{\mu}  & 0
\end{pmatrix} 
\,,\quad 
\mathbb{J}(2\sqrt{\mu},0) = 
\begin{pmatrix}
0 & 1 \\
-2 \sqrt{\mu}  & 0
\end{pmatrix} 
\end{equation}
with eigenvalues $\pm \sqrt[4]{4\mu}$ and $\pm \sqrt[4]{4 \, |\mu|} \, i$, respectively. Hence, the origin is a saddle and $(2\sqrt{\mu},0)$ is a center equilibrium point. This is also supported by the result that eigenvalues do not change under translational change of coordinates.

\begin{figure}[htbp]
	\begin{center}
		A)\includegraphics[scale=0.33]{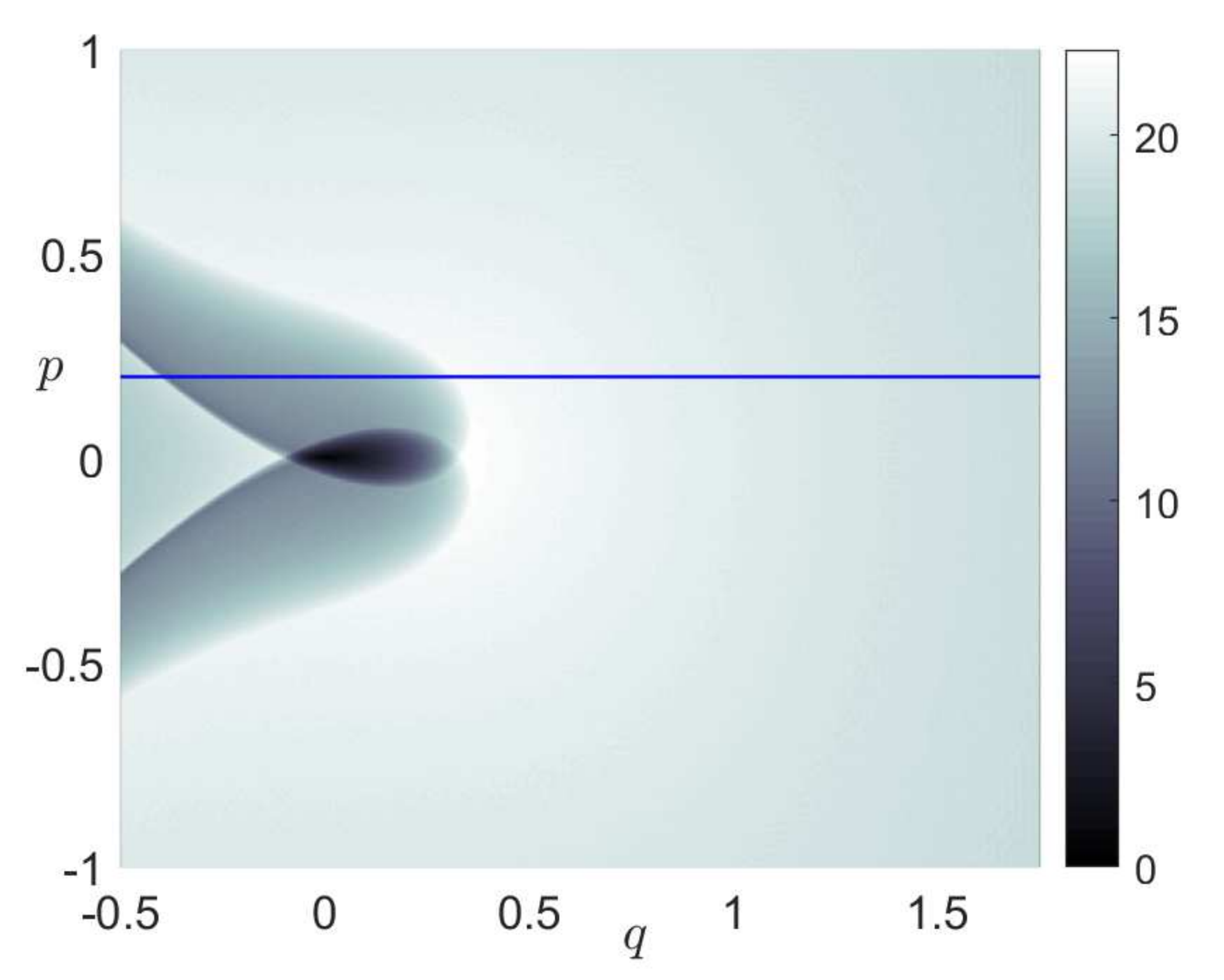}
		B)\includegraphics[scale=0.32]{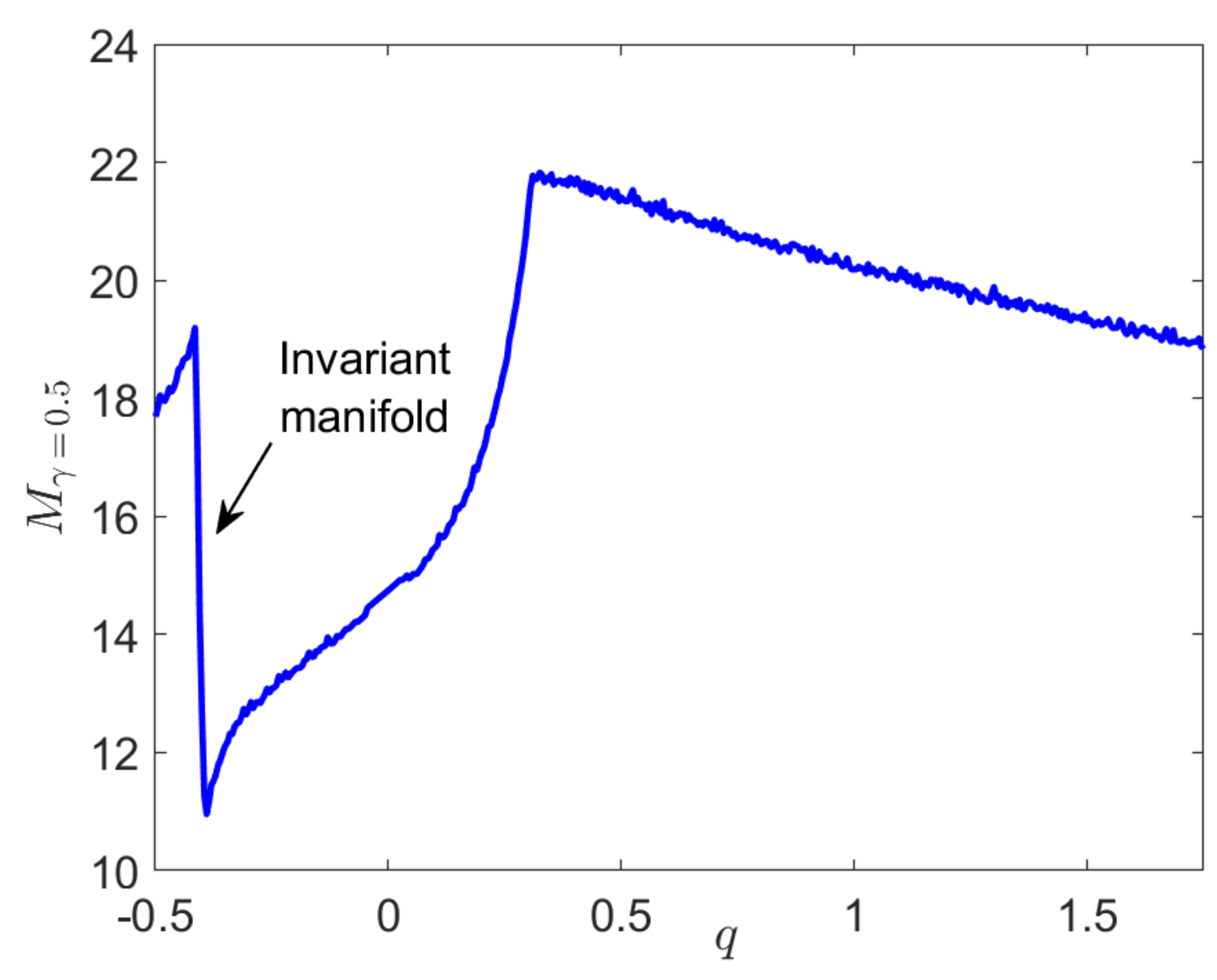}
		C)\includegraphics[scale=0.32]{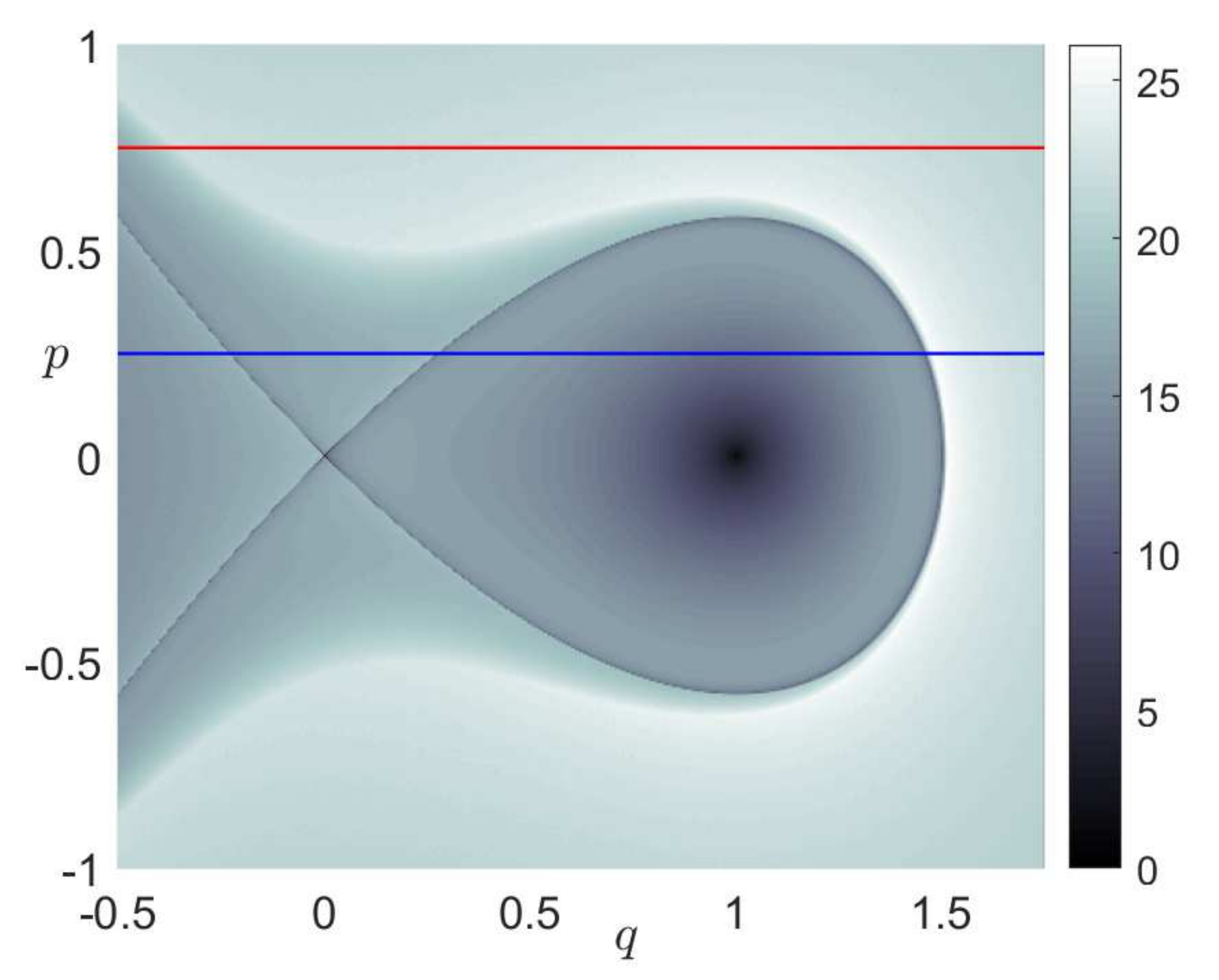}
		D)\includegraphics[scale=0.32]{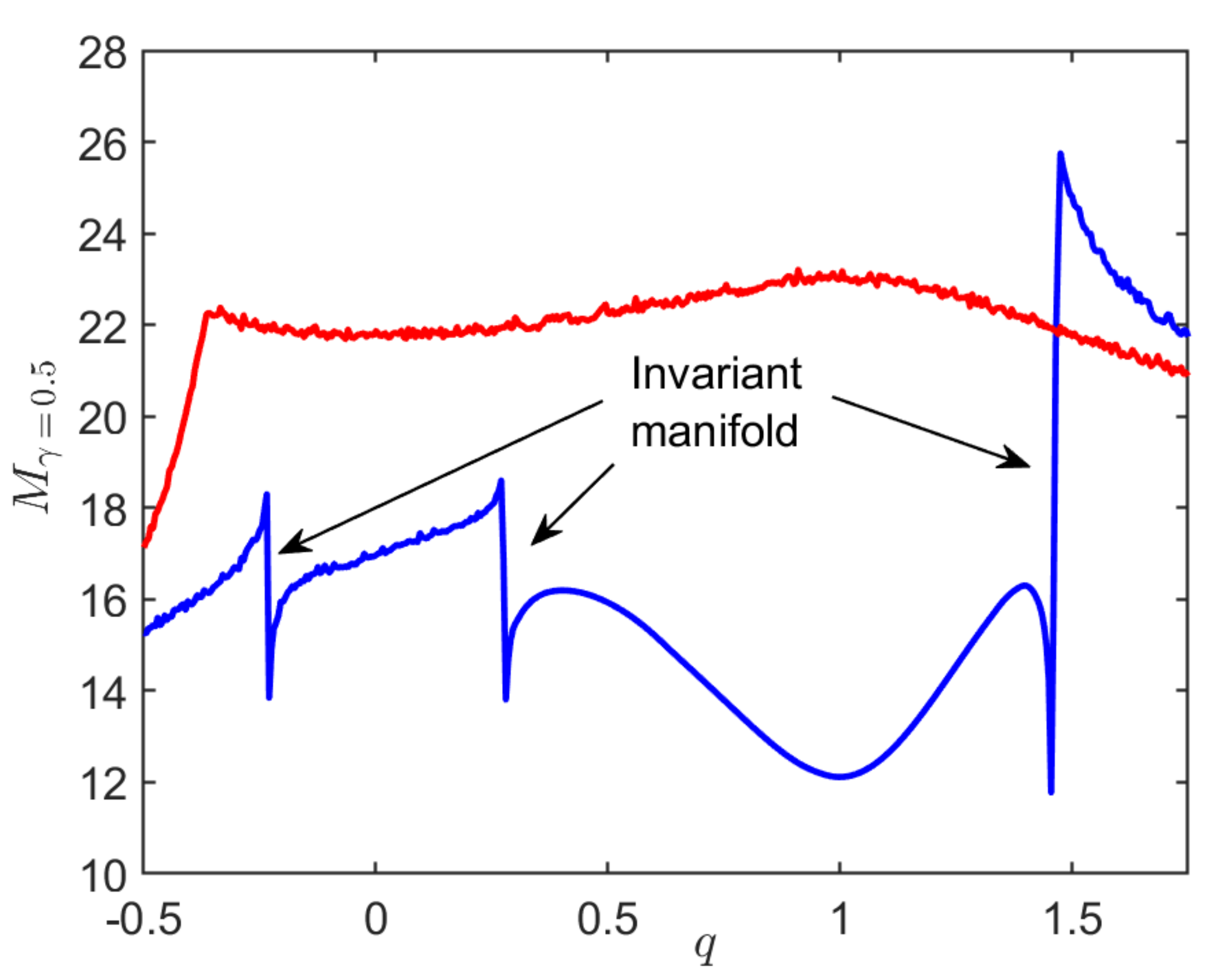}
	\end{center}
	\vspace{-3ex}
	\caption{Comparison between the output of variable time LDs using the $\gamma$-norm definition with $\gamma = 1/2$ and $\tau = 8$ (left column) and the $M_\gamma$ function value along lines parallel to the $q$-axis (right column). A) and B) correspond to $\mu = 0$ and the line of initial conditions is $p = 0.2$; C) and D) use $\mu = 0.25$ and the lines of initial conditions are $p = 0.25$ and $p = 0.75$. }
	\label{ld_bif_sn_transf}
\end{figure}

In Fig.~\ref{ld_bif_sn_transf}, we show the variable integration time LD contour maps for Eq.~\eqref{eq:hameq2} at two different values of the bifurcation parameter, $\mu = 0$ and $\mu = 0.25$. These plots illustrate how the stable and unstable manifolds of the saddle equilibrium point are identified by the singular features, points where LDs are non-differentiable, in the LD contour maps. These singular features can be visualized by taking one dimensional slices of the LD contour map, where jump discontinuities mark the initial conditions on an invariant manifold.

\paragraph{Introducing a parameter to control ``depth'' of the PES.}

The depth of the well on the PES is controlled by the cubic term that appears in the Hamiltonian~\eqref{eq:ham_saddle_origin_1dof}. Therefore, we introduce a parameter that allows us to vary the amplitude of this term, and consequently the strength of the nonlinearity in the vector field~\eqref{eq:hameq2}. Thus, the Hamiltonian becomes
\begin{equation}
H(q,p) = \frac{1}{2} \, p^2 - \sqrt{\mu} \, q^2 + \frac{\alpha}{3} \, q^3 \equiv T(p) + V(q) \;,
\label{hameq_1dof_gen}
\end{equation}
where $\mu \in \mathbb{R}^{+} \cup \lbrace0\rbrace$ and $ \alpha \in \mathbb{R}^{+}$ are the two parameters, $T$ is the kinetic energy of the DoF and $V$ its potential energy. Hamilton's equations are given by
\begin{equation}
\begin{cases}
\dot{q} = \dfrac{\partial H}{\partial p} = p \\[.2cm]
\dot{p} = -\dfrac{\partial H}{\partial q} = 2\sqrt{\mu} \, q - \alpha \, q^2
\end{cases}
\;.
\label{eq:hameq_saddlenode_1dof}
\end{equation}
The equilibria are $(0,0)$ and $\left(2\sqrt{\mu}/ \alpha,0\right)$, and the Jacobian of the vector field is given by
\begin{equation}
J(q,p) = 
\begin{pmatrix}
0 & 1 \\
2 \sqrt{\mu} -2 \alpha q & 0
\end{pmatrix}.
\end{equation}
We determine the stability of the equilibria by evaluating the Jacobian which gives
\begin{equation}
J(0,0) = 
\begin{pmatrix}
0 & 1 \\
2 \sqrt{\mu} & 0
\end{pmatrix}
\,,\quad 
J\left(2\sqrt{\mu}/ \alpha,0\right) = 
\begin{pmatrix}
0 & 1 \\
-2 \sqrt{\mu} & 0
\end{pmatrix}.
\end{equation}
The eigenvalues of the Jacobian at the equilibrium point $(0,0)$ are $\pm \sqrt[4]{4\mu}$ with corresponding eigenvectors $(1,\pm \sqrt[4]{4\mu})$. The eigenvalues for the Jacobian at the equilibrium point $\left(2\sqrt{\mu}/ \alpha,0\right)$ are $\pm \sqrt[4]{4\mu} \, i$ with eigenvectors $(1,\pm \sqrt[4]{4\mu} \, i)$. Clearly, the eigenvalues of the Hamiltonian~\eqref{eq:ham_saddle_origin_1dof} are perserved, and $(0,0)$ is a saddle, while $\left(2\sqrt{\mu}/ \alpha,0\right)$ is a center. We note here that the eigenvalues and eigenvectors of the Jacobian evaluated at the equilibrium points only depend on $\mu$. 

``Depth'' of the well on the PES, referred to as well-depth from here on, is determined by the difference between the potential energy of the saddle and the potential energy of the center (minimum of the well) equilibrium points. The potential energy function is given by
\begin{equation}
V(q) =  - \sqrt{\mu} \, q^2 + \frac{\alpha}{3} \, q^3 \; ,
\label{pe_1dof}
\end{equation}
and this difference is given by
\begin{equation}
\mathcal{F} \equiv V(0) - V\left(\frac{2\sqrt{\mu}}{\alpha}\right) = - \frac{4 \sqrt{\mu^3}}{3 \alpha^2} \; .
\end{equation}
Hence, for a fixed $\mu$, the well-depth is increased by decreasing $\alpha$. In fact, if we denote $\mathcal{D} = 2\sqrt{\mu} / \alpha$ as the distance between the saddle and the center equilibrium point, we get
\begin{equation}
\mathcal{F} = \dfrac{\sqrt{\mu}}{3} \mathcal{D}^2,
\end{equation}
therefore, as $\alpha$ increases, the well-depth approaches zero faster than the distance between the saddle and the center equlibrium points. Furthermore, the rate at which the well-depth changes as we change the distance between equilibria is given by
\begin{equation}
\dfrac{d \mathcal{F}}{d \mathcal{D}} = \dfrac{2\sqrt{\mu}}{3} \mathcal{D} = \frac{1}{3} \lambda_0^2 \, \mathcal{D} \; .
\end{equation}
which is proportional to the product of the square of the eigefrequency associated to the saddle equilibrium and the distance between equilibria.

\begin{figure}[htbp]
	\begin{center}
		a) \includegraphics[scale=0.3]{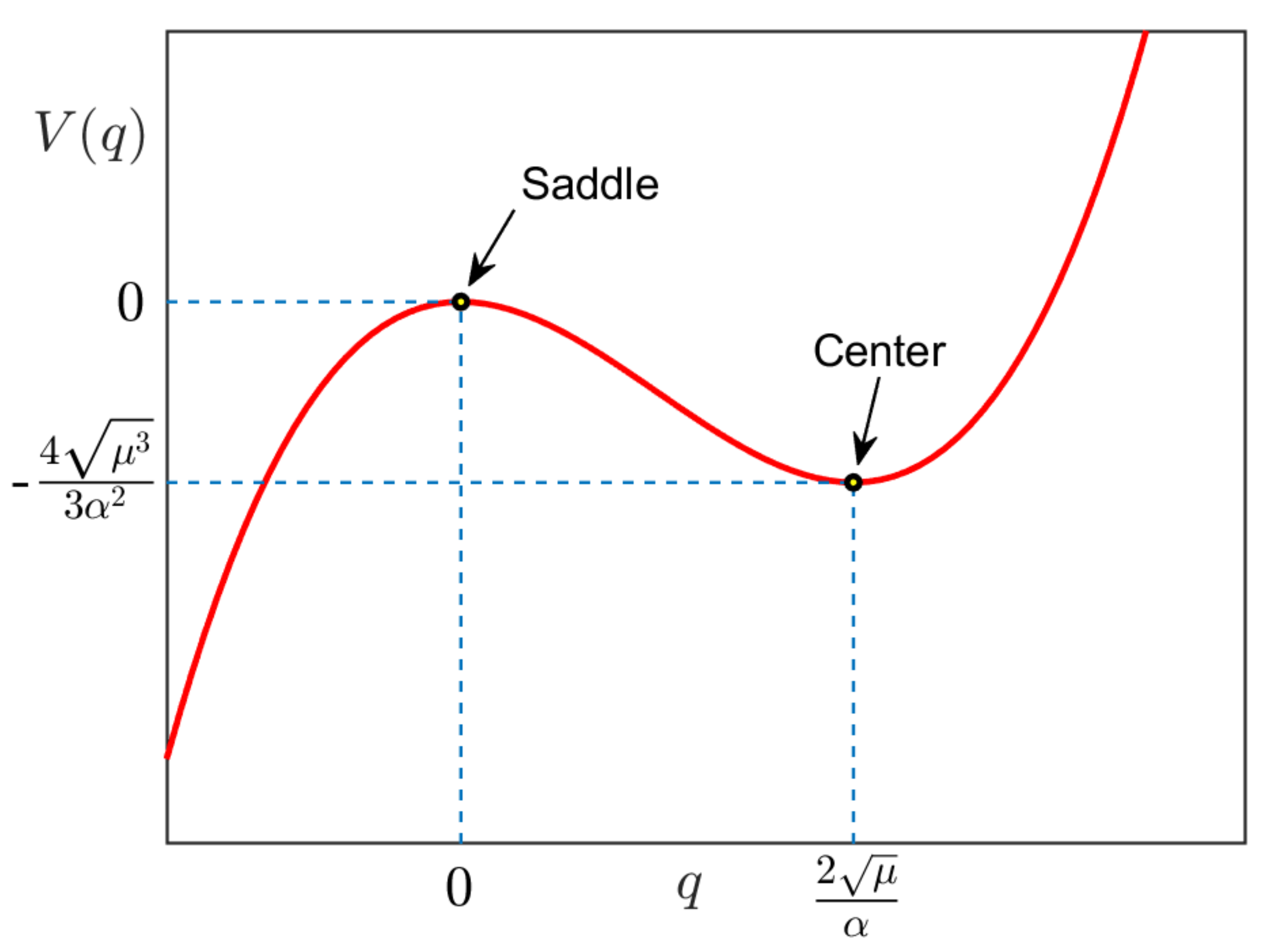}
		b) \includegraphics[scale=0.29]{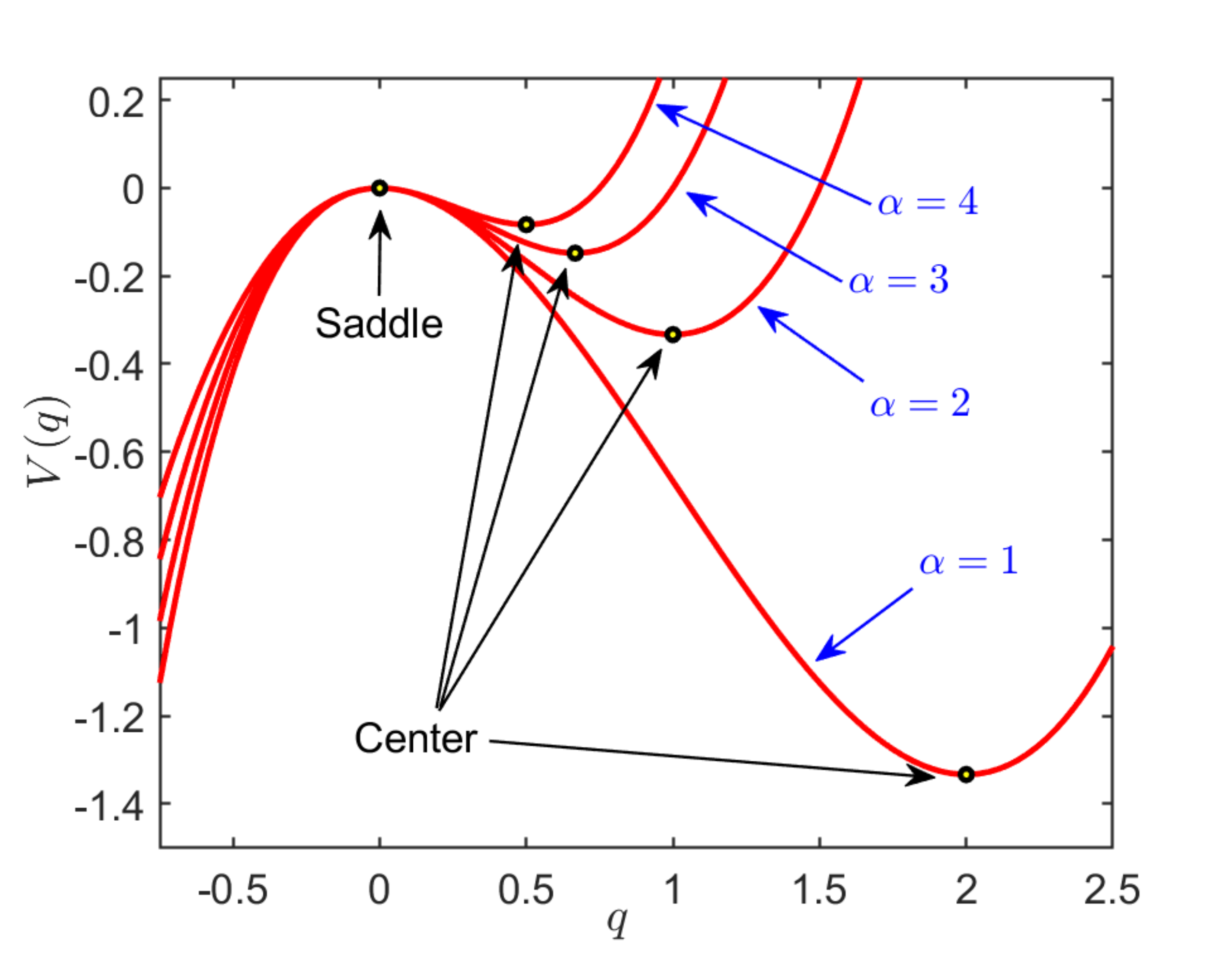}
	\end{center}
	\vspace{-3ex}
	\caption{A) Potential energy function described by Eq. \eqref{pe_1dof} illustrating the ``depth'' parameter together with the horizontal distance from the saddle to the center equilibrium. B) Evolution of he potential energy function for a fixed value of $\mu = 1$, as $\alpha$ is varied.}
	\label{pes_1dof_alpha}
\end{figure}

In Fig.~\ref{pes_1dof_alpha}(a) we illustrate the well-depth together with the distance between the saddle and the center. In Fig.~\ref{pes_1dof_alpha}(b) shows how the potential energy function changes, for a fixed value $\mu = 1$, as the well-depth parameter $\alpha$ (dstrength of the nonlinearity) is increased.

\begin{figure}[htbp]
	\begin{center}		
		\subfigure[]{\includegraphics[width=0.49\textwidth]{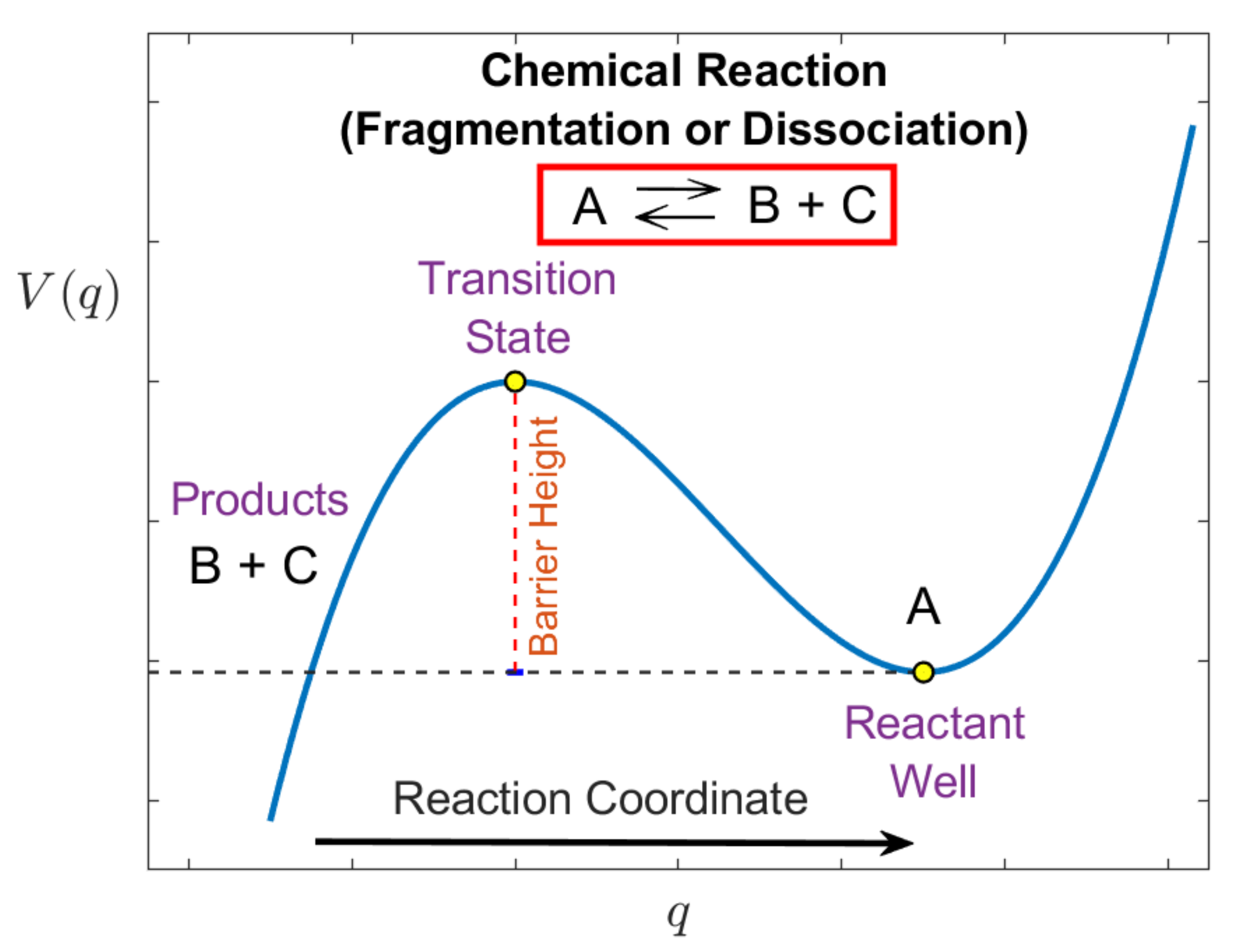}}
		\subfigure[]{\includegraphics[width=0.49\textwidth]{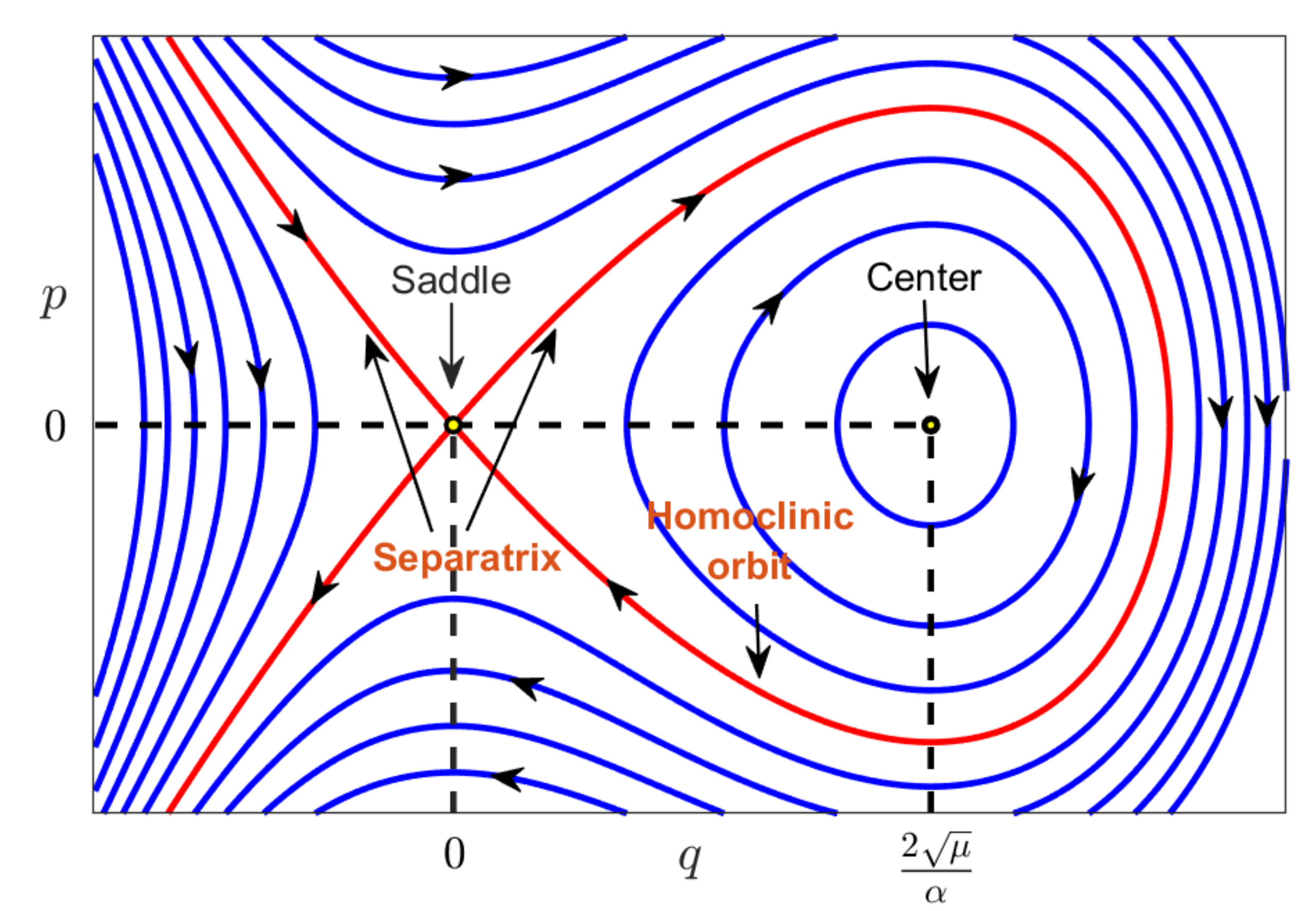}}
		\subfigure[]{\includegraphics[width=0.49\textwidth]{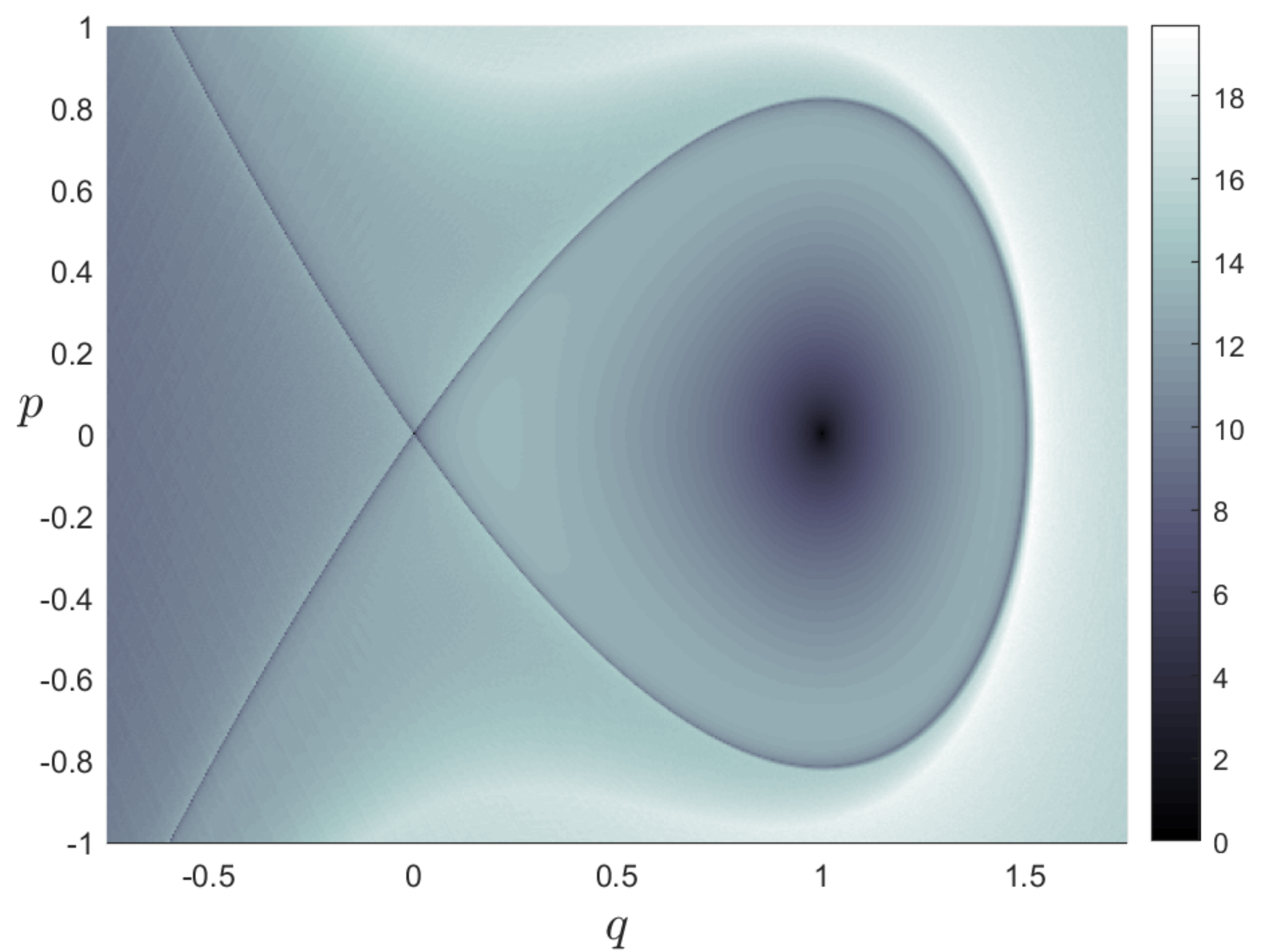}}
		\subfigure[]{\includegraphics[width=0.49\textwidth]{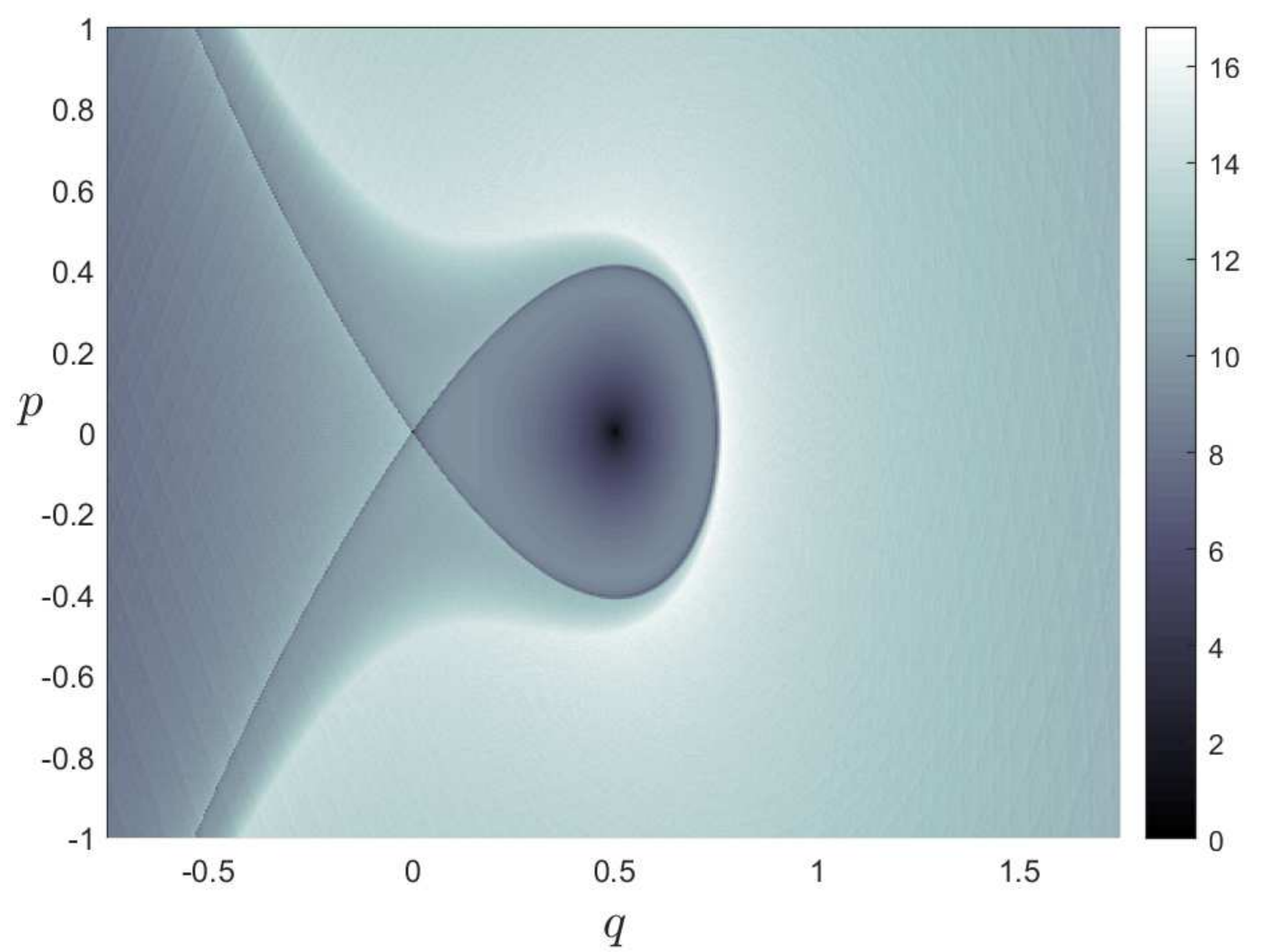}}
	\end{center}
	\vspace{-3ex}
	\caption{(a) Cubic potential energy function as a basic model for dissociation chemical reactions. The shape of this potential is the building block (normal form) for saddle-node bifurcations in phase space. (b) Phase portrait showing the location of the saddle and center equilibria in terms of the model parameters $\alpha$ and $\mu$. (c-d) Variable time LDs calculated using the $p$-norm definition with $p = 1/2$ and $\tau = 5$ for the dynamical system described in Eq. \eqref{eq:hameq_saddlenode_1dof} using (c) $\mu = 1$ and $\alpha = 2$ (d) $\mu = 1$ and $\alpha = 4$.}
	\label{ld_ps_1dof_alpha}
\end{figure}

Recall that, as we discussed in the introduction, in the context of chemical reaction dynamics the escape from a potential well problem can be identified for instance with dissociation or fragmentation reactions, where
a chemical transformation takes place if a bond of a molecule A breaks up, giving rise to two products B and C. In this situation, the equilibrium conformation of the given molecule A is represented by a potential well in a PES, and dissociation into B and C takes place if the system has sufficient energy to croos the potential barrier that separates bounded (vibration) from unbounded (bond breakup) motion. This setting can be modeled for instance by the potential energy function in Eq. \eqref{pe_1dof}, and is illustrated in Fig. \ref{ld_ps_1dof_alpha}(a), where a locally cubic potential energy function about the potential barrier describes this phenomenon, which could be the result of a saddle-node bifurcation that has occurred in the phase space of the Hamiltonian system. In Fig. \ref{ld_ps_1dof_alpha}(b) we show the phase portrait for the dynamical system~\eqref{eq:hameq_saddlenode_1dof} which shows the equilibrium points and the homoclinic orbit. Fig. \ref{ld_ps_1dof_alpha}(c-d) display the changes in the phase space structure using the variable integration time LDs for different values of the well-depth parameter $\alpha$.

\subsection{Two degrees-of-freedom Hamiltonian}
\label{sec:HSN_2DOF}

In this section we introduce the normal form for the two DoF Hamiltonian that undergoes a saddle-node bifurcation by extending the one DoF Hamiltonian discussed above. To do so, we add another degree of freedom in the form of a harmonic oscillator with mass $m = 1$ and frequency $\omega$, known as a bath mode in the terminology of chemical reaction dynamics, and parametrize the coupling of the reaction and bath modes. Thus, the Hamiltonian becomes 
\begin{equation}
H(q,x,p,p_x) = \dfrac{1}{2} \left(p^2 + p_x^2 \right) - \sqrt{\mu} \, q^2 + \frac{\alpha}{3} \,q^3 + \dfrac{\omega^2}{2} x^2 + \dfrac{\varepsilon}{2} \left(x-q\right)^2 \; ,
\label{ham_2dof}
\end{equation}
where $\alpha > 0$ is the well-depth parameter and measures the strength of the nonlinearity, $\omega > 0$ is the frequency of the harmonic oscillator or the bath mode, $\mu \geq 0$ is the bifurcation parameter, and $\varepsilon \geqslant 0$ is the strength of the coupling between the reaction and the bath mode. Identifying this Hamiltonian's kinetic energy, $T(p,p_x)$, potential energy, $V(q,x)$, and assuming $H(q,x,p,p_x) = T(p,p_x) + V(q,x)$, we get that
\begin{equation}
T(p,p_x) = \frac{1}{2}\left(p^2 + p_x^2\right)  \quad,\quad V(q,x) = - \sqrt{\mu} \, q^2 + \frac{\alpha}{3} \,q^3 + \dfrac{\omega^2}{2} x^2 + \dfrac{\varepsilon}{2} \left(x-q\right)^2
\label{pes_2dof}
\end{equation}

The corresponding Hamilton's equations are given by:
\begin{equation}
\left\{
\begin{aligned}
\dot{q} &= \dfrac{\partial H}{\partial p} =  p \\[.1cm]
\dot{x} &= \dfrac{\partial H}{\partial p_x} = p_x \\[.1cm]
\dot{p} &= -\dfrac{\partial H}{\partial q} =  2\sqrt{\mu} \, q - \alpha \, q^2 + \varepsilon (x - q) \\[.1cm]
\dot{p}_x &= -\dfrac{\partial H}{\partial x} = -\omega^2 x + \varepsilon (q-x) 
\end{aligned}
\right.
\label{hameq_2dof}
\end{equation}
For this Hamiltonian system, the phase space is four dimensional, and since energy is conserved, the trajectories evolve on a three dimensional energy surface. The equilibria for this system are located at $\mathbf{x}_1^e = (0,0,0,0)$ and $\mathbf{x}_2^e = \left(q_e,x_e,0,0\right)$
where
\begin{equation}
q_e = \frac{2\sqrt{\mu}}{\alpha} - \frac{\omega^2\varepsilon}{\alpha(\omega^2 + \varepsilon)} \quad,\quad x_e = \frac{\varepsilon}{\omega^2 +\varepsilon}\, q_e
\label{cSpace_eqCoords}
\end{equation}
The energy of the system at the equilibrium points is:
\begin{equation}
H(\mathbf{x}_1^e) = 0 \;,\quad H(\mathbf{x}_2^e) = \left(-2\sqrt{\mu} + \frac{\omega^4 \varepsilon - 2\omega^2\varepsilon^2}{\left(\omega^2 + \varepsilon\right)^2} \right) \frac{q_e^2}{6}
\label{energy_eqpts}
\end{equation}
One can easily show that there exists a critical value of the coupling strength (also interpreted as a perturbation to the one DoF model) given by
\begin{equation}
\varepsilon_c = \dfrac{2\sqrt{\mu} \, \omega^2}{\omega^2 - 2\sqrt{\mu}} \;,
\label{crit_epsi}
\end{equation}
for which the Hamiltonian~\eqref{hameq_2dof} has only one equilibrium point at the origin. We note here that this critical value is independent of the nonlinearity strength parameter $\alpha$ and requires $\omega^2 > 2\sqrt{\mu}$ to be satisfied. Moreover, it is characterized by a functional relationship between the squares of the eigenfrequencies of the reactive and bath modes of the uncoupled system ($\varepsilon = 0$). We will analyze the influence of the perturbation strength $\varepsilon$ on the geometry of the phase space structures up until this critical condition in Section \ref{sec:RD}.

\begin{figure}[htbp]
	\centering	
		\subfigure[]{\includegraphics[scale=0.31]{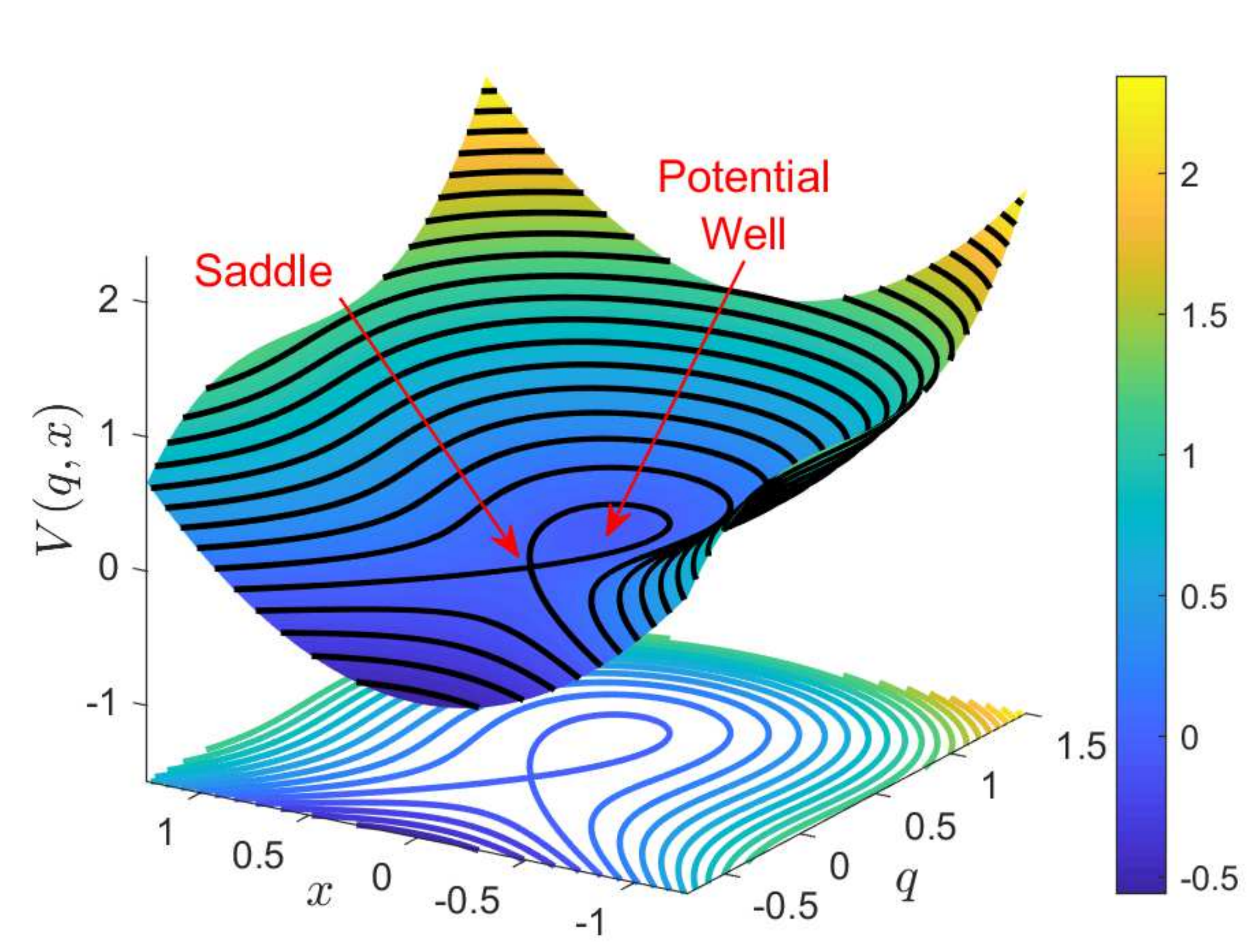}}
		\subfigure[]{\includegraphics[scale=0.29]{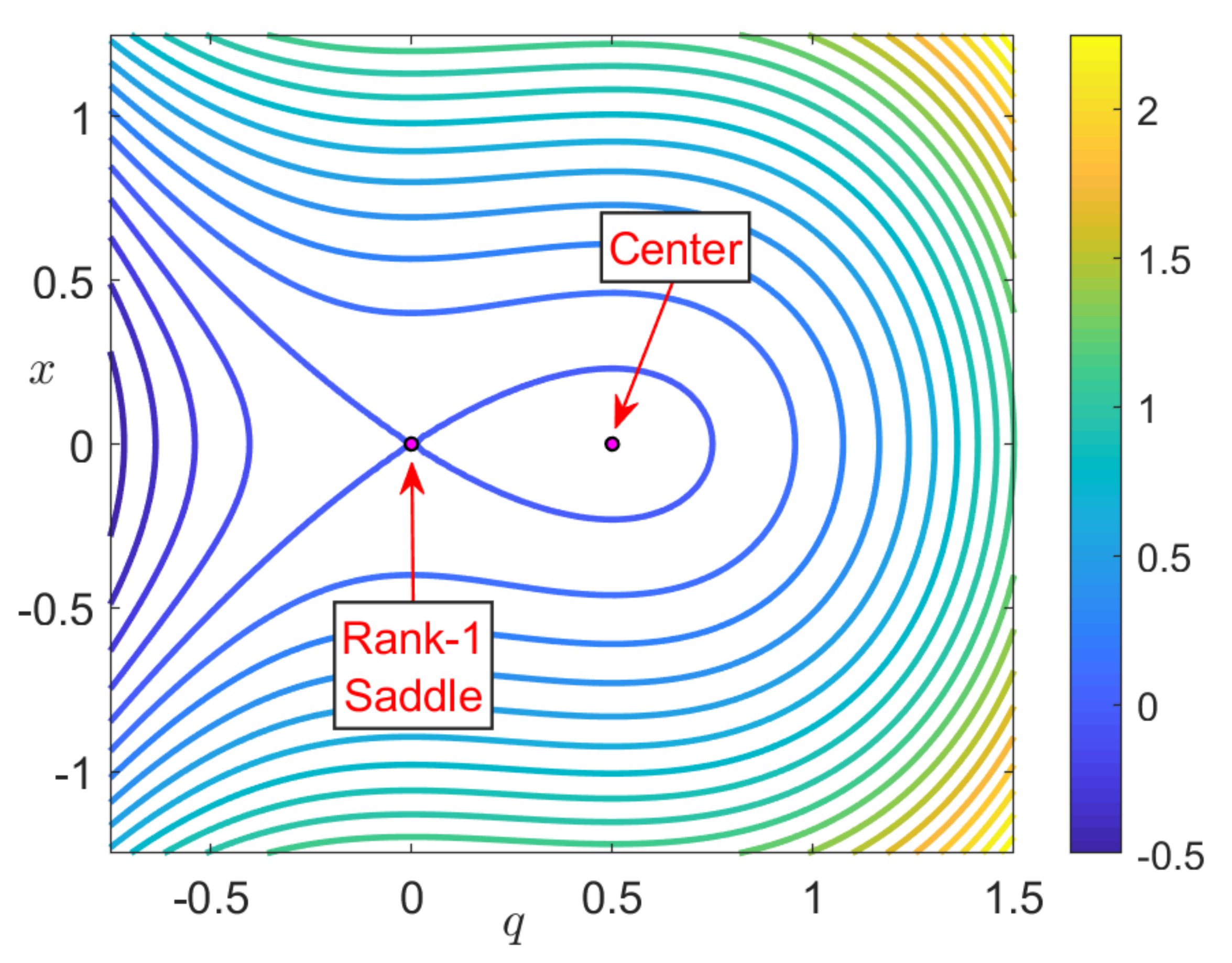}}\\
		\vspace{-2.5ex}
		\subfigure[]{\includegraphics[scale=0.31]{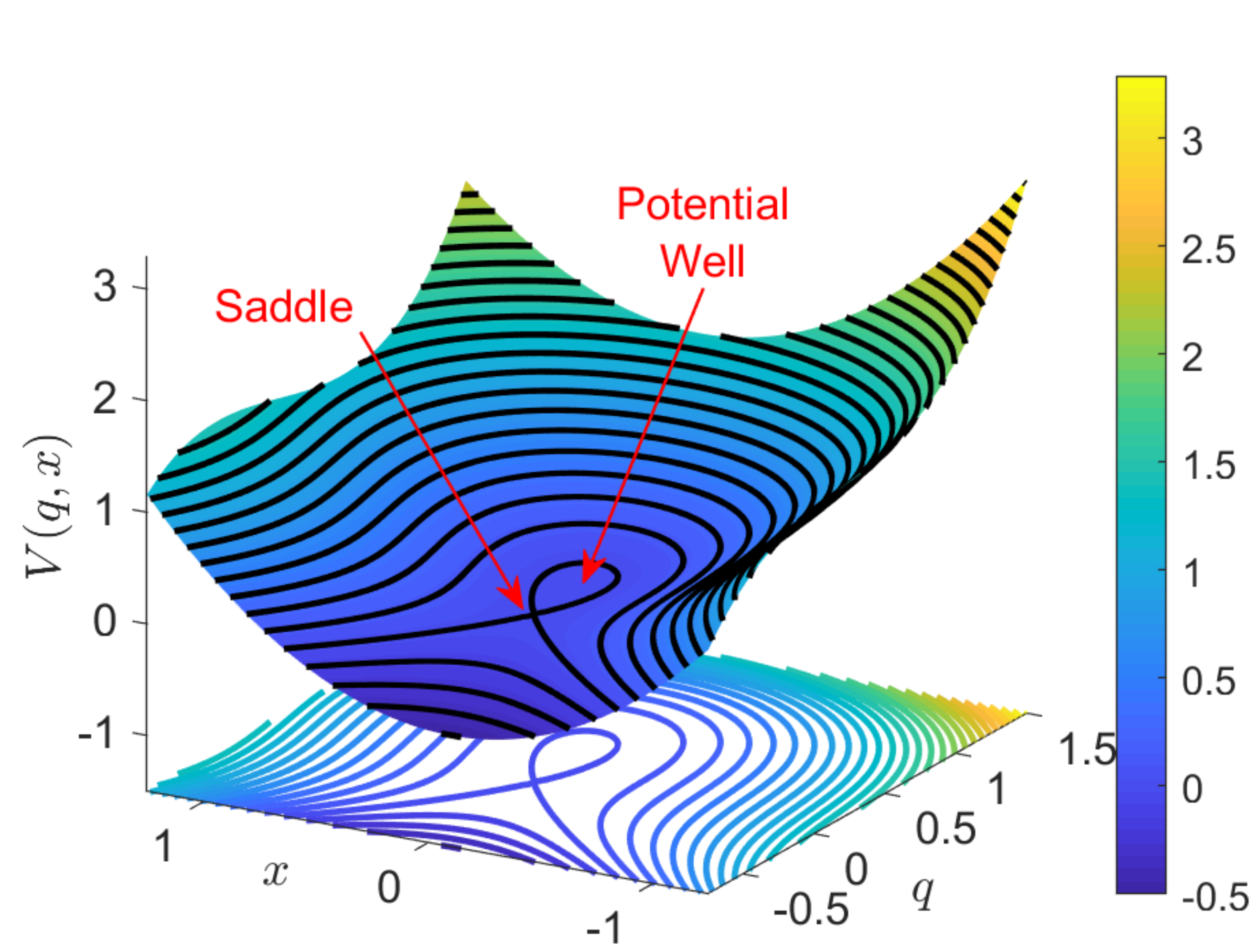}}
		\subfigure[]{\includegraphics[scale=0.29]{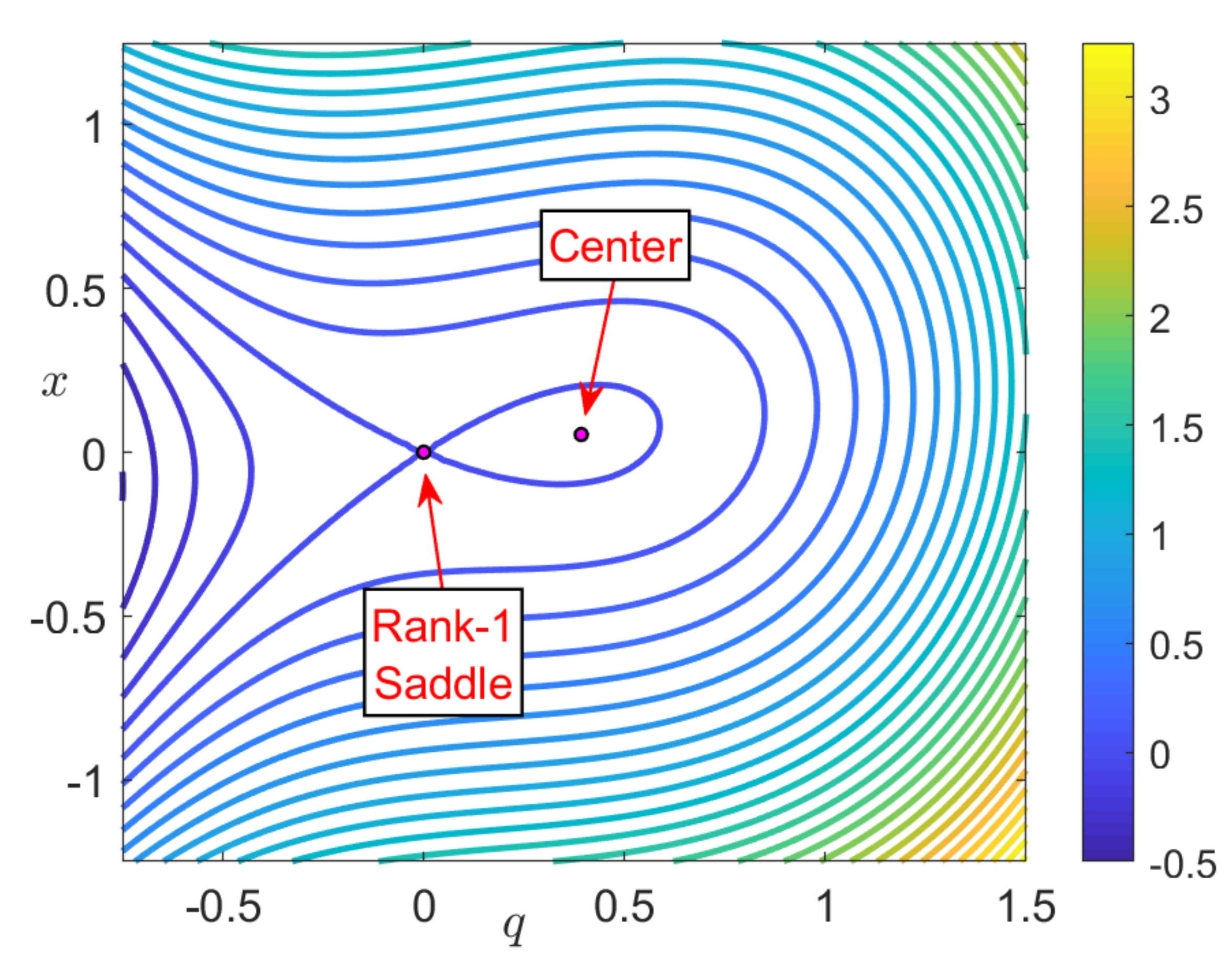}}\\
		\vspace{-2.5ex}
		\subfigure[]{\includegraphics[scale=0.31]{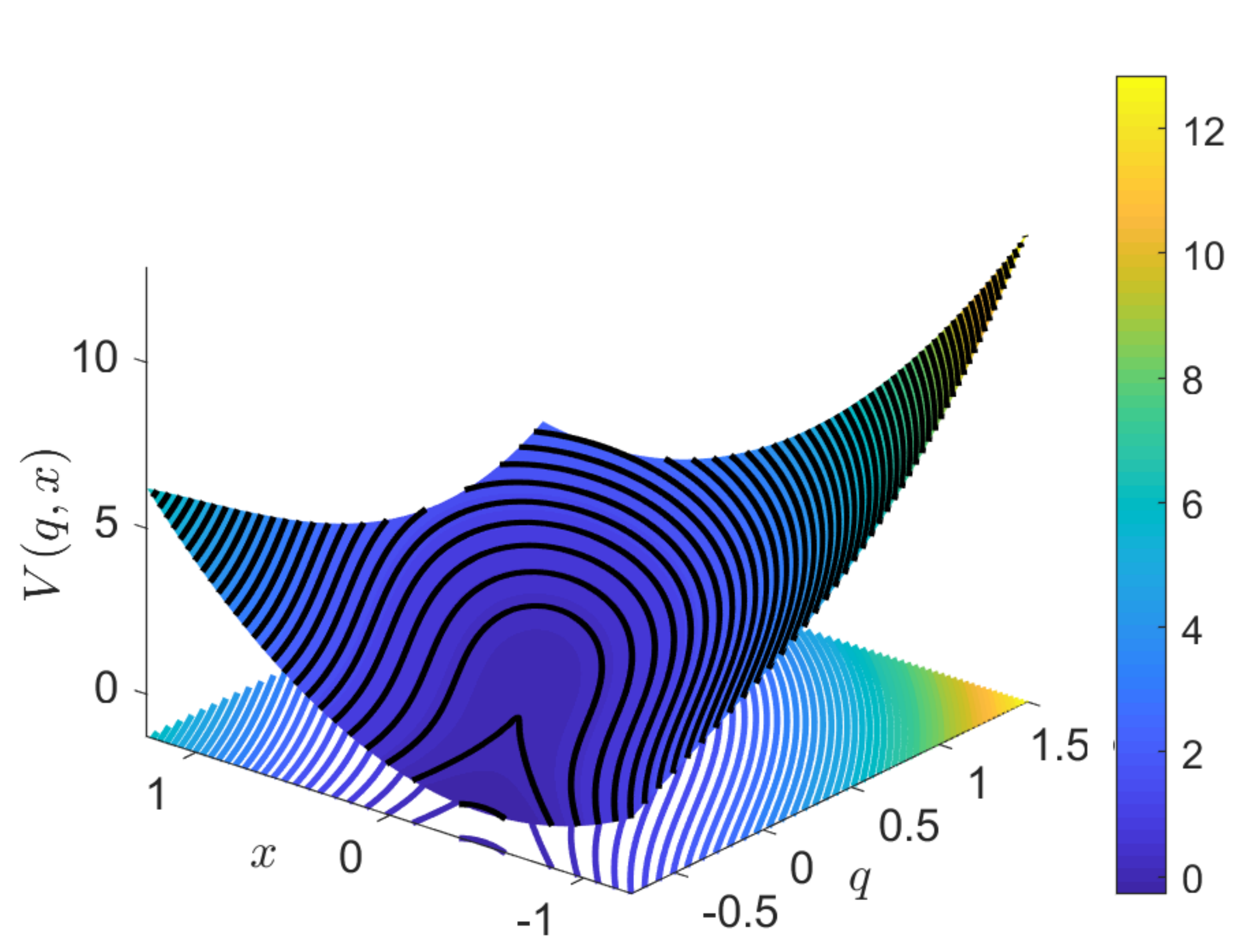}}
		\subfigure[]{\includegraphics[scale=0.29]{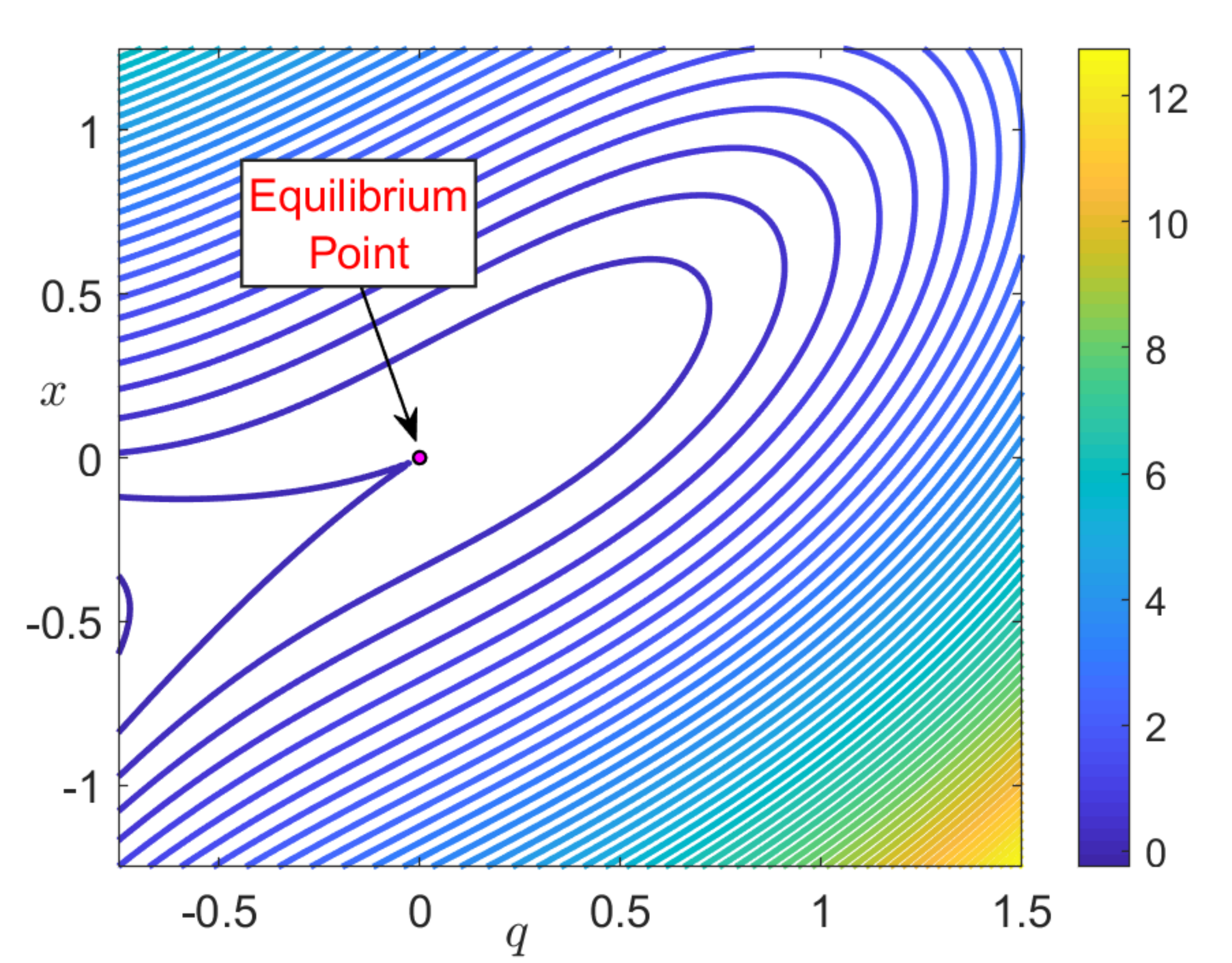}}\\
		\vspace{-2ex}
	\caption{Geometry of the PES (left-column) and equipotentials in configuration space  (right column) for the model parameters $\mu = 0.25$, $\alpha = 2$ and $\omega = 1.25$. (a) and (b) correspond to $\varepsilon = 0$; (c) and (d) represent the case $\varepsilon = 0.25$; (e) and (f) are for the critical coupling strength $\varepsilon = 25/9$.}
	\label{figpes_2dof}
\end{figure}

We illustrate in Fig.~\ref{figpes_2dof} how the geometry of the PES~\eqref{pes_2dof} and the equipotentials in configuration space change as we vary the coupling strength (perturbation parameter) $\varepsilon$. For this visualization, we have used the following values of the model parameters $\mu = 0.25$, $\alpha = 2$, and $\omega = 1.25$. The reason for doing so is that they satisfy the condition $\omega^2 > 2\sqrt{\mu}$ and consequently a critical value of the perturbation parameter exists, given by $\varepsilon_c = 25/9$, at which the two equilibrium points `collide' at the origin resulting in a \textbf{saddle-node bifurcation}. The dynamics after this collision is beyond the scope of this study and therefore we will only focus on the description of the system dynamics for different values of the coupling strength when $\varepsilon < \varepsilon_c$.  We observe in Fig.~\ref{figpes_2dof}(a,b) that the location of the well when the DoF are uncoupled $(\varepsilon = 0)$ lies on the $q$-axis. As $\varepsilon$ is increased, Fig.~\ref{figpes_2dof}(c,d) show that its effect is to \textit{tilt} the PES with respect to the configuration space plane, and the position of the center equilibrium point moves off the $q$-axis towards the origin. Finally, the situation for which the coupling strength reaches the critical value $\varepsilon_c$  is shown in Fig.~\ref{figpes_2dof}(e,f) when the collision has happened and there is only one equilibrium point at the origin.

\paragraph{Linear stability analysis of the equilibrium point at the origin.}
We show here that the equilibrium point $\mathbf{x}_1^e$ is a rank-1 saddle on the PES and that $\mathbf{x}_2^e$ is a center corresponding to the bottom of the potential well. This linear stability analysis will be useful later for computing the rank-1 saddle's NHIM, which carries its influence to higher energies, and also to determine the NHIM's invariant stable and unstable manifolds that act as codimension one impenetrable barriers on the constant energy surface. The global geometry of the invariant manifolds is paramount for quantifying reaction rates such as escape rate from potential well.

We determine the stability of the equilibria by linearizing Eq.~\eqref{hameq_2dof}, which gives the Jacobian matrix
\begin{equation}
\mathbb{J}(q,x,p,p_x) = 
\begin{pmatrix}
\dfrac{\partial^2 H}{\partial q \partial p} & \dfrac{\partial^2 H}{\partial x \partial p} & \dfrac{\partial^2 H}{\partial p^2} & \dfrac{\partial^2 H}{\partial p_x \partial p}  \\[.4cm]
\dfrac{\partial^2 H}{\partial q \partial p_x} & \dfrac{\partial^2 H}{\partial x \partial p_x} & \dfrac{\partial^2 H}{\partial p \partial p_x} & \dfrac{\partial^2 H}{\partial p_x^2}  \\[.4cm]
-\dfrac{\partial^2 H}{\partial q^2} & -\dfrac{\partial^2 H}{\partial x \partial q} & -\dfrac{\partial^2 H}{\partial p \partial q} & -\dfrac{\partial^2 H}{\partial p_x \partial q}  \\[.4cm]
-\dfrac{\partial^2 H}{\partial q \partial x} & -\dfrac{\partial^2 H}{\partial x^2} & -\dfrac{\partial^2 H}{\partial p \partial x} & -\dfrac{\partial^2 H}{\partial p_x \partial x} 
\end{pmatrix} = 
\begin{pmatrix}
0 & 0 & 1 & 0 \\
0 & 0 & 0 & 1 \\
2\sqrt{\mu} - 2\alpha q - \varepsilon & \varepsilon & 0 & 0 \\
\varepsilon & -\omega^2 - \varepsilon & 0 & 0 \\
\end{pmatrix}
\end{equation}
The stability of $\mathbf{x}_1^e = (0,0,0,0) $ is given by the eigenvalues of the Jacobian
\begin{equation}
\mathbb{J}(\mathbf{x}_1^e) = 
\begin{pmatrix}
0 & 0 & 1 & 0 \\
0 & 0 & 0 & 1 \\
2\sqrt{\mu} - \varepsilon & \varepsilon & 0 & 0 \\
\varepsilon & -\omega^2 - \varepsilon & 0 & 0 \\
\end{pmatrix}
\label{eqn:jacobian_2dof}
\end{equation}
which has the characteristic equation
\begin{equation}
\det\left(\mathbb{J}(\mathbf{x}_1^e) - \beta \, \mathbb{I}\right) = \left(\beta^2 - 2\sqrt{\mu} + \varepsilon\right) \left(\beta^2 + \omega^2 + \varepsilon\right) - \varepsilon^2 = 0
\label{char_eq}
\end{equation}
where $\beta$ are the eigenvalues and $\mathbb{I}$ the identity matrix.

When the reaction and bath modes are decoupled, that is $\varepsilon = 0$ in~\eqref{char_eq}, the eigenvalues are given by $\pm \lambda_0$ and $\pm \omega_0 \, i$, where 
\begin{equation}
\lambda_0 = \sqrt[4]{4\mu} \;,\quad \omega_0 = \omega
\end{equation}
Therefore, the origin is a rank-1 saddle equilibrium point, since the linearized system has exactly one pair of real eigenvalues, $\lambda_0$ and $-\lambda_0$ and the saddle plane is spanned by their eigenvectors. We know from the Moser's generalization of the Lyapunov Subcenter Theorem\cite{wiggins2013normally,wiggins2003applied} that when the energy of the system is above that of the rank-1 saddle, there is a two dimensional plane spanned by the eigenvectors of $\pm \omega_0 \, i$, known as the center invariant manifold. This invariant manifold with normal hyperbolicity is referred to as the \textit{normally hyperbolic invariant manifold} (NHIM) and has the topology of a $(2N-3)$-sphere, that is $S^{2N-3}$. In the two DoF setting, a NHIM is simply an unstable periodic orbit whose geometry is topologically equivalent to a circle, i.e. $S^1$. 

As the energy of the system is increased above the energy of the rank-1 saddle equilibrium point, a bottleneck opens in the phase space connecting dynamically different phase space regions and allowing trajectories to move between them. This phenomenon results in a phase space transport mechanism. This framework of understanding chemical reactions is realized by computing the stable and unstable manifolds associated with the unstable periodic orbit. These invariant manifolds have cylindrical geometry, that is $\mathbb{R} \times \S^1$, and their global behavior is referred to as \textit{tube dynamics}. The cylindrical manifolds on the constant energy surface are impenetrable barriers (since they are two dimensional on the three dimensional energy surface) separating the reactive and non-reactive trajectories in the phase space. Thus, they determine the initial conditions that will pass through the bottleneck in some future time (or had passed through it in the past) during their evolution~\cite{wiggins_impenetrable_2001}. The unstable periodic orbit provides us with the scaffolding to construct the dividing surface that separates the trapped motion in the well region of the PES and the escape to infinity of particle trajectories through the phase space bottleneck. The local (linearized) dynamics mediated by these phase space structures is shown in Appendix~\ref{sec:app_viz}. We will resume the discussion on computing the NHIM and its invariant manifolds in \S:\ref{sec:RD_2DOF}.

When $\varepsilon \neq 0$, the reaction and bath modes are coupled and the Hamilton's equations~\eqref{hameq_2dof} are non-integrable. This means we can expect chaotic trajectories to appear that might lead to reaction by escaping the potential well. However, the geometry of the NHIM and its invariant manifolds still governs this reacting and non-reacting behavior. To simplify the eigenvalue problem arising in the stability analysis, we observe that Eq.~\eqref{char_eq} can be rewritten in terms of the eigenvalues of the uncoupled system as
\begin{equation}
\left(\beta^2 - \lambda_0^2 + \varepsilon\right) \left(\beta^2 + \omega_0^2 + \varepsilon\right) - \varepsilon^2 = \beta^4 + \left(\omega_0^2 - \lambda_0^2 + 2\varepsilon\right) \beta^2 + \left( \omega_0^2 - \lambda_0^2\right)\varepsilon - \lambda_0^2 \, \omega_0^2 = 0
\end{equation}
Introducing $\xi = \beta^2$ the solutions are
\begin{equation}
\xi =\dfrac{\lambda_0^2 - \omega_0^2}{2} - \varepsilon \pm \sqrt{\left(\dfrac{\lambda_0^2 + \omega_0^2}{2}\right)^2 + \varepsilon^2} \;\;.
\label{eqn:eig_ham_2dof}
\end{equation}
We note here that the two possible values of $\xi$ have opposite signs for all values $\varepsilon > 0$ of the coupling strength when $\omega_0^2 \leq \lambda_0^2$, and for $0 < \varepsilon < \varepsilon_c$ whenever the condition $\omega_0^2 > \lambda_0^2$ is satisfied, where the critical coupling strength $\varepsilon_c$ is given by Eq. \eqref{crit_epsi}. If we denote $\xi_{1,\varepsilon}$ and $\xi_{2,\varepsilon}$ as the positive and negative roots respectively in Eq. \eqref{eqn:eig_ham_2dof}, then the eigenvalues of the Jacobian matrix are $\pm \lambda_\varepsilon$ and $\pm \omega_\varepsilon i$ where
\begin{equation}
\lambda_\varepsilon = \sqrt{\xi_{1,\varepsilon}} \;,\quad \omega_\varepsilon = \sqrt{|\xi_{2,\varepsilon}|}
\label{eigenvalues_2dof}
\end{equation}
This shows that in this case the equilibrium point at the origin is a rank-1 saddle. In particular, when the coupling between the reaction and the bath mode is weak, that is $\varepsilon \ll 1$, we have 
\begin{equation}
\lambda_\varepsilon \approx \sqrt{\lambda_0^2 - \varepsilon} \quad,\quad \omega_\varepsilon \approx \sqrt{\omega_0^2 + \varepsilon}
\end{equation}
which confirms that the rank-1 saddle structure of the equilibrium point at the origin persists under small perturbations.

We finish the linear stability analysis of the equilibrium point at the origin with the computation of the eigenvectors. Let us denote the eigenvector of the matrix $J(\mathbf{x}_1^e)$ by $\mathbf{v} = \left[v_1,v_2,v_3,v_4\right]^T$, then the eigenvalue $\beta$ satisfies
\begin{equation}
J(\mathbf{x}_1^e) \mathbf{v} = \beta \mathbf{v} \quad\Leftrightarrow\quad 
\begin{cases}
\beta v_1 = v_3 \\[-.1cm]
\beta v_2 = v_4 \\[-.1cm]
\beta v_3 = \left(\lambda_0^2 - \varepsilon\right) v_1 + \varepsilon v_2 \\[-.1cm]
\beta v_4 = \varepsilon v_1 - \left(\omega_0^2 + \varepsilon\right) v_2
\end{cases}
\end{equation}
The eigenvectors associated with the real eigenvalues $\beta = \pm \lambda_{\varepsilon}$ correspond to the tangent directions of the unstable and stable manifolds, respectively, of the NHIM at the rank-1 saddle, and are given by
\begin{equation}
\mathbf{u}_{\pm} = \left[1,\frac{\varepsilon}{\varepsilon + \left(\lambda^2_\varepsilon + \omega^2_0\right)},\pm\lambda_{\varepsilon},\pm\frac{\lambda_{\varepsilon} \, \varepsilon}{\varepsilon + \left(\lambda^2_\varepsilon + \omega^2_0\right)}\right]^T.
\label{saddle_eigenv}
\end{equation}
The eigenvectors associated with the complex eigenvalues $\beta = \pm \omega_\varepsilon i$ span the center subspace of the NHIM at the rank-1 saddle and are given by
\begin{equation}
\mathbf{w}_{\pm} = \left[\frac{\varepsilon}{\varepsilon - \left(\lambda^2_0 + \omega^2_\varepsilon\right)},1,\pm\frac{\omega_{\varepsilon} \, \varepsilon}{\varepsilon - \left(\lambda^2_0 + \omega^2_\varepsilon\right)}\, i,\pm\omega_{\varepsilon}\, i\right]^T.
\label{center_eigenv}
\end{equation}
where $\omega_\varepsilon$ is the magnitude of the complex eigenvalue in~\eqref{eigenvalues_2dof}. Therefore, we can write the solution to the linearized system at the rank-1 saddle as
\begin{equation}
\mathbf{x}(t) = C_1 e^{\lambda_\varepsilon t} \mathbf{u}_{+} + C_2 e^{-\lambda_\varepsilon t} \mathbf{u}_{-} + 2 Re\left(\eta e^{i\omega_\varepsilon t} \mathbf{w}_{+}\right)
\label{geneq_lin_ham}
\end{equation}
where $C_1,C_2 \in \mathbb{R}$ and $\eta = \eta_1 + \eta_2 \, i \in \mathbb{C}$ are constants to be determined from an initial condition. We will use this general solution to the linear system to select an initial guess to search for the NHIM using the differential correction and numerical continuation method~\cite{Koon2011}. 

For the sake of completeness, we show next the linear stability analysis for the equilibrium point $\mathbf{x}_2^e$. The Jacobian is given by
\begin{equation}
\mathbb{J}(\mathbf{x}_2^e) = 
\begin{pmatrix}
0 & 0 & 1 & 0 \\
0 & 0 & 0 & 1 \\
\gamma -2\sqrt{\mu} - \varepsilon & \varepsilon & 0 & 0 \\
\varepsilon & -\omega^2 - \varepsilon & 0 & 0 \\
\end{pmatrix} \quad,\quad \text{where }\; \gamma = \dfrac{2\omega^2\varepsilon}{\omega^2 + \varepsilon} \label{jacobian_center}
\end{equation}
The characteristic equation $\det\left(J(\mathbf{x}_2^e) - \beta \, \mathbb{I}\right) = 0$ becomes
\begin{equation}
\begin{aligned}
\left(\beta^2 + \lambda_0^2 - \gamma + \varepsilon \right) \left(\beta^2 + \omega_0^2 + \varepsilon\right) - \varepsilon^2 & = 0 \\ 
\beta^4 + \left(\omega_0^2 + \lambda_0^2 - \gamma + 2\varepsilon\right)\beta^2 + \left(\omega_0^2 + \lambda_0^2 - \gamma\right)\varepsilon + \omega_0^2\left(\lambda_0^2 - \gamma\right) & = 0
\label{char_eq_center}
\end{aligned}
\end{equation}
Introducing $\xi = \beta^2$ the solutions are
\begin{equation}
\xi = \dfrac{\gamma - \left(\lambda_0^2 + \omega_0^2\right)}{2} - \varepsilon \pm \sqrt{\left(\dfrac{\gamma + \omega_0^2 - \lambda_0^2}{2}\right)^2 + \varepsilon^2}
\end{equation}
It can be easily shown that both solutions are negative, and consequently $\mathbb{J}(\mathbf{x}_2^e)$ has two pairs of complex eigenvalues. This implies that the equilibrium point $\mathbf{x}_2^e$ is a center.

\section{Results and Discussion: Influence of well-depth on reaction dynamics}
\label{sec:RD}

In this section we discuss the implications of changing the well-depth in terms of the geometry of the invariant manifolds that mediate reacting trajectories (or escape from the potential well). This is done by identifying the invariant manifolds using Lagrangian descriptors and a numerical continuation with globalization method (see Appendix~\ref{sec:appB}). The geometry of these invariant manifolds and character of changes affected by varying the potential well-depth are involved in the computation of reaction fraction (or escape rates). 

\subsection{One Degree-of-Freedom Hamiltonian}
\label{sec:RD_1DOF}

Here we return to the 1 DoF Hamiltonian~\eqref{hameq_1dof_gen} where the {\it reaction} is defined as the change in the sign of the configuration coordinate $q$. 
For the rank-1 saddle equilibrium point at $(0,0)$, the energy is $H(0,0) = H_c = 0$ which we call the \textit{critical energy}, and that of the center equilibrium point at $(2\sqrt{\mu}/\alpha,0)$, at the bottom of the potential well, is $H(2\sqrt{\mu}/\alpha,0) = H_w = - 4 \sqrt{\mu^3}/3 \alpha^2$. In this model system, the phase space is two-dimensional and since the total energy is conserved, the trajectories evolve on the one-dimensional constant energy lines or isoenergetic contours of the Hamiltonian in the 2D phase space. Therefore, the system is completely integrable, and for a fixed energy $H_0$ the trajectories evolve on the one-dimensional curve
\begin{equation}
H_0 = \frac{1}{2} \, p^2 - \sqrt{\mu} \, q^2 + \frac{\alpha}{3} q^3
\end{equation}
The value of total energy $H_0$ with respect to $H_c$ and $H_w$ characterizes the nature of trajectories as follows
We discuss the nature of the trajectories depending on the energy $H_0$ of the system:

\begin{itemize}
\item \textbf{Case $H_0 < H_w$:} The initial conditions that satisfy $q < -\sqrt{\mu}/\alpha$, which lie to the left of the potential energy barrier at the origin. Initial conditions in this configuration space initially climb the potential and at some point their velocity reverses direction and they roll down the potential going to infinity. In the phase space, they come from infinity and fly-by the rank-1 saddle on their way towards infinity. They can not cross the barrier and thus do not lead to reaction shown by the blue curve in Fig.~\ref{fig:saddlenodeham1dofparamset1}.

\item \textbf{Case $H_w \leq H_0 < 0$:} Now two types of trajectories are possible. For the initial conditions that satisfy $-\sqrt{\mu}/\alpha \leq q < 0$, trajectories will show same fly-by behavior as in the previous case. If initial conditions satisfy $0 < q_0 < 3 \sqrt{\mu}/\alpha$, then they are confined in the potential well on the right of the potential barrier. Since they are bounded by the homoclinic orbit, trajectories will never escape the well, and thus do not lead to reaction and shown as green curves in Fig.~\ref{fig:saddlenodeham1dofparamset1}. 

\begin{figure}[htbp]
	\centering
	\subfigure[]{\includegraphics[width=0.42\textwidth]{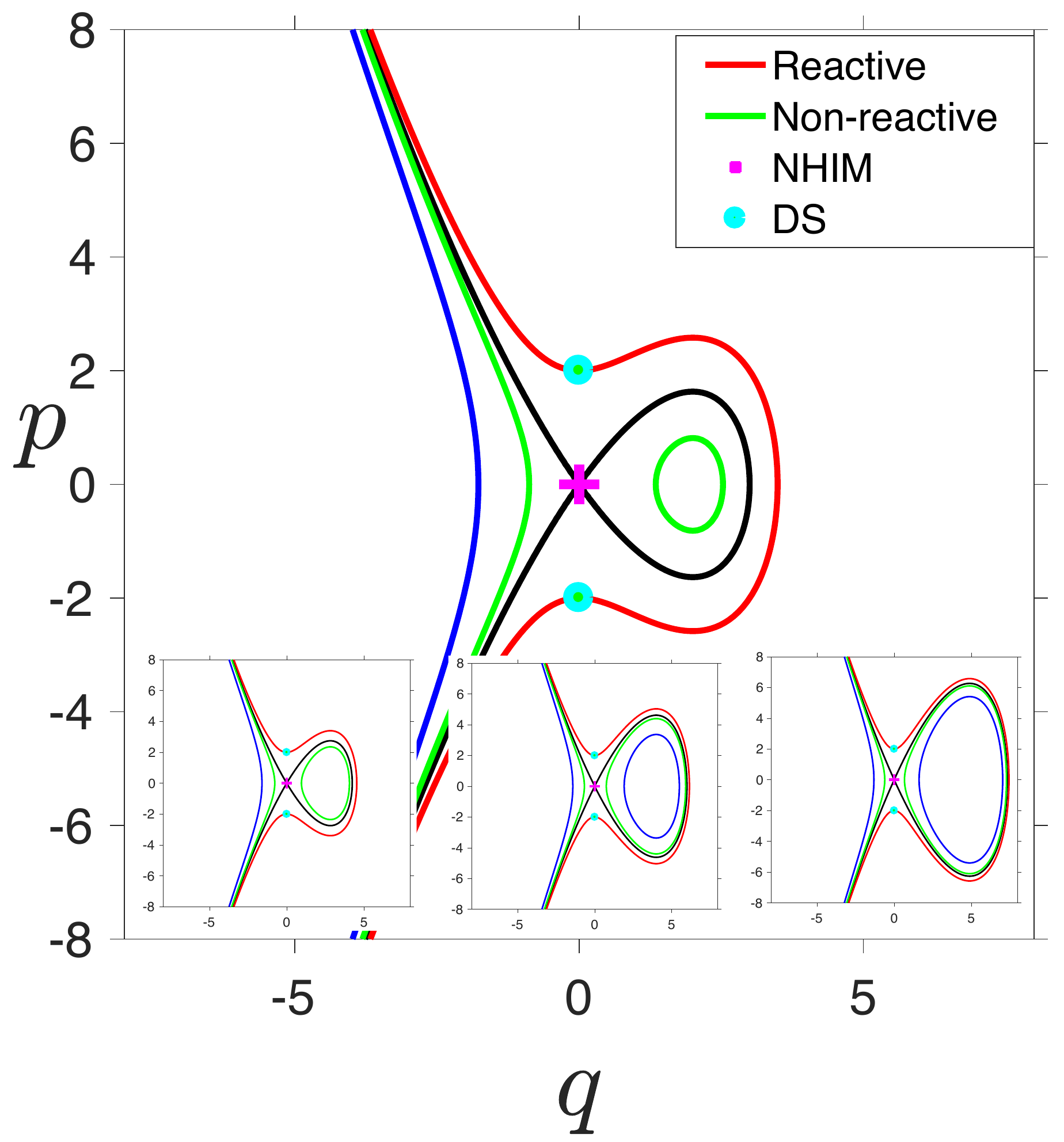}}
	\subfigure[]{\includegraphics[width=0.42\textwidth]{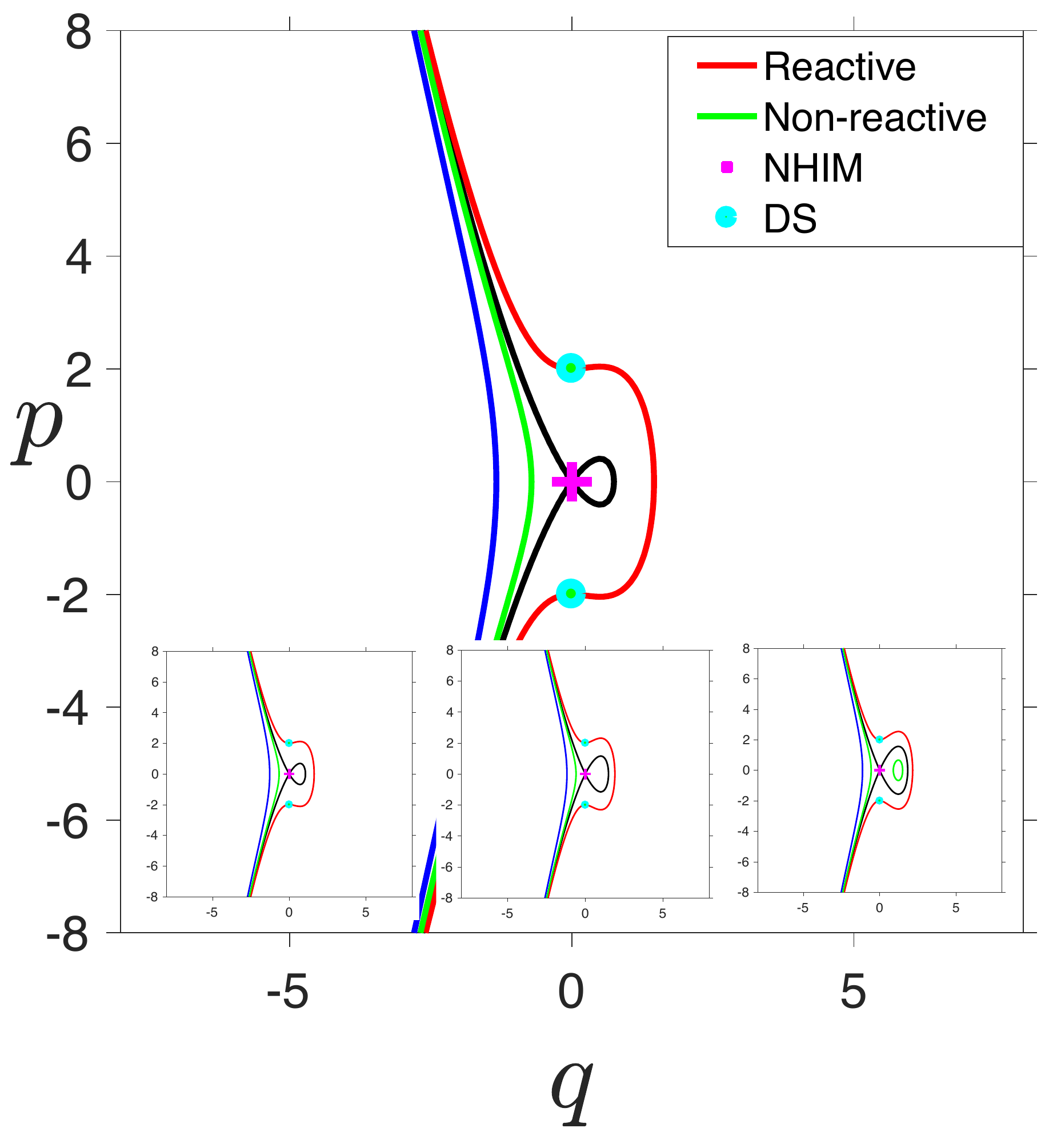}}
	\vspace{-2ex}
	\caption{Trajectory behaviors dictated by the potential energy~\eqref{pe_1dof}. (a) $\mu = 1$ and $\alpha = 1$ with the insets $\mu = 2,4,6$ for $\alpha = 1$ (b) $\mu = 1$ and $\alpha = 4$ with the insets $\mu = 2,4,6$ for $\alpha = 4$. Here we are showing the reactive and nonreactive trajectories (red and blue, respectively) partitioned by the homoclinic orbit (shown in black) formed by the unstable and stable manifolds of the NHIM (shown as a magenta cross), and the reactive trajectory (red) is at energy $H_0 = 2$, while the two non-reactive trajectories are at $H_0 = -1 , -5$. The dividing surface that the reactive trajectory must cross in order for reaction to occur is marked by green asterisks.}
	\label{fig:saddlenodeham1dofparamset1}
\end{figure}

\item \textbf{Case $H_0 = H_c = 0$:} When the total energy of the initial conditions is equal to the potential energy of the barrier, resulting trajectories approach the barrier asymptotically in forward and backward time. The initial conditions that start on the left of the potential barrier and asymptotically approach it in forward and backward time form pieces of the stable and unstable manifolds of the saddle equilibrium point. The initial conditions that start on the right of the potential barrier also approach the barrier as time goes to infinity. These trajectories combine to form the homoclinic orbit and shown as a black curves in Fig.~\ref{fig:saddlenodeham1dofparamset1}.  

\item \textbf{Case $H_0 > H_c = 0$:} The energy of the system is above that of the barrier so trajectories can escape from the well, that is the configuration space coordinate $q$ can change sign and lead to reaction. The dividing surface that trajectories must cross once in a given direction\cite{wiggins2016} (locally no recrossing), that is when escaping from the potential well, is given by 
\begin{equation}
\mathcal{D} = \left\lbrace (q, p) \in \mathbb{R}^2 \; | \; q = 0 \;,\; \frac{p^2}{2}  = H_0 > 0  \right\rbrace = \left\lbrace \left(0, \pm \sqrt{2H_0}\right) \right\rbrace \;. \label{eqn:snham_1dof_ds}
\end{equation}
We note that in this case the dividing surface for a fixed energy and given direction of crossing is a point, or has a geometry of $\mathbb{S}^0$, thus it can partition the one-dimensional (isoenergetic) curves into ``reactant'' and ```product'' regions. To further characterize the reaction dynamics, we resort to the linear stability analysis near the saddle equilibrium point at the origin. The eigenvalues of the saddle equilibrium point are $\pm \sqrt[4]{4\mu}$ and the corresponding eigenvectors $\left(\pm 1/ \sqrt[4]{4\mu},1\right)$. Therefore, the origin is a NHIM since the dynamics is hyperbolic in directions normal to it~\cite{wig2016}. Furthermore, we remark that the stability of the NHIM only depends on the $\mu$ parameter. The eigenvectors tell us how the unstable and stable manifolds of the NHIM are oriented and can be used to numerically globalize the linear approximation. However, we can compute the stable and unstable manifolds of the saddle equilibrium point analytically by noting that they lie on the zero level curve of the Hamiltonian (total energy). Thus, the stable and unstable manifolds are given by
\begin{equation}
\mathcal{W}^s(0,0) = \Gamma \cup \mathcal{W}^s_l(0,0) \;,\; \; \mathcal{W}^u(0,0) = \Gamma \cup \mathcal{W}^u_l(0,0)
\end{equation}
where $\Gamma$ is the homoclinic orbit
\begin{equation}
\Gamma = \left\lbrace  (q, p) \in \mathbb{R}^2 \; | \; q > 0 \;,\; H(q,p) = 0 \right\rbrace
\end{equation}
and
\begin{equation}
\begin{split}
\mathcal{W}^s_l(0,0) &= \left\lbrace  (q, p) \in \mathbb{R}^2 \; | \; q < 0 \:,\; p > 0 \;,\; H(q,p) = 0 \right\rbrace \\
\mathcal{W}^u_l(0,0) &= \left\lbrace  (q, p) \in \mathbb{R}^2 \; | \; q < 0 \:,\; p < 0 \;,\; H(q,p) = 0 \right\rbrace
\end{split}
\end{equation}
We note here that the geometry of the invariant manifolds is $\mathbb{R} \times \mathbb{S}^0$, that is codimension-1 in $\mathbb{R}^2$, and hence can form the impenetrable barrier between the reactive and non-reactive trajectories as shown in Fig.~\ref{fig:saddlenodeham1dofparamset1}.
\end{itemize}

\subsection{Two degree-of-freedom Hamiltonian}
\label{sec:RD_2DOF}

We return to the 2 DoF Hamiltonian~\eqref{ham_2dof} which introduces the bath degree-of-freedom as a harmonic oscillator into the reaction dynamics. We show the changes in the geometry of the invariant manifolds due to the changes in the well-depth which leads to a saddle-node bifurcation. This is done using Lagrangian descriptors along with numerical continuation and globalization for computing the NHIM and its invariant manifolds.

Let us consider a fixed energy $H_0$ and since the model has 2 DoF we know that the dynamics is on a three-dimensional energy surface given by
\begin{equation}
\begin{split}
\mathcal{S}(H_0) &= \left\lbrace (q,x,p,p_x) \in \mathbb{R}^4 \; \big|\; \dfrac{1}{2} \left(p^2 + p_x^2 \right) - \sqrt{\mu} \, q^2 + \frac{\alpha}{3} \,q^3 + \dfrac{\omega^2}{2} x^2 + \dfrac{\varepsilon}{2} \left(x-q\right)^2 = H_0 \right\rbrace
\end{split}
\end{equation}

\begin{figure}[!ht]
	\begin{center}		
	\subfigure[]{ \includegraphics[scale=0.33]{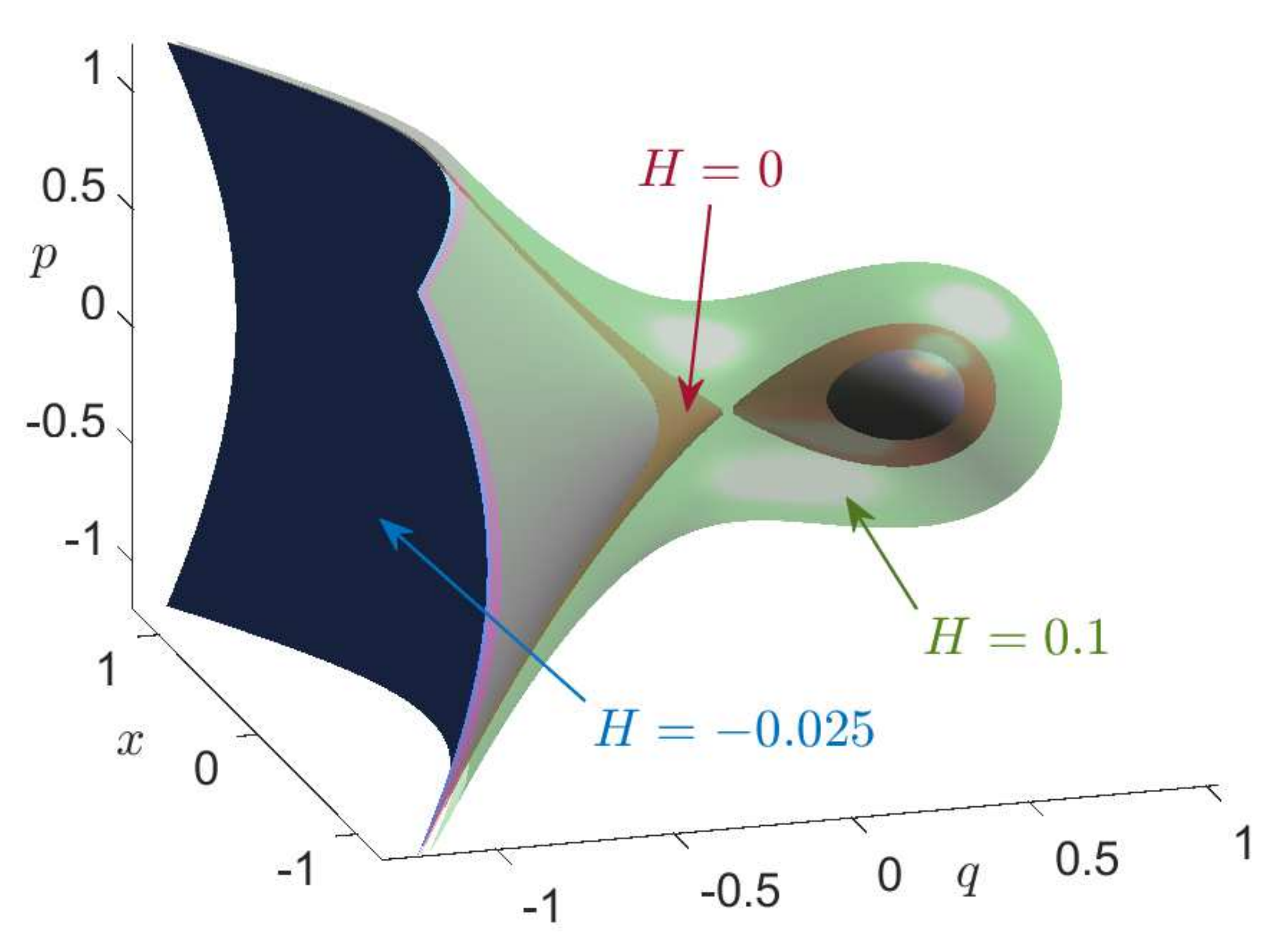}}
	\subfigure[]{ \includegraphics[scale=0.31]{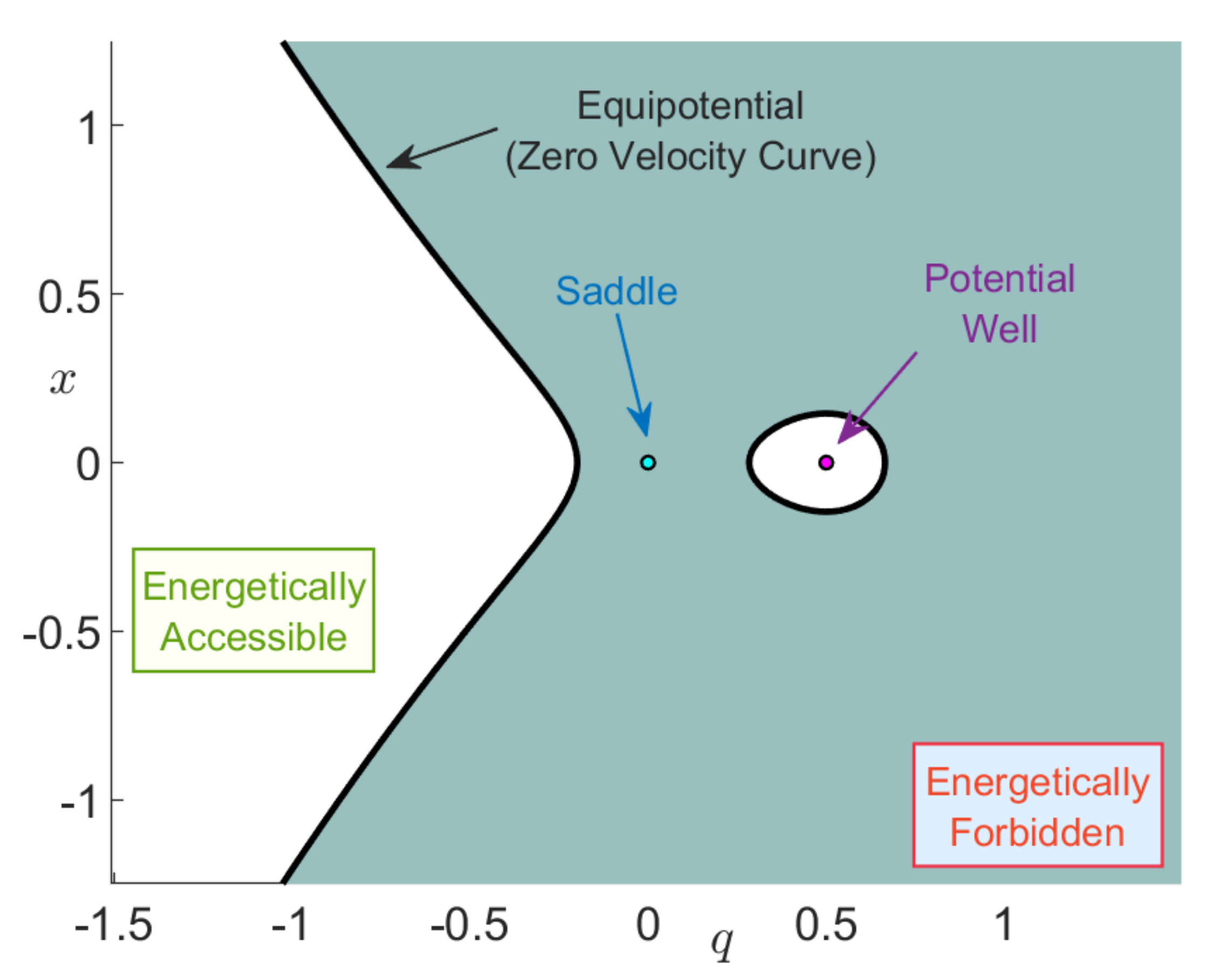}}
	\subfigure[]{ \includegraphics[scale=0.33]{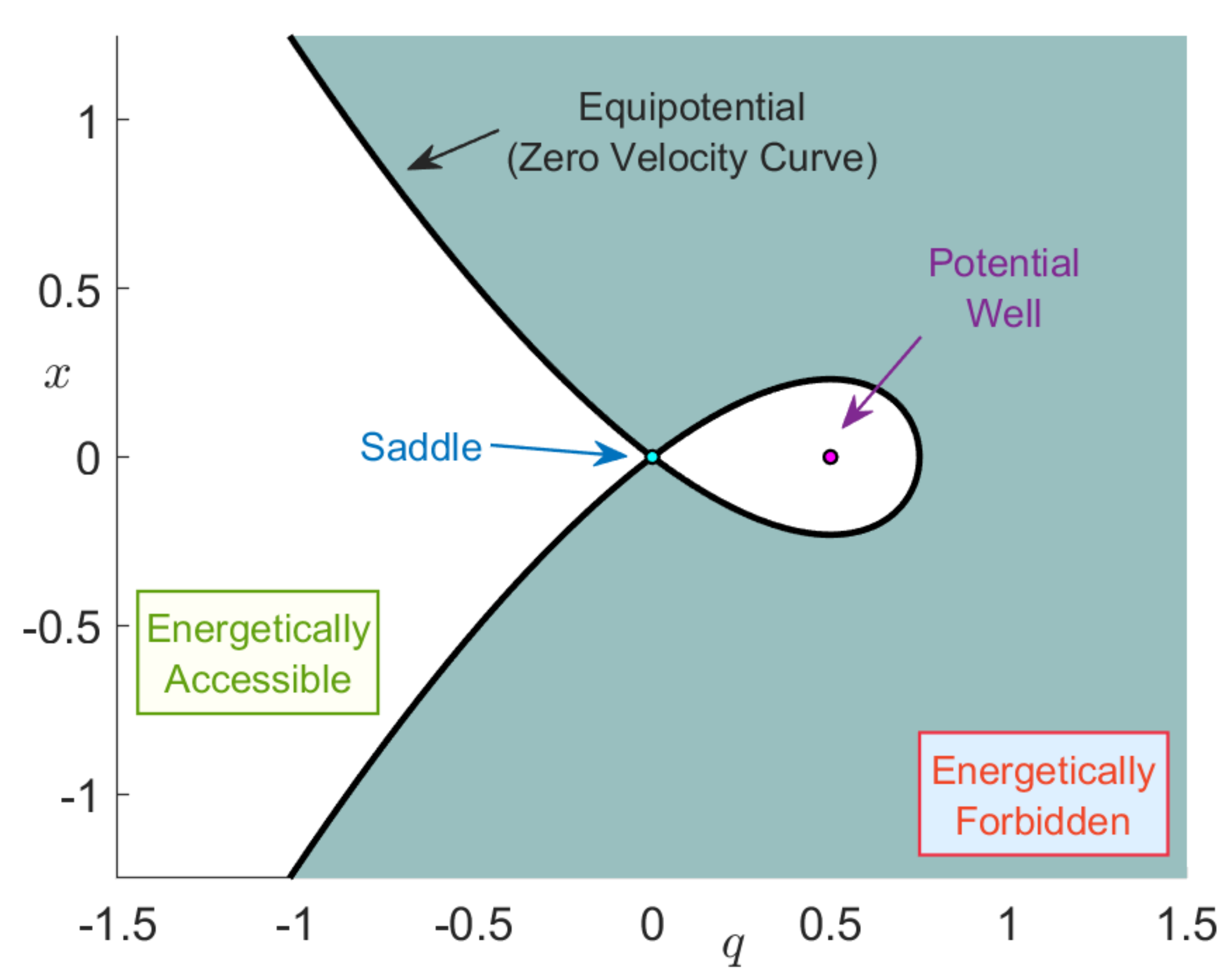}}
	\subfigure[]{ \includegraphics[scale=0.32]{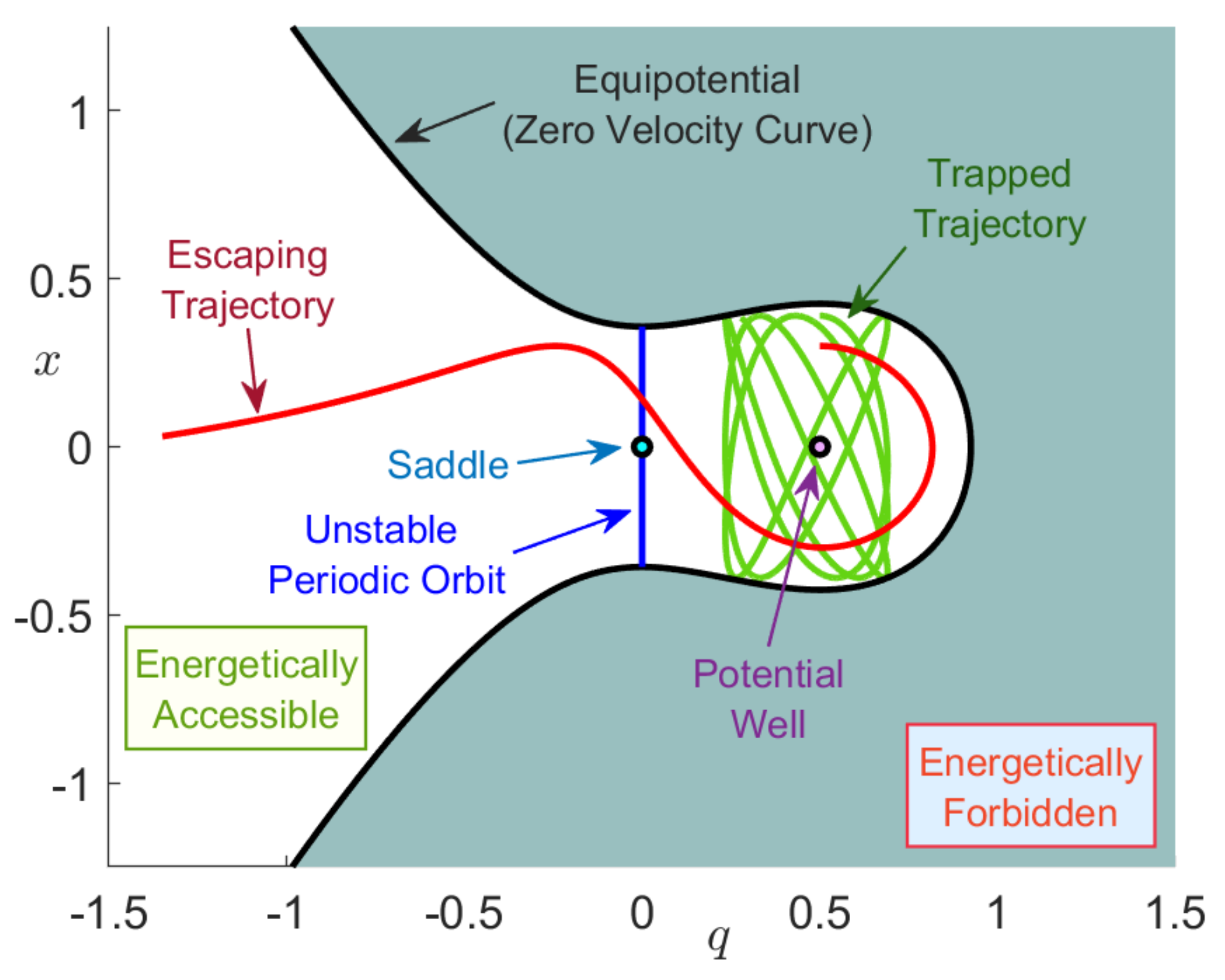}}
	\end{center}
	\vspace{-2ex}
	\caption{(a) Phase space energy surface, inside which motion takes place, for three different values of the energy; (b) Configuration space projection for an energy $H = -0.025$ below the barrier energy; (c) for the rank-1 saddle energy $H = 0$; (d) for the energy $H = 0.1$ which is above the barrier energy. The parameter values chosen are $\mu = 0.25$, $\alpha = 2$, $\omega = 1.25$ and $\varepsilon = 0$. We have marked the energy accessible regions in white and the forbidden regions in dark grey. The equipotential in black marks the zero velocity curve for which the kinetic energy of the system is zero.}
	\label{EnergySurf_Hills}
\end{figure}

The projection of the energy surface onto the $(q,x)$ configuration space 
is the region of energetically possible motion for a fixed energy $H_0$, and is given  by
\begin{equation}
\begin{split}
\mathcal{C}(H_0) &= \left\lbrace (q,x) \in \mathbb{R}^2 \; \big|\; V(q,x) \leqslant H_0 \right\rbrace \\ 
&= \left\lbrace (q,x) \in \mathbb{R}^2 \; \big|\; - \sqrt{\mu} \, q^2 + \frac{\alpha}{3} \,q^3 + \dfrac{\omega^2}{2} x^2 + \dfrac{\varepsilon}{2} \left(x-q\right)^2 \leqslant H_0 \right\rbrace
\end{split}
\label{eqn:hillsreg}
\end{equation}
This projection denotes configurations with positive kinetic energy and has been known in classical mechanics as the \textit{Hill's region}. The boundary of $M(H_0)$ is defined as the locus of points in the $(q,x)$ plane where the kinetic energy is zero, that is $(H_0 - V(x,y)) = 0 $, and is called the \textit{zero velocity curve}. The trajectories are only able to move on the side of the curve where the kinetic energy is positive, shown as white regions in Fig.~\ref{EnergySurf_Hills}(b-d). 

To identify the invariant manifolds, we take two dimensional slices of the energy surface and determine the intersection of the invariant manifolds with these low-dimensional slices. In particular, we calculate Lagrangian desriptor and compare with qualitative understandings from Poincar\'e surface-of-section. The isoenergetic surfaces-of-section are
\begin{align}
\mathcal{U}_{qp}^{+} &= \left\lbrace (q,x,p,p_x) \in \mathbb{R}^4 \; \big| \; x = 0 \; ,\; p_x(q,x,p;H_0) \geq 0 \right\rbrace \label{sos_qp}\\ 
\mathcal{U}_{xp_x}^{+} &= \left\lbrace (q,x,p,p_x) \in \mathbb{R}^4 \; \big| \; q = q_e \; , \; p(q,x,p_x;H_0) \geq 0 \right\rbrace \label{sos_xpx}
\end{align}
where $q_e$ is the configuration space coordinate of the equilibrium point at the bottom of the well on the PES~\eqref{cSpace_eqCoords}.

We return to the Hamiltonian~\eqref{ham_2dof} where the ``reaction'' and ``bath'' DoF are uncoupled, that is $\varepsilon = 0$. Therefore, the system is integrable and the trajectories are regular, thus the separable Hamiltonian is
\begin{equation}
H(q,x,p,p_x) = \underbrace{\frac{1}{2} p^2 - \sqrt{\mu} q^2 + \frac{\alpha}{3} q^3}_{H_r(q,p)} + \underbrace{\frac{1}{2} \, p_x^2 + \frac{\omega^2}{2} x^2}_{H_b(x,p_x)}
\end{equation}
where $H_r$ is the Hamiltonian for the reaction and $H_b$ is the Hamiltonian for the bath DoF.

\begin{figure}[!ht]
	\centering
	\subfigure{A) \includegraphics[scale=0.37]{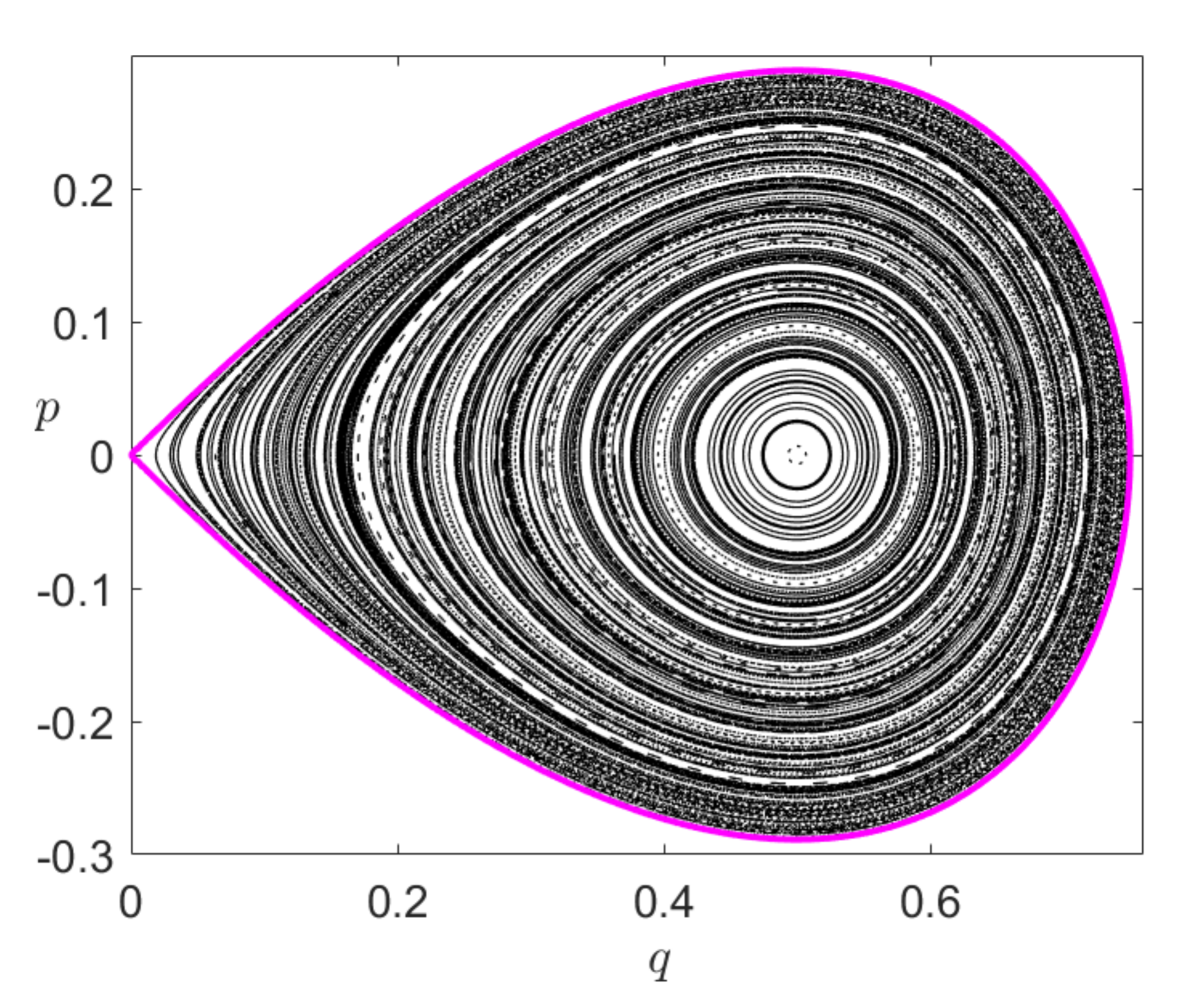}}
	\subfigure{B) \includegraphics[scale=0.375]{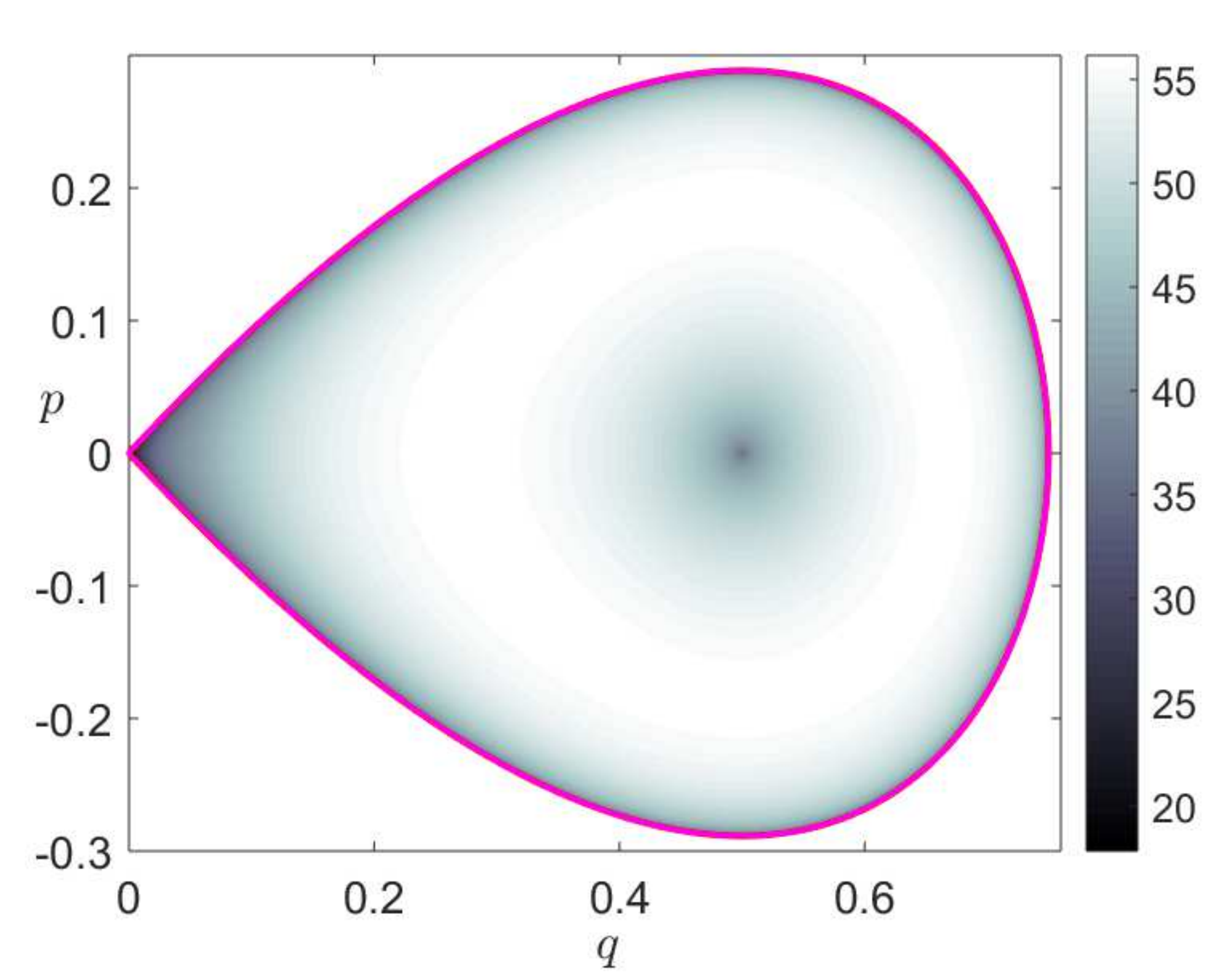}}
	\vspace{-3ex}
	\caption{Quasiperiodic trajectories describing regular motion in the potential well region of the PES for the uncoupled ($\varepsilon = 0$) Hamiltonian system with energy $H_0 = 0$. The model parameters chosen for this calculation are $\mu = 0.25$, $\alpha = 2$ and $\omega = 1.25$; A) Poincar\'e map; B) LDs obtained for an integration time $\tau = 20$. The curve in magenta depicts the energy boundary.}
	\label{LD_PS_uncoupled}
\end{figure}

Given a fixed total energy of the system $H_0$, the \textit{necessary condition} for the reaction to take place is when the total energy is above that of the barrier of the PES located at the origin, which is zero. For $H_0 \leq 0$ the energy surface divides phase space into two disconnected regions as illustrated in Fig.~\ref{EnergySurf_Hills} so that we have bounded motion in the potential well region. In Fig.~\ref{LD_PS_uncoupled}, we compare the bounded trajectories (regular dynamics) of the system for $H_0 = 0$ by computing LDs and Poincar\'e section on the surface-of-section, $\mathcal{U}_{qp}^{+}$~\eqref{sos_qp}. Both methods clearly recover, the trajectories on tori, as known for integrable Hamiltonian systems~\cite{Meyer2009}, that fills the energy surface. We observe that the high values (white regions in~\ref{LD_PS_uncoupled}) in the LD contour map recovers the quasiperiodic trajectories, which is a consequence of the relationship between the convergence of time averages of LDs with the Ergodic Partition Theorem as explained in Appendix~\ref{sec:appA}. 

\begin{figure}[!ht]
	\centering
	\subfigure[]{\includegraphics[width=0.48\linewidth]{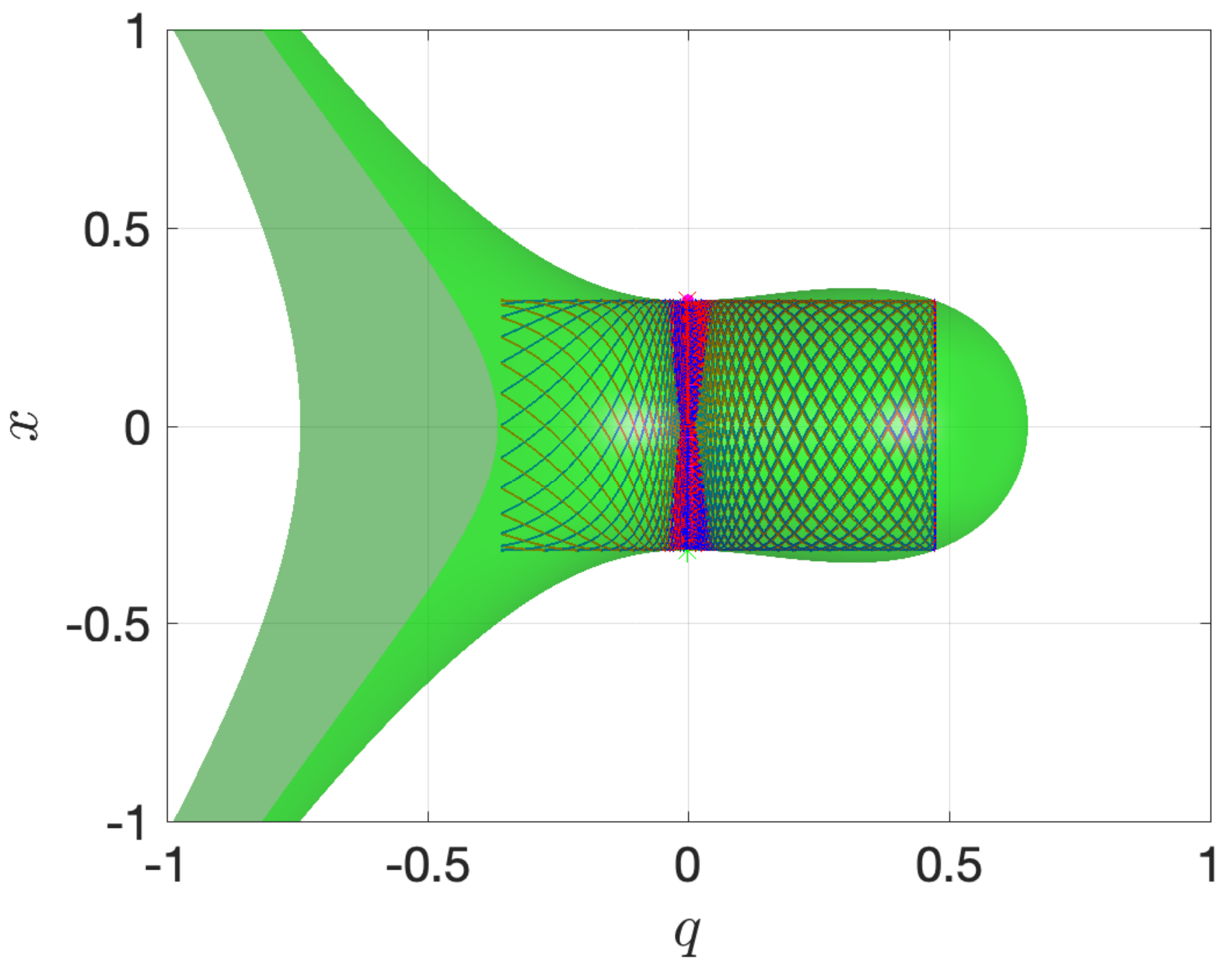}}
	\subfigure[]{\includegraphics[width=0.48\linewidth]{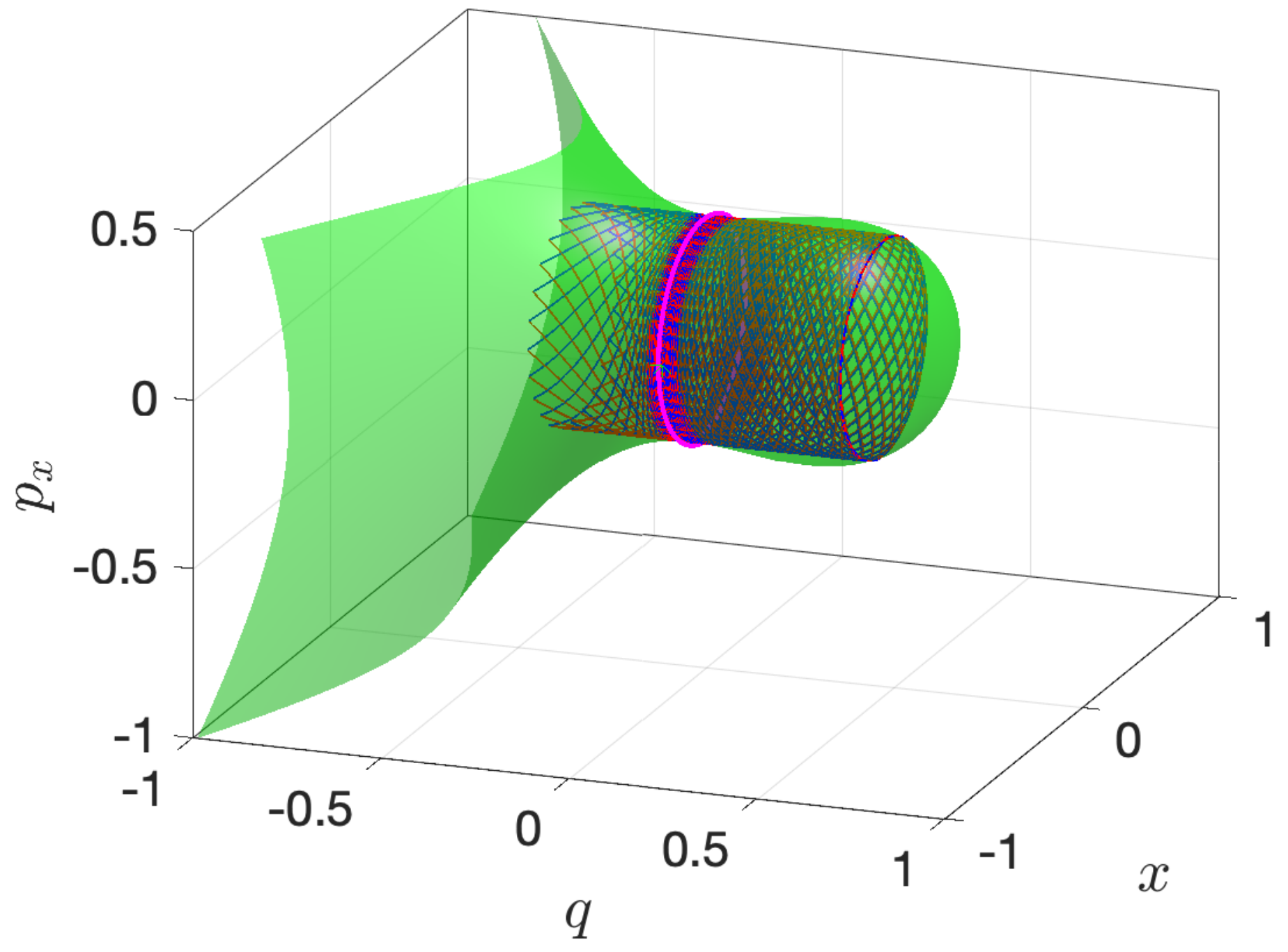}}
	\subfigure[]{\includegraphics[width=0.45\linewidth]{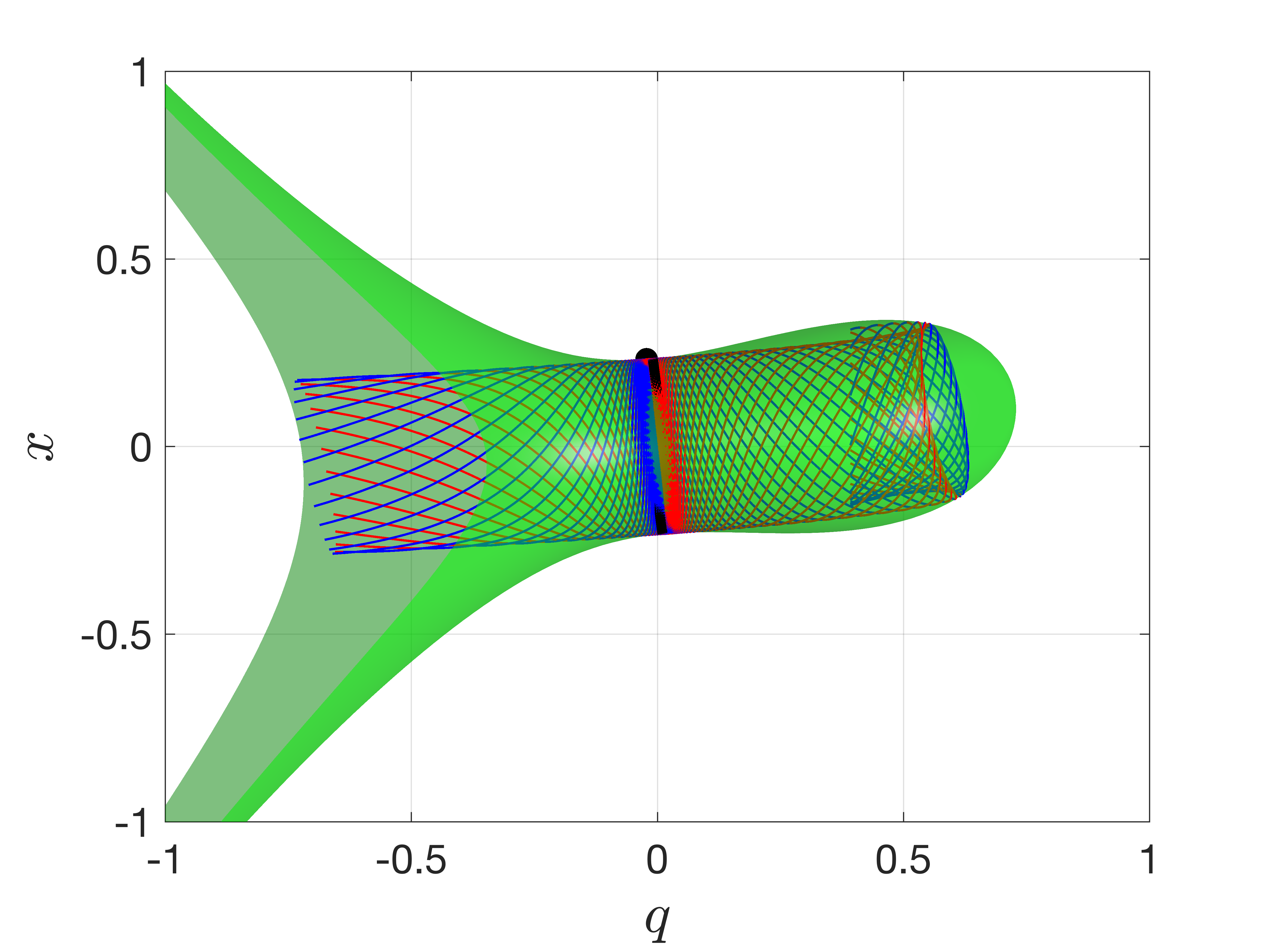}}
	\subfigure[]{\includegraphics[width=0.45\linewidth]{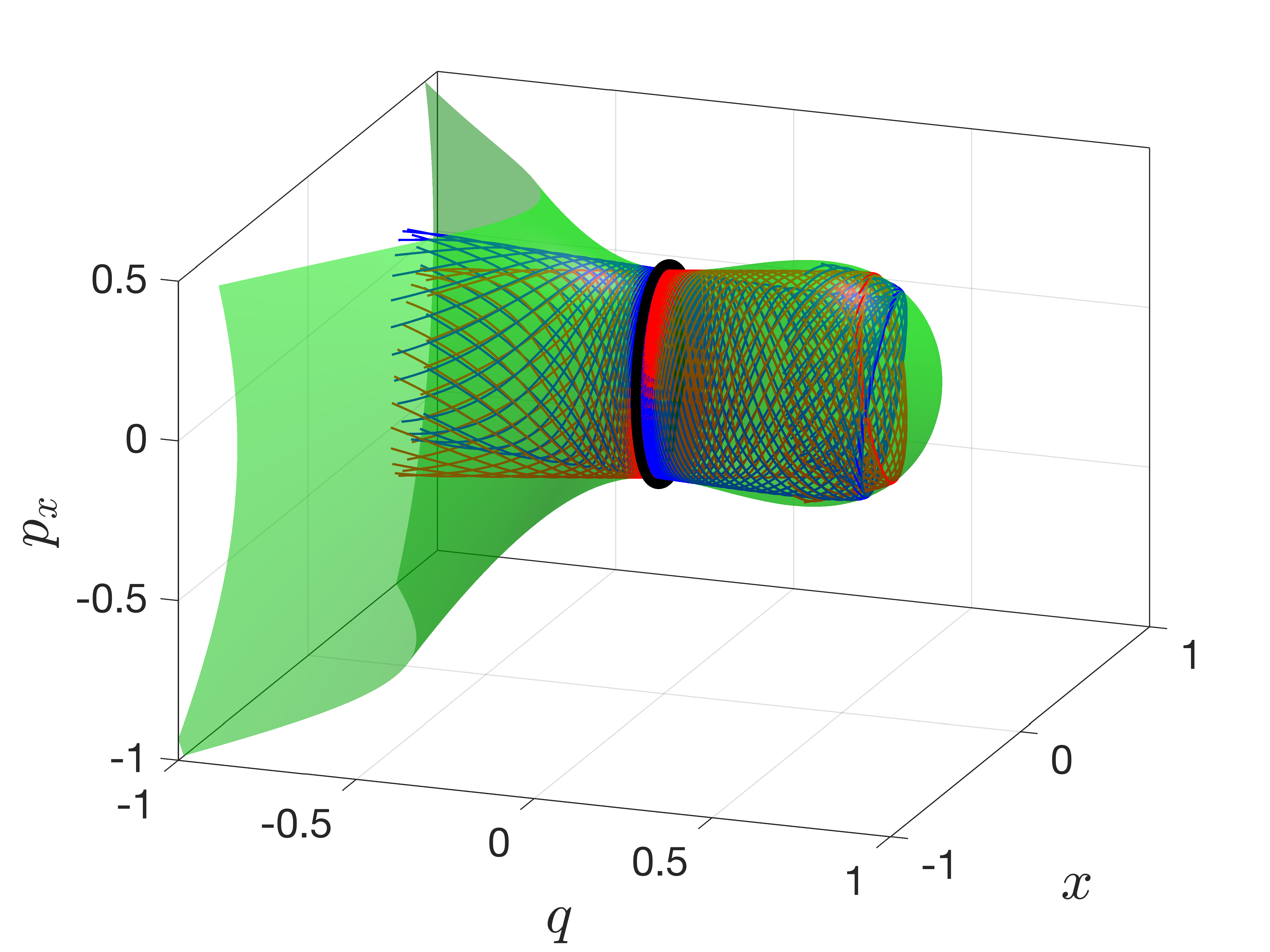}}
	\caption{Configuration space and phase space view of the energy surface and the cylindrical manifolds of the unstable periodic orbit associated with the rank-1 saddle in the bottleneck. Blue and red denote the stable and unstable manifolds respectively, and the green surface is the isosurface of total energy $H_0 = 0.05$. The magenta curve represents the unstable periodic orbit. (a-b) $\epsilon = 0$ and (c-d) $\epsilon = 0.25$.}
	\label{fig:set2_manifolds_upo_energysurf}
\end{figure}

When the energy of the system is above the barrier, that is $H_0 > 0$, the topology of the energy surface changes and a phase space bottleneck opens up in the barrier region allowing the escape from the well as shown in Fig.~\ref{EnergySurf_Hills}(d). Thus, reaction can take place by crossing the bottleneck and passing from the $q > 0$ to $q < 0$ or vice versa.	Therefore, $q=0$ is a natural choice for defining a dividing surface (DS) separating reactants (bounded motion in the potential well region) from products (escape to infinity) in phase space or vice versa. Hence, the isoenergetic DS is given by
\begin{equation}
\mathcal{D}(H_0) = \left\lbrace (q,x,p,p_x) \in \mathbb{R}^4 \; | \; q = 0 \;,\; 2H_0 = p^2 + p_x^2 + \omega^2 x^2 \right\rbrace
\end{equation}
which has the geometry of a 2-sphere, that is $\mathbb{S}^2$ in $\mathbb{R}^3$. To be precise, it is an ellipsoid with semi-major axis $\sqrt{2H_0}$ in the $p$, $p_x$ axis, and $\sqrt{2H_0}/\omega$ in the $x$ axis. This ellipsoid has two hemispheres, known as the forward and backward DS with the form
\begin{equation}
\begin{split}
\mathcal{D}_f(H_0) &= \left\lbrace (q,x,p,p_x) \in \mathbb{R}^4 \; | \; q = 0 \;,\; p = -\sqrt{2H_0 -  p_x^2 - \omega^2 x^2} \right\rbrace \\
\mathcal{D}_b(H_0) &=\left\lbrace (q,x,p,p_x) \in \mathbb{R}^4 \; | \; q = 0 \;,\; p = +\sqrt{2H_0 -  p_x^2 - \omega^2 x^2} \right\rbrace
\end{split}
\end{equation}
Forward ``reaction'' occurs when trajectories cross $\mathcal{D}_f$ and back ``reaction'' when trajectories cross $\mathcal{D}_b$. The forward and backward DS is joined at the equator along the normally hyperbolic invariant manifold given by
\begin{equation}
\mathcal{N}(H_0) = \left\lbrace (q,x,p,p_x) \in \mathbb{R}^4 \; | \; q = p = 0 \;,\; 2H_0 = p_x^2 + \omega^2 x^2 \right\rbrace
\end{equation}
which has the topology of $S^1$. To be precise, it is an ellipse with semiaxis $\sqrt{2H_0}$ in the $p_x$ direction and $\sqrt{2H_0}/\omega$ in the $x$ direction. As discussed earlier, for a 2 DoF Hamiltonian the NHIM is an unstable periodic orbit which extends the influence of the rank-1 saddle equilibrium point in the bottleneck of the PES (a configuration space concept) into phase space. We note that for $H_0 > 0$ the NHIM is the correct phase space structure that anchors the barriers to the reaction and carry the effect of the saddle equilibrium point to a range of energies as given by Moser's generalization of Lyapunov Subcenter Manifold Theorem~\cite{wiggins2003applied}. The stable and unstable manifolds of the unstable periodic orbit are 
\begin{equation}
\mathcal{W}^s = \Gamma \cup \mathcal{W}^s_l \quad , \quad \mathcal{W}^u = \Gamma \cup \mathcal{W}^u_l 
\end{equation}
where $\Gamma$ is the homoclinic orbit
\begin{equation}
\Gamma = \left\lbrace  (q,x,p,p_x) \in \mathbb{R}^4 \; | \; q > 0 \;,\; H_r(q,p) = 0 \;,\; H_b(x,p_x) = H_0  \right\rbrace
\end{equation}
and the left branches are
\begin{equation}
\begin{split}
\mathcal{W}^s_l &= \left\lbrace  (q,x,p,p_x) \in \mathbb{R}^4 \; | \; q < 0 \:,\; p > 0 \;,\; H_r(q,p) = 0 \;,\; H_b(x,p_x) = H_0 \right\rbrace \\[.2cm]
\mathcal{W}^u_l &= \left\lbrace  (q,x,p,p_x) \in \mathbb{R}^4 \; | \; q < 0 \:,\; p < 0 \;,\; H_r(q,p) = 0 \;,\; H_b(x,p_x) = H_0 \right\rbrace
\end{split}
\end{equation}
We note that the stable and unstable invariant manifolds have the structure of a cartesian product of a curve in the $(q,p)$ saddle space and an ellipse in the $(x,p_x)$ center space, and thus become cylindrical (or \textit{tube}) manifolds. We show the energy surface in Fig.~\ref{EnergySurf_Hills}(a) at 3 energy values for $\epsilon = 0$ where the bottleneck only opens for $H_0 > 0$. The NHIM computed using differential correction and continuation, and its invariant manifolds computed using globalization are shown in Fig.~\ref{fig:set2_manifolds_upo_energysurf}(a) for $H_0 = 0.05$ and $\epsilon = 0$. 

\begin{figure}[!ht]
	\begin{center}		
	A) \includegraphics[scale=0.45]{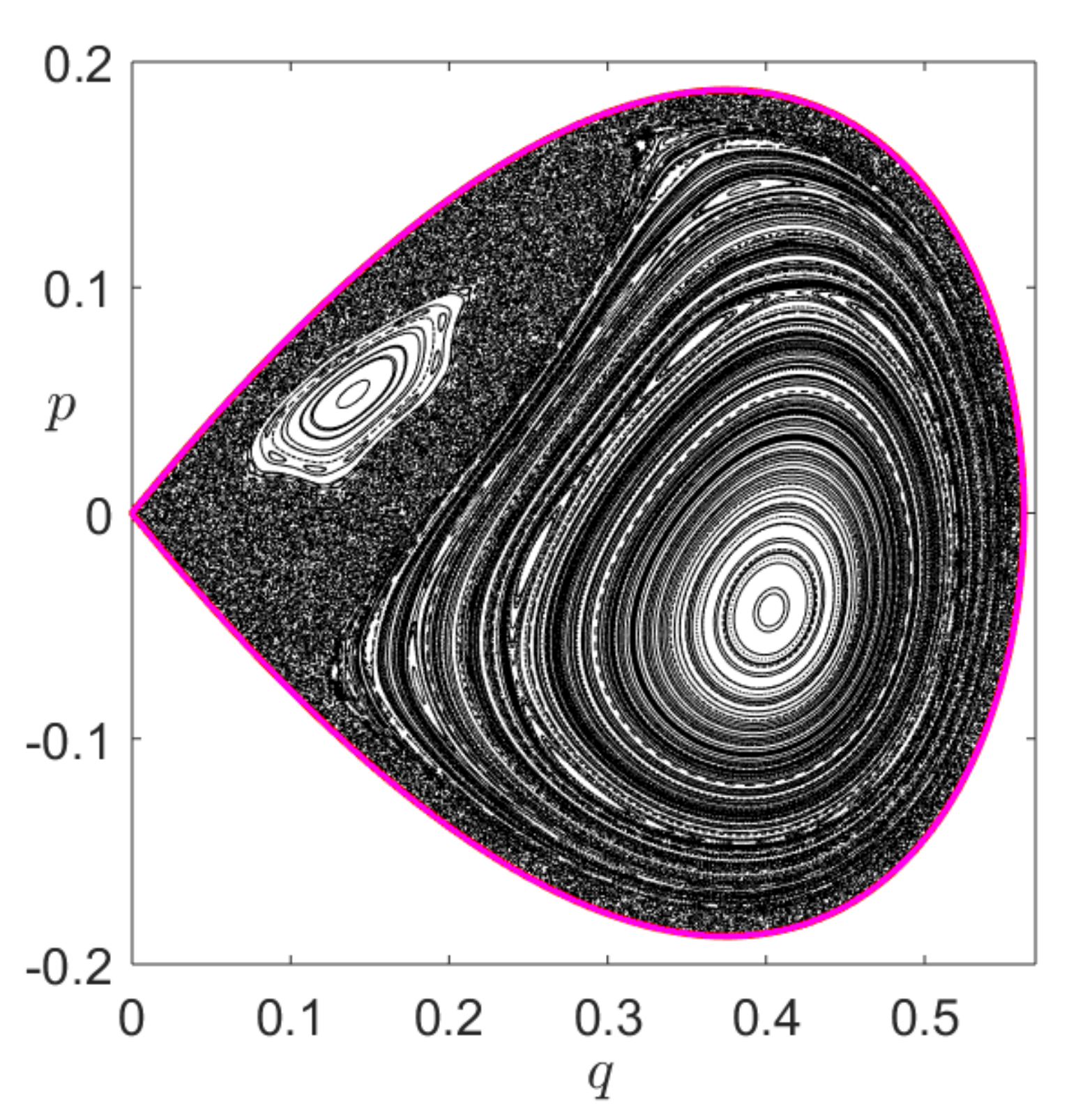}	
	B) \includegraphics[scale=0.45]{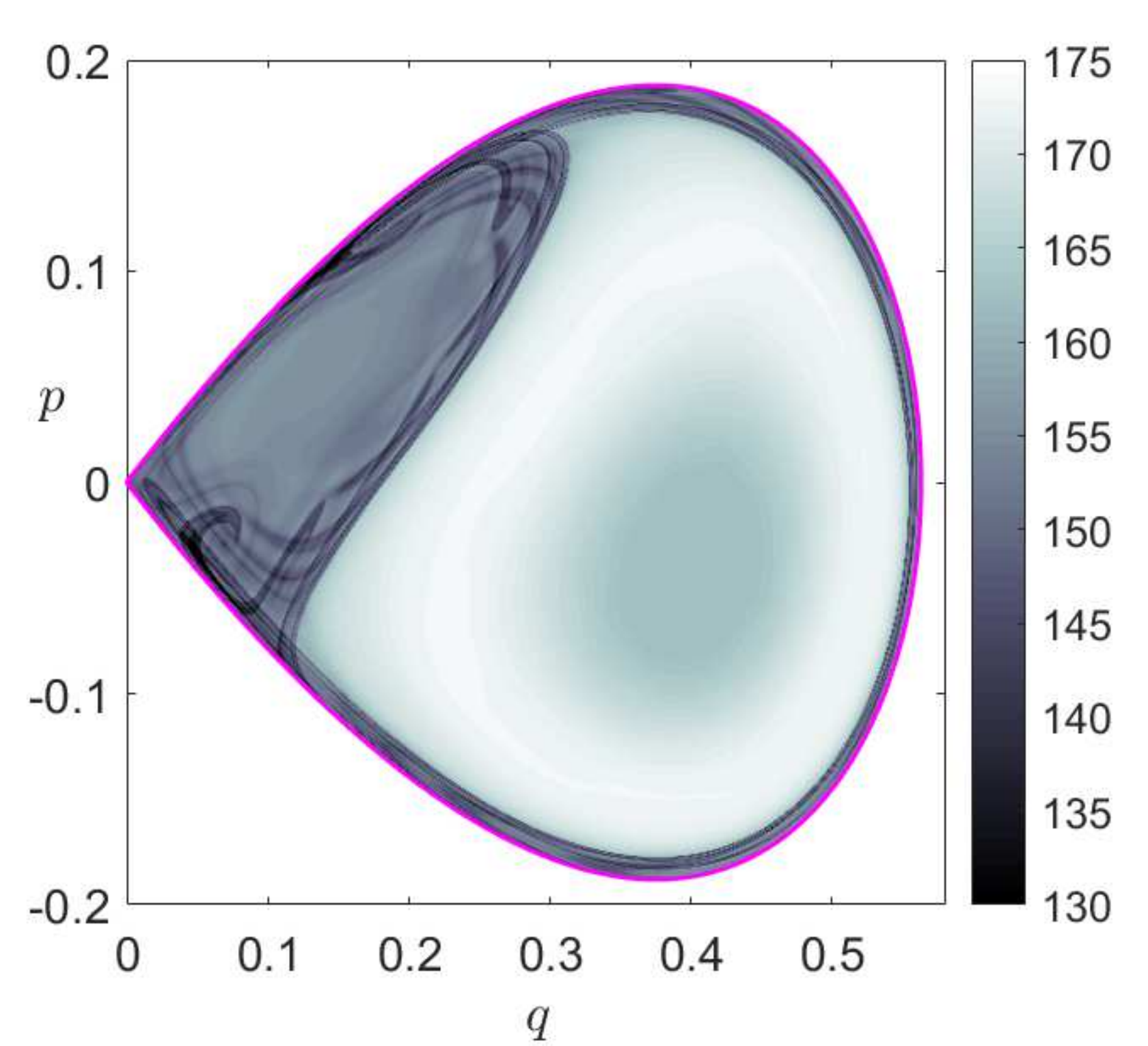}
	C) \includegraphics[scale=0.42]{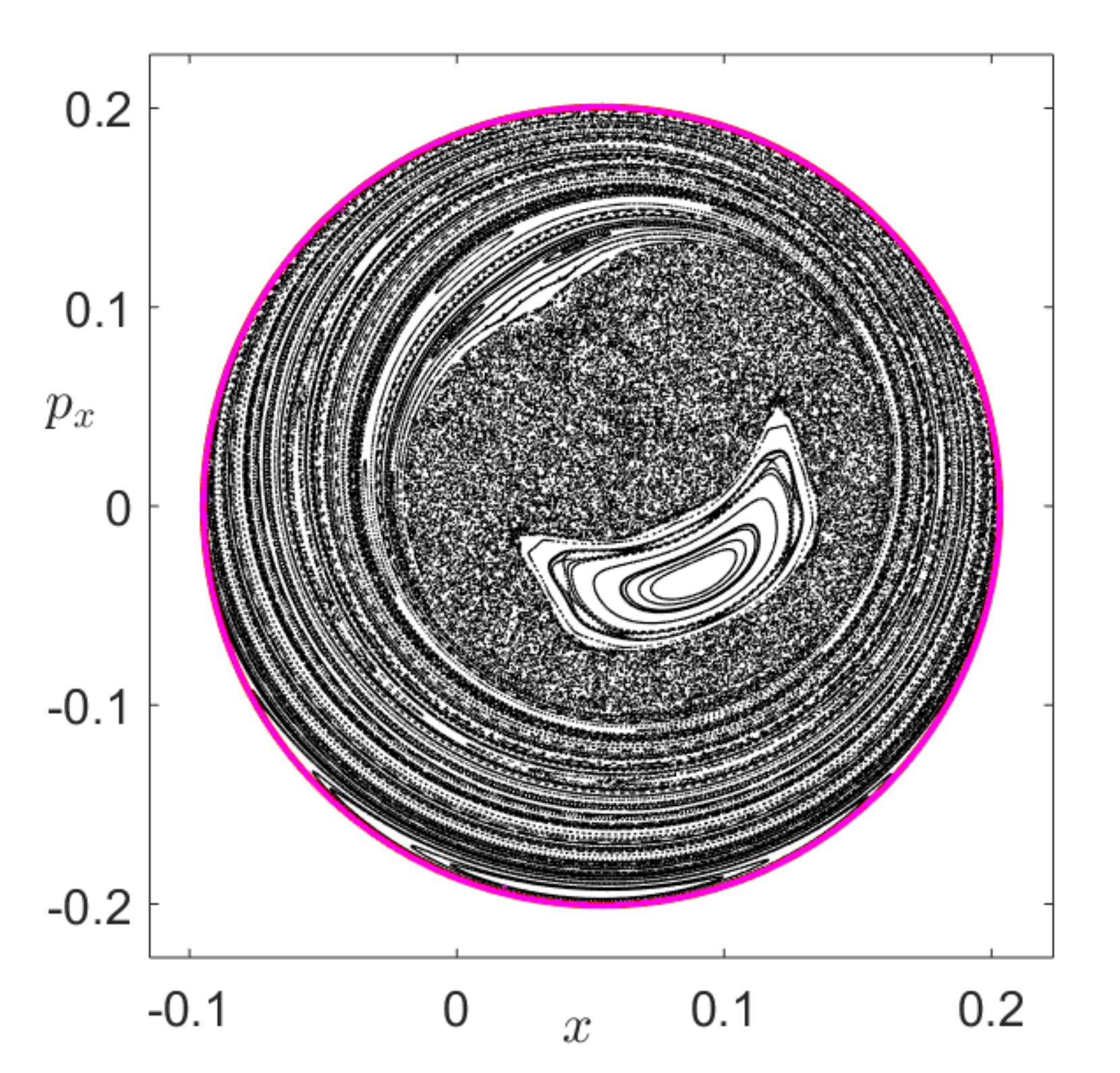}
	D) \includegraphics[scale=0.42]{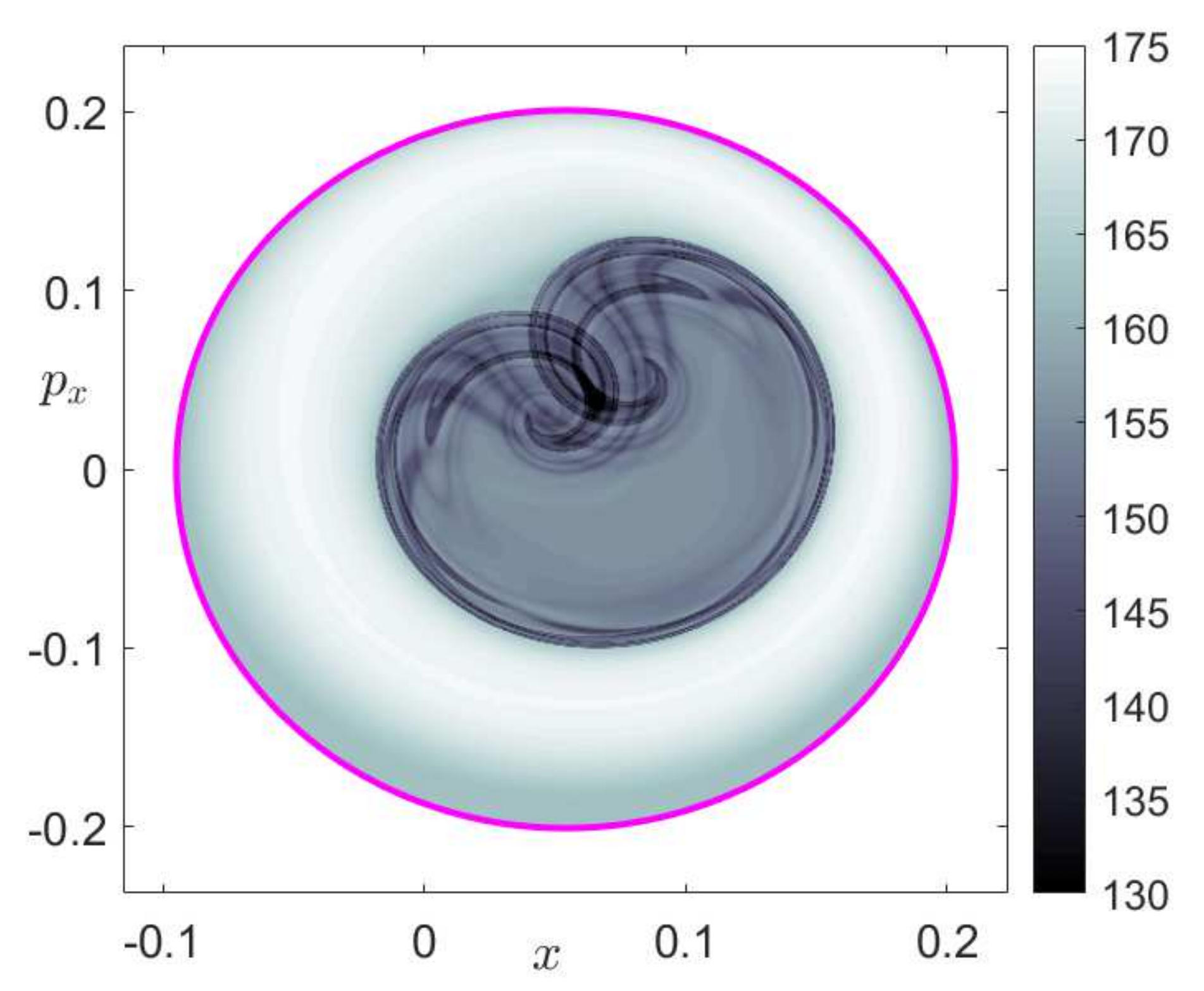}
	\end{center}
	\vspace{-3ex}
	\caption{Phase space structures of the coupled Hamiltonian with model parameters $\mu = 0.25$, $\alpha = 2$, $\omega = 1.25$ and $\varepsilon = 0.25$. The total energy of the system is $H_0 = 0$ (barrier energy). A) Poincar\'e map on the surface of section $\mathcal{U}^{+}_{qp}$; B) LDs calculated for $\tau = 75$ on the surface of section $\mathcal{U}^{+}_{qp}$; C) Poincar\'e map on the surface of section $\mathcal{U}^{+}_{xp_x}$; D) LDs calculated for $\tau = 75$ on the surface of section $\mathcal{U}^{+}_{xp_x}$. We have marked the energy boundary with a magenta curve.}
	\label{ps_vs_ld_1}
\end{figure}

When $\varepsilon \neq 0$, the ``reaction'' and ``bath'' modes are coupled and for $\varepsilon$ small, we can think about the resulting dynamics as a perturbation of the uncoupled case . Given an energy of the system below that of the barrier, that is $H_0 \leq 0$, the phase space bottleneck is closed and trajectories are trapped in the potential well region. However, due to the perturbation, the unstable regular motion on the tori in the uncoupled system is destroyed as given by the KAM theorem. Therefore, chaotic motion arises in some regions of the phase space due to a non-zero coupling of the ``reaction'' and ``bath'' modes. We illustrate the system's behavior for the energy $H_0 = 0$ (energy of the rank-1 saddle) in Fig.~\ref{ps_vs_ld_1} using LDs and Poincar\'e maps on the surfaces of section $\mathcal{U}^{+}_{qp}$~\eqref{sos_qp} and $\mathcal{U}^{+}_{xp_x}$~\eqref{sos_xpx}. We observe that there is a distinct correlation between the qualitative dynamics revealed by the Lagrangian descriptor (LD) contour maps and Poincar\'e sections. That is, chaotic regions of phase space which appear as a sea of points in the Poincar\'e section and hide the underlying structures of stable and unstable manifolds are completely resolved by LD contour maps where the tangled geometry of the manifolds is revealed by the points where the function attains a local minimum as has also been shown in~\cite{Naik2019b}. The capability of LDs to identify the phase space structures relevant in chemical reaction dynamics can also be found in recent literature~\cite{craven2016deconstructing,demian2017,Naik2019a}. In addition, the generation of Poincar\'e section relies on tracking the crossing of a 2D surface which can not be guaranteed in high dimensional phase space, while LDs just accumulate a positive scalar quantity along the trajectory and thus have potential to reveal high dimensional phase space structures.

\begin{figure}[!ht]
	\begin{center}		
	\includegraphics[scale=0.45]{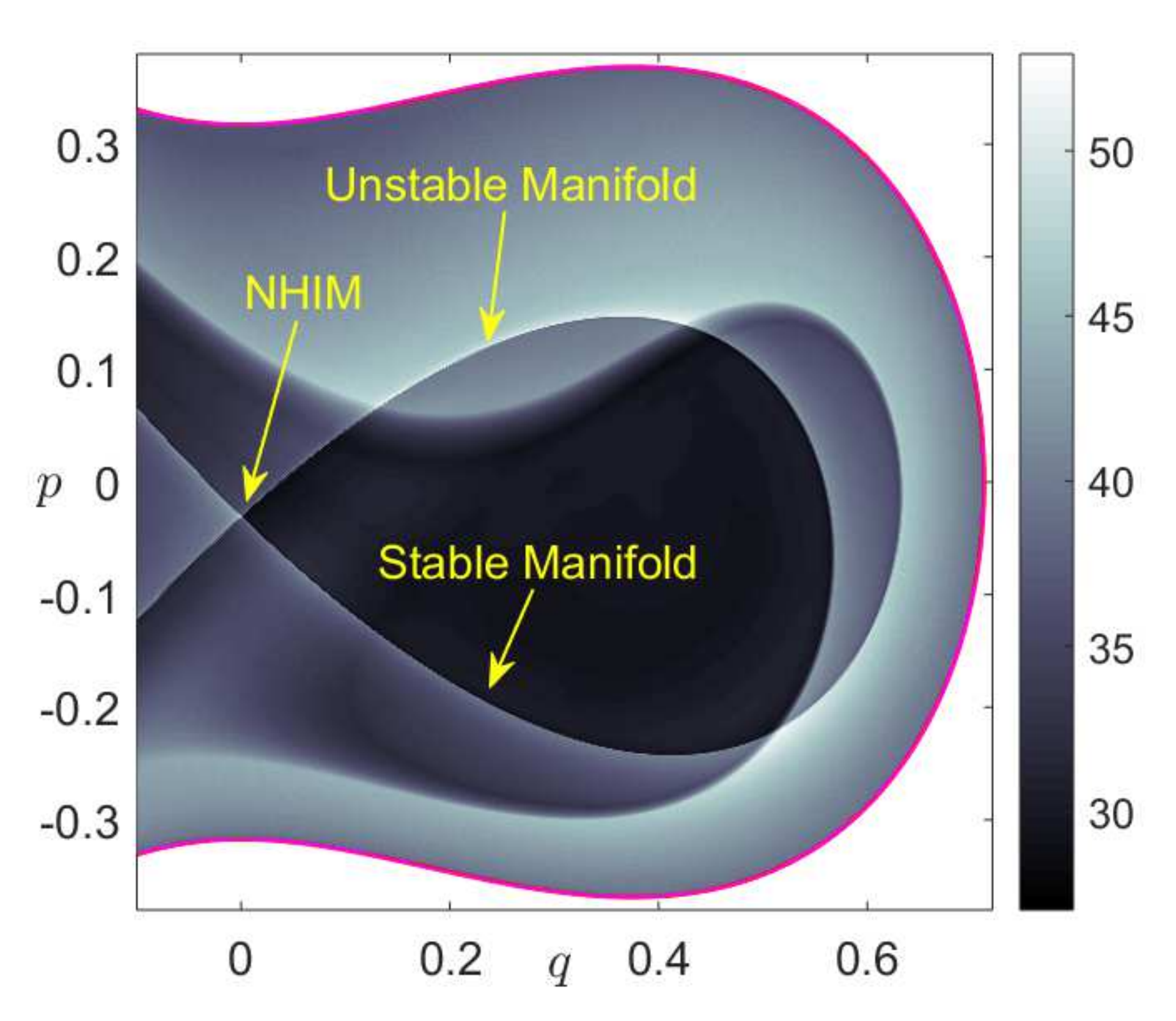}
	\end{center}
	\vspace{-3ex}
	\caption{Variable time LDs calculated on the surface of section $\mathcal{U}^{+}_{qp}$ for $\tau = 10$. The NHIM and its stable and unstable manifolds are revealed as points where the LD field in non-differentiable and attains a local minimum. We have marked the energy boundary with a magenta curve.}
	\label{LD_NHIM_detect}
\end{figure}

\begin{figure}[!ht]
	\begin{center}		
		\subfigure[]{\includegraphics[scale=0.42]{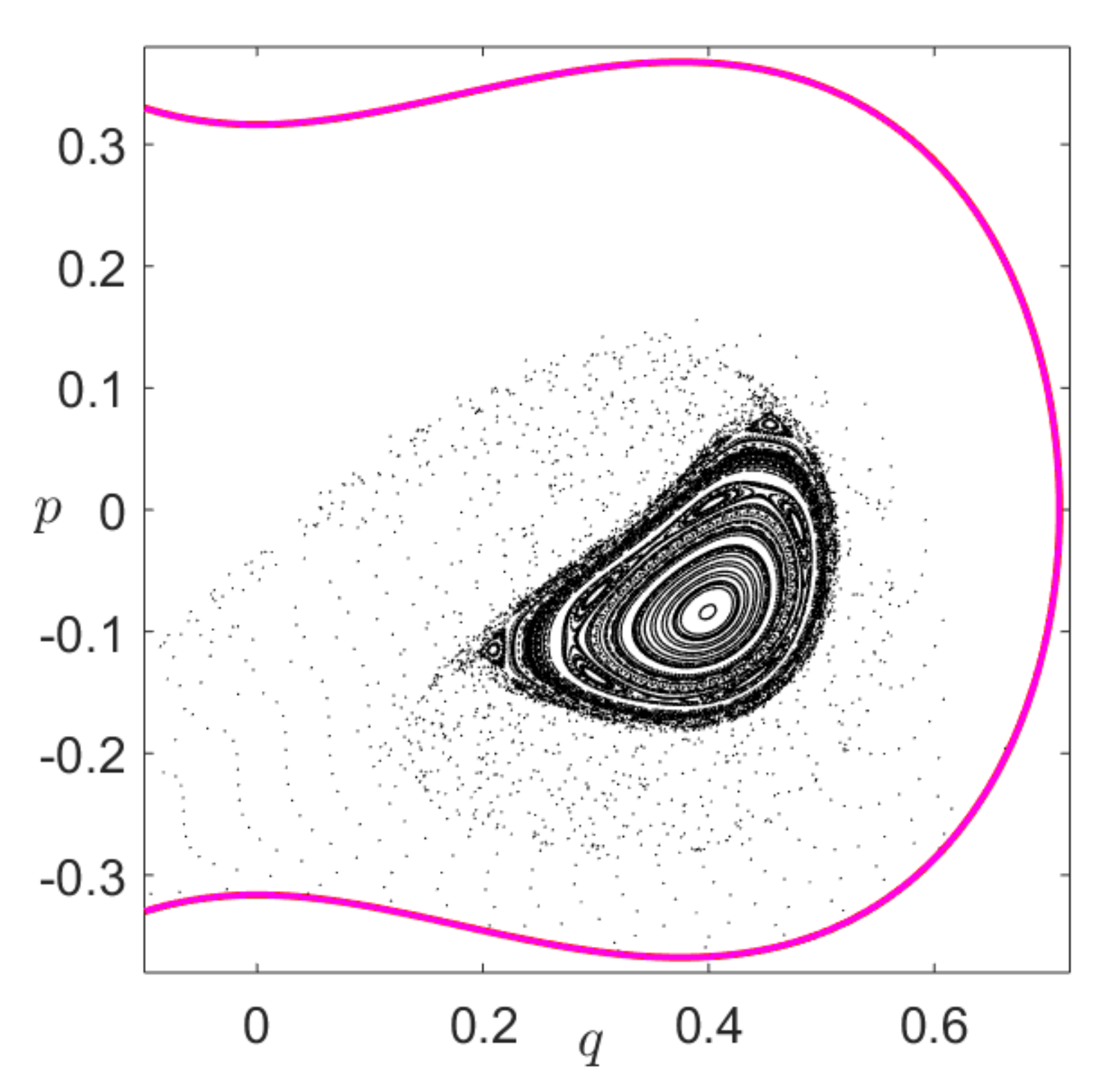}}
		\subfigure[]{\includegraphics[scale=0.42]{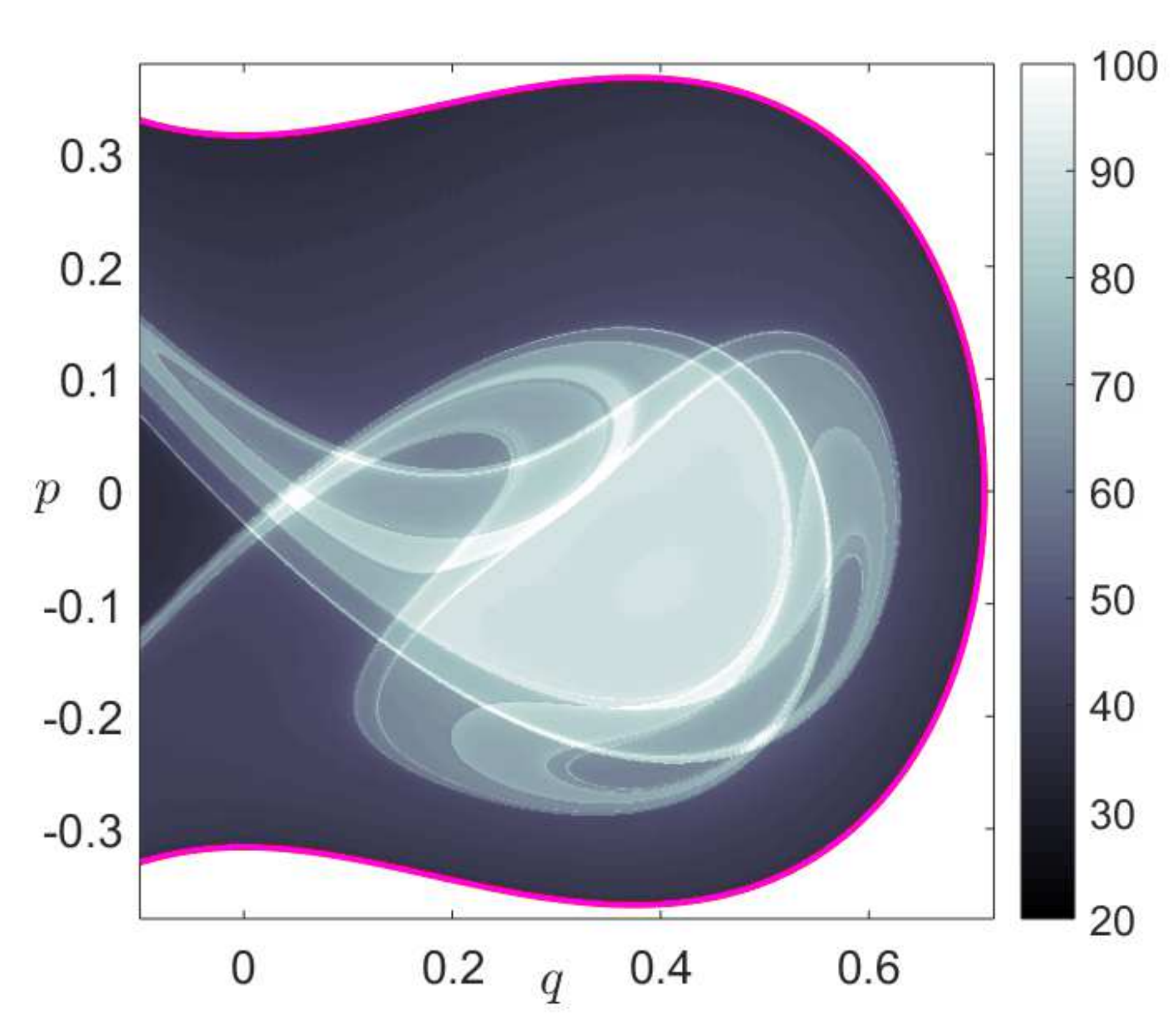}}
		\subfigure[]{\includegraphics[scale=0.42]{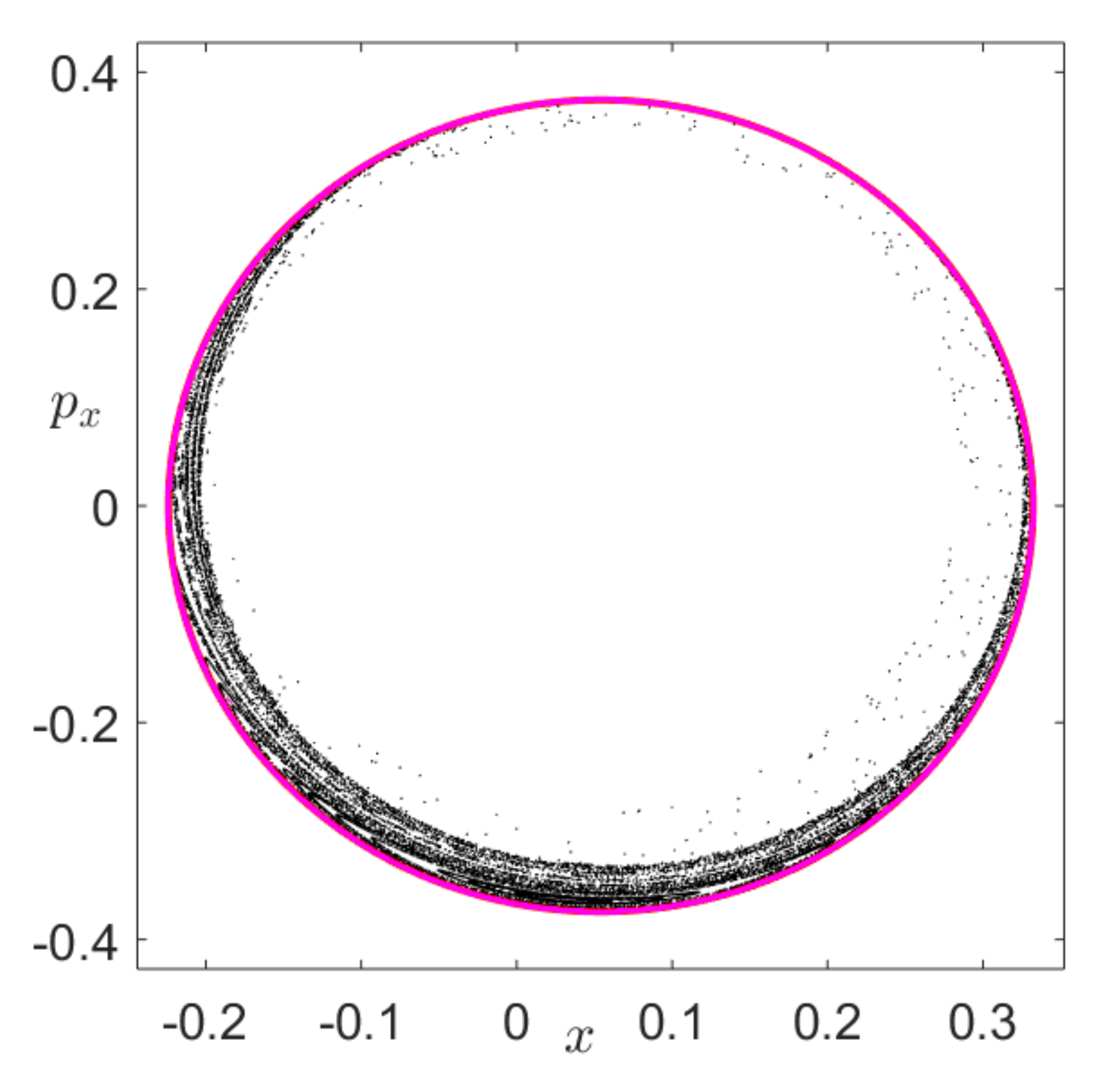}}     
		\subfigure[]{\includegraphics[scale=0.425]{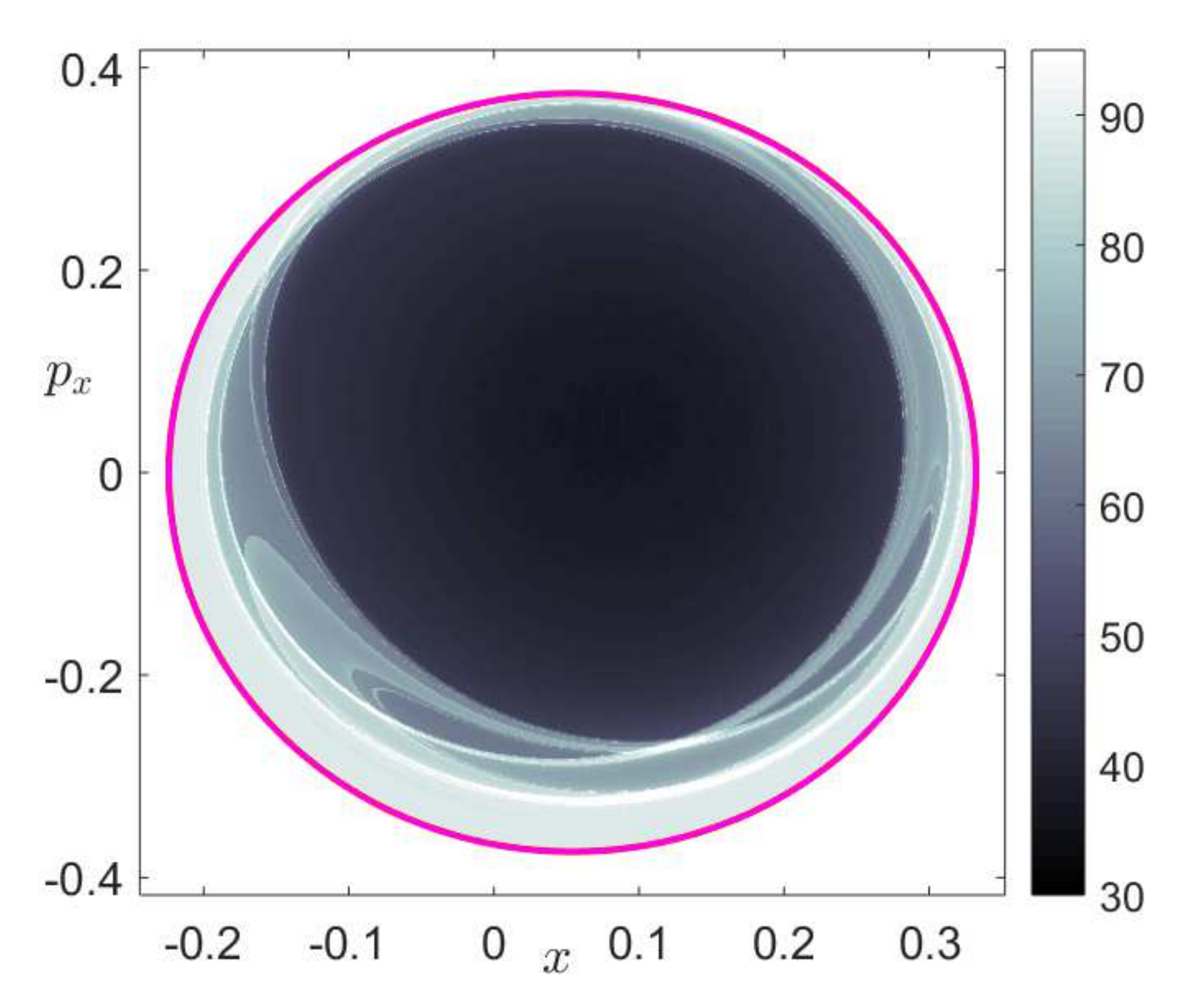}}
	\end{center}
	\vspace{-3ex}
	\caption{Phase space structures of the coupled Hamiltonian with model parameters $\mu = 0.25$, $\alpha = 2$, $\omega = 1.25$ and $\varepsilon = 0.25$. The total energy of the system is $H_0 = 0.05$,	which is above the barrier energy. (a) Poincar\'e map on the surface of section $\mathcal{U}^{+}_{qp}$ (b) LDs calculated for $\tau = 30$ on the surface of section $\mathcal{U}^{+}_{qp}$ (c) Poincar\'e map on the surface of section $\mathcal{U}^{+}_{xp_x}$ (d) LDs calculated for $\tau = 30$ on the surface of section $\mathcal{U}^{+}_{xp_x}$. We have marked with a magenta curve the energy boundary.}
	\label{ps_vs_ld_2}
\end{figure}

Now let us consider the dynamics when the total energy is above that of the rank-1 saddle, that is $H_0 > 0$, and in particular we will set $H_0 = 0.05$ in our analysis. In this situation, a phase space bottleneck opens up in the barrier region, as shown in Fig.~\ref{EnergySurf_Hills}. In order to describe the structures that mediate reaction dynamics, that is the NHIM (or unstable periodic orbit in this case) and its stable and unstable manifolds, we use differential correction and continuation along with globalization as described in Appendix \ref{sec:appB}. In Fig.~\ref{fig:set2_manifolds_upo_energysurf}(c-d), we have shown the cylindrical manifolds along with the energy surface and the unstable periodic orbit for a fixed energy. In order to recover the homoclinic tangle geometry of the invariant manifolds, we calculate LDs on the surfaces of section $\mathcal{U}^{+}_{qp}$ and $\mathcal{U}^{+}_{xp_x}$, and compare with the direct numerical construction of these invariant manifolds. This LD based diagnostic is similar to performing a ``phase space tomography'' of the high dimensional phase space structures using a low dimensional slice. In Fig.~\ref{LD_NHIM_detect} we show the computation of variable time LD for an integration time $\tau = 10$ on the slice $\mathcal{U}^{+}_{qp}$. We observe that LD clearly identify, by means of points which attain a minimum, the location of the stable and unstable manifolds and the NHIM at the intersection of the invariant manifolds. Since we are using a small integration time of $\tau = 10$ to compute LDs in Fig. \ref{LD_NHIM_detect}, the complete geometry of the homoclinic tangle is not fully revealed. 
Therefore, in order to recover a more complete and intricate dynamical picture of the homoclinic tangle, the integration time to compute LDs has to be increased. This is shown in Fig. \ref{ps_vs_ld_2}(b), where $\tau = 30$ and we observe that the regular motion obtained in the middle of the Poincar\'e section displayed in Fig.~\ref{ps_vs_ld_2}(a) corresponds to trajectories that remain trapped in the potential well region and never escape. The trapped dynamics on the tori are also captured by the LD contour map shown in \ref{ps_vs_ld_2}(b) which also reveals the homoclinic tangle of the stable and unstable manifolds and the resulting lobe dynamics~\cite{beigie_dynamics_1992}. We note that in all these computations we are using the variable time definition of LDs, since the open potential surface causes trajectories to escape to infinity through the bottleneck in finite time and resulting in NaN values in the LD contour map. This issue is discussed in Appendix~\ref{sec:appA} and illustrated in Fig.~\ref{ld_fixTime}, and would hide the important underlying phase space structures making the interpretation of results difficult. In addition, when we analyze the dynamics using Poincar\'e sections, we can not ensure that trajectories return to the surface of section when escaping to infinity is possible. This will result in blank regions in the Poincar\'e sections as shown in Figs.~\ref{ps_vs_ld_2}(a) and (c). However, we note that trapped trajectories in the potential well corresponding to regular (motion on the tori) and chaotic motion are highlighted as expected in the Poincar\'e sections. These trajectories are non-reactive and will remain so until they satisfy the \textit{sufficient} condition for reaction which is entering the cylindrical manifolds.

\begin{figure}[htbp]
	\begin{center}		
	\subfigure{A) \includegraphics[scale=0.35]{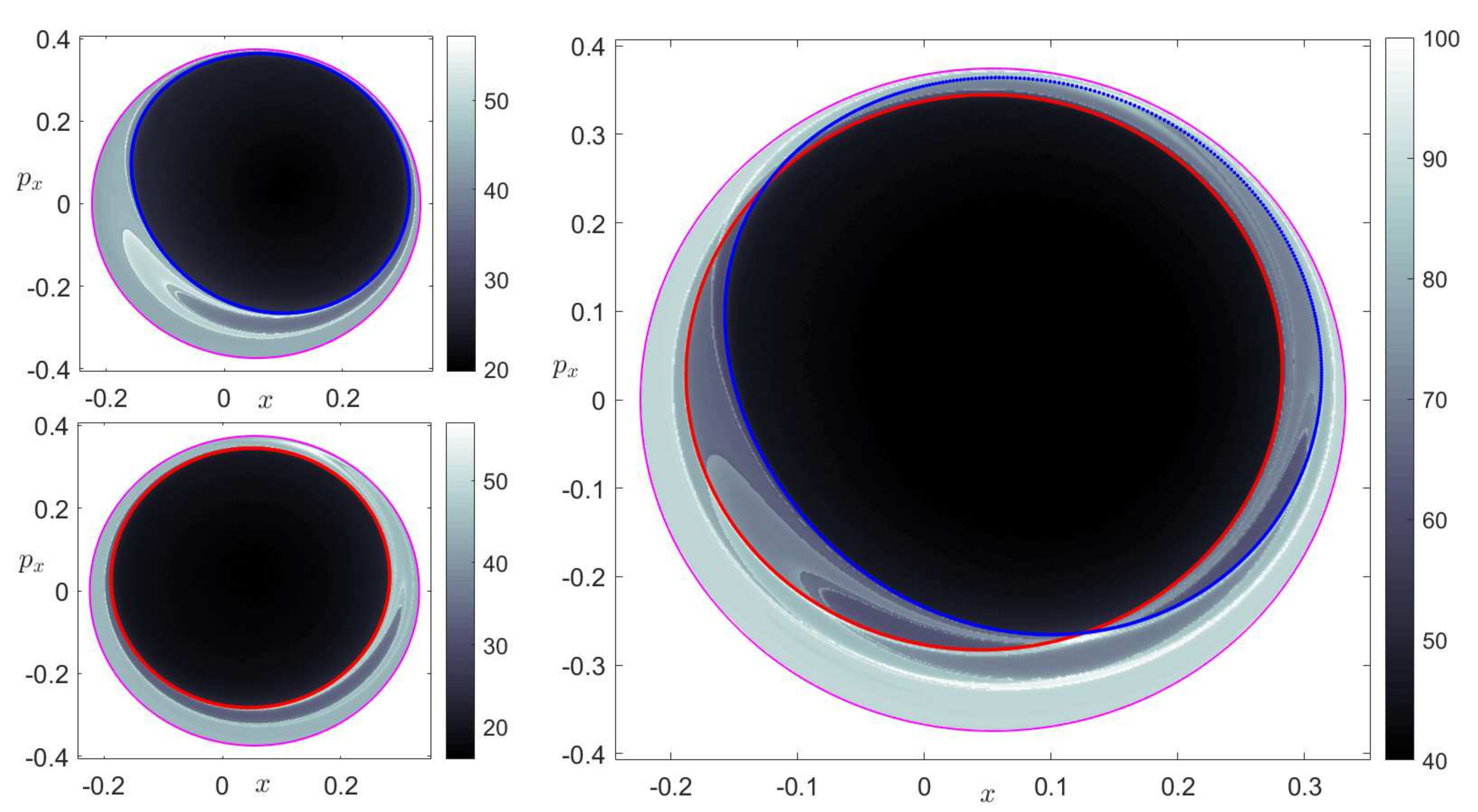}}
	\subfigure{B) \includegraphics[scale=0.37]{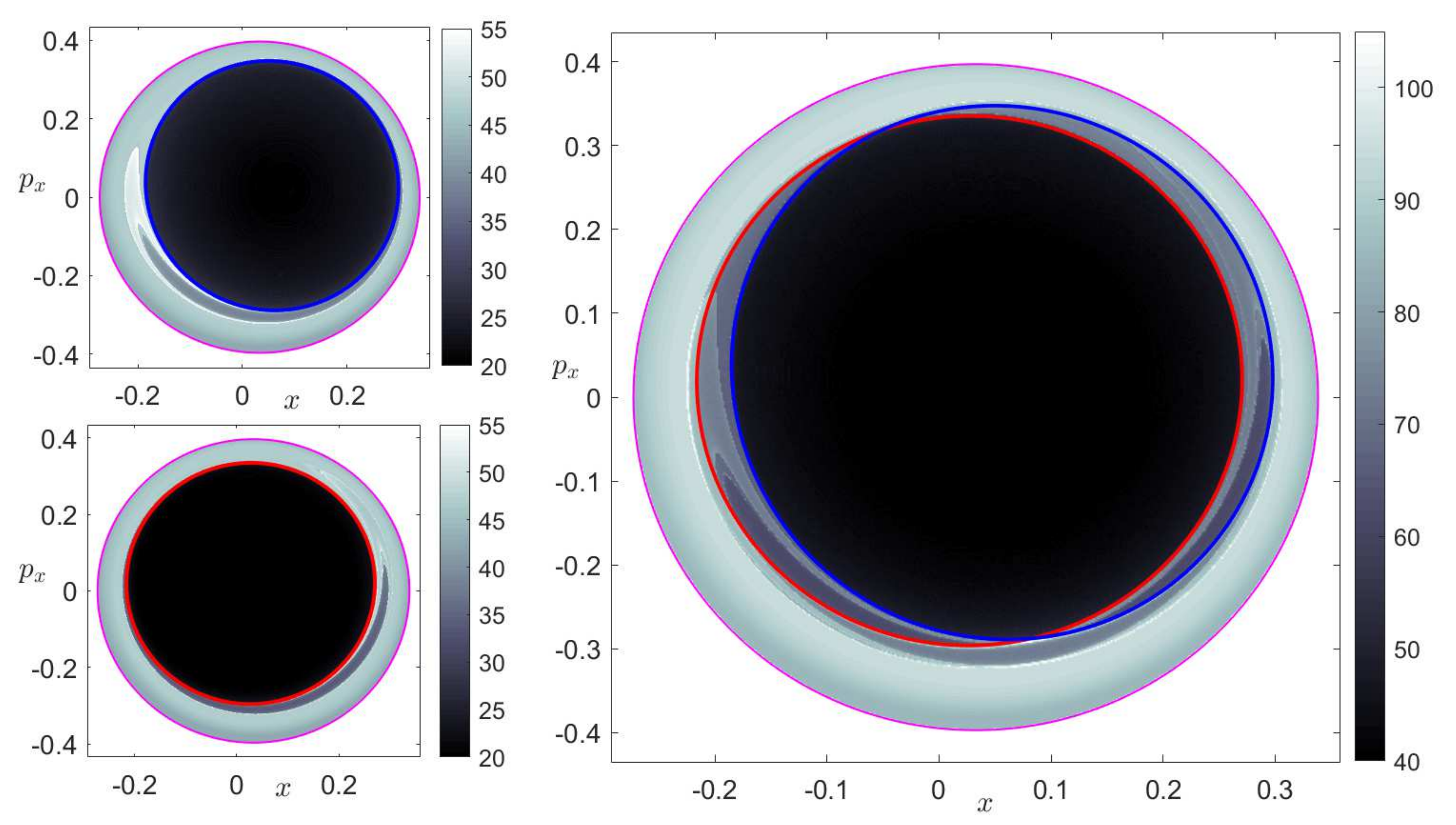}}
	\end{center}
	\vspace{-4ex}
	\caption{Phase space structures revealed by LDs using $\tau = 30$, for the coupled Hamiltonian with parameters $\mu = 0.25$, $\alpha = 2$, $\omega = 1.25$ and total energy $H_0 = 0.05$. A) Coupling strength $\varepsilon = 0.25$. On the top/bottom left we show the forward/backward LD and, superimposed, the first intersection of the stable/unstable manifold with $\mathcal{U}^{+}_{xp_x}$ as a blue/red curve. On the right, the total LD is depicted (addition of forward and backward LD) together with the first intersection of the stable and unstable manifolds, known in chemistry as reactive islands since trajectories inside these regions will escape the potential well through the bottleneck in forward/backward time respectively. The magenta curve represents the energy boundary. B) same analysis using $\varepsilon = 0.125$.}
	\label{LD_Manifolds}
\end{figure}

Identifying the regions inside the cylindrical (tube) stable and unstable manifolds can be done using globalization or using the LD contour map on appropriate isoenergetic surfaces. We compute LD contour maps to recover the intersections of these tube manifolds with the surface of section~\eqref{sos_xpx}. The intersection of the invariant manifolds with the isoenergetic surface of section yields topological ellipses, known in the chemical reactions literature as \textit{reactive islands}~\cite{deleon1991a,deleon1991b,deleon1992,patra2018detecting} and is involved in calculating reaction rates/fraction. In order to illustrate the capability of LDs to recover the reactive island structure, we have compared the LD contour map and manifold intersection with the surface of section $\mathcal{U}^{+}_{xp_x}$ obtained using globalization in Fig.~\ref{LD_Manifolds}. We have shown the first intersection of the stable and unstable manifold with the surface of section as a blue/red curve superimposed on the LD contour map. We observe that the minima in the LD contour maps and the manifolds intersections are in agreement, thus verifying the LD based identification of reaction islands. We also observe that successive fold and resulting intersection of the tube manifolds with the surface of section are also revealed by the points with minima in the LD values. We note here that the first intersection of the manifolds encloses a large phase space volume in the potential well which indicates that a large portion of the potential well escapes to infinity through the bottleneck. This is despite the fact that we have chosen a very small value for the energy of the system, $H_0 = 0.05$, compared to the energy of the barrier. This is a consequence of using a high coupling strength, $\varepsilon = 0.25$, which makes the vase-like shaped potential well region small and narrow as the center equilibrium point at the bottom of the potential well approaches the rank-1 saddle equilibrium point located at the origin as $\varepsilon$ is increased. Movies illustrating the change in both the shape of the energy surface and the equipotentials in configurations space as we vary the coupling strength can be found at \href{https://youtu.be/kYklCrSwtls}{here} and \href{https://youtu.be/0FNChWVM6nU}{here}, respectively. We can see that the effect of increasing the coupling strength from zero is to \textit{tilt and squeeze} the vase-like shape of the energy surface that qualitatively increases the number of reactive trajectories. This action of tilting and squeezing of the vase-like container (boundary defined by the total energy surface) is to pour out its contents (the reactive trajectories) and hence the increase in reaction fraction. A quantitative investigation of this phenomenon is current work in progress and beyond the scope of this article.

\section{Conclusions}
\label{sec:concl}

In this article we have presented new results, by means of using a normal form Hamiltonian that models a saddle-node bifurcation, to investigate the influence on phase space dynamics of decreasing the potential well depth of a PES. The decrease in potential well depth causes the saddle and center equilibrium points to collide \textemdash an effect that takes place in the configuration space. The resulting saddle-node bifurcation, when the well depth is decreased, manifests in the phase space as tilting and squeezing of the energy surface. This change in geometry of the energy surface due to decreasing well depth leads to more reacting (or escaping) trajectories for a given energy. This is supported by the increase in the area of the first intersection of the invariant manifolds with an appropriate surface of section. The trajectory diagnostic method of Lagrangian descriptors also identifies these changes in the geometry of the phase space structures. This method also provides an approach for revealing the influence of potential well depth on high dimensional phase space structures. Detecting the qualitative changes in the geometry of the invariant manifolds using Lagrangian descriptors gets us closer to the goal of achieving a complete high-dimensional ``phase space tomography'' for realistic molecular systems.

\section*{Acknowledgments}

We acknowledge the support of EPSRC Grant No. EP/P021123/1 and Office of Naval Research Grant No. N00014-01-1-0769. The authors would like to acknowledge the EPSRC sponsored Chemistry and Mathematics in Phase Space (CHAMPS) workshop held March 19, 2019 at the University of Bristol where they had the opportunity to discuss with Florentino Borondo about the role of the saddle-node bifurcation in the isomerization of lithium cyanide.

\bibliography{SNreac}

\appendix

\section{Lagrangian Descriptors}
\label{sec:appA}

The computational tool that we use in this work to explore the template of geometrical structures governing phase space transport is the method of Lagrangian descriptors (LDs). This mathematical technique is a trajectory-based diagnostic that was originally developed in the context of Lagrangian transport studies in geophysical fluid dynamics\cite{madrid2009,mancho2013lagrangian}. The fundamental idea behind this methodology is to integrate a positive scalar function along particle trajectories of a dynamical system of general time-dependence in the form:
\begin{equation}
\dfrac{d\mathbf{x}}{dt} = \mathbf{v}(\mathbf{x},t) \;,\quad \mathbf{x} \in \mathbb{R}^{n} \;,\; t \in \mathbb{R} \;,
\label{gtp_dynSys}
\end{equation}
where the vector field $\mathbf{v}(\mathbf{x},t) \in C^{r}  (r \geq 1)$ in $x$ and continuous in time. 

Lagrangian Descriptors were first introduced in \cite{madrid2009,mancho2013lagrangian} by means of a scalar function, referred to as the function $M$, to identify distinguished hyperbolic trajectories, i.e. moving saddles, of a dynamical system with general time dependence. In this original approach, the function $M$ was based on the computation of the arclength of a trajectory starting at an initial condition $\mathbf{x}(t_0) = \mathbf{x}_{0}$ as it evolves forward and backward for a specified time period $\tau > 0$. In this context, LDs were defined as:
\begin{equation}
M(\mathbf{x}_{0},t_0,\tau) = \int_{t_0-\tau}^{t_0+\tau}\|\mathbf{v}(\mathbf{x}(t;\mathbf{x}_0),t)\| \; dt \;,
\label{M_classDef} 
\end{equation} 	
where $\|\cdot\|$ stands for the Euclidean distance and $\mathbf{v}(\mathbf{x},t)$ is the vector field of the dynamical system defined in Eq. \eqref{gtp_dynSys}. The connection between the function $M$ and invariant manifolds in phase space has been primarily demonstrated through numerical simulation experiments. Hyperbolic trajectories and their stable and unstable manifolds are revealed by the function $M$ through sharp changes, which we call \textit{singular features}, in the values of the $M$ field where the gradient becomes very large and changes abruptly. Moreover, it is important to highlight that if the function $M$ is broken into forward and backward integration:
\[
M(\mathbf{x}_{0},t_0,\tau) = M^{(f)}(\mathbf{x}_{0},t_0,\tau) + M^{(b)}(\mathbf{x}_{0},t_0,\tau)
\]
where we have that:
\[
M^{(f)} = \int_{t_0}^{t_0+\tau}\|\mathbf{v}(\mathbf{x}(t;\mathbf{x}_0),t)\| \; dt \quad,\quad M^{(b)} = \int_{t_0-\tau}^{t_0}\|\mathbf{v}(\mathbf{x}(t;\mathbf{x}_0),t)\| \; dt
\]
then $M^{(f)}$ detects the stable manifolds in phase space and $M^{(b)}$ does the same for unstable manifolds. Other structures such as tori-like invariant manifolds are highlighted due to the relation of the time averages of the function $M$ with the Ergodic Partition Theory \cite{mezic1999}. This can be done by defining the time average:
\begin{equation}
\overline{M}(\mathbf{x}_{0},t_0,\tau) = \dfrac{1}{2\tau}\int_{t_0-\tau}^{t_0+\tau}\|\mathbf{v}(\mathbf{x}(t;\mathbf{x}_0),t)\| \; dt \;,
\end{equation} 
and analyzing its convergence as $\tau \to\infty$. A detailed description of the detection and visualization of phase space structures with the function $M$ can be found in \cite{mancho2013lagrangian,lopesino2017}. It is important to note here that in the context of chemical reaction dynamics, this definition of LDs would measure phase space arclength of trajectories. In particular, in the Transition State Theory literature \cite{craven2015lagrangian,junginger2016transition,craven2017lagrangian,junginger2017chemical},  configuration space arclength has been used for the computation of LDs. 

Recently, a rigorous mathematical foundation for Lagrangian descriptors has been established in \cite{lopesino2017} by means of introducing an alternative definition  based on the $p$-norm of the components of the vector field that define the dynamical system in Eq. \eqref{gtp_dynSys}. Consider the scalar function:
\begin{equation}
M_p(\mathbf{x}_{0},t_0,\tau) = \int^{t_0+\tau}_{t_0-\tau} \sum_{i=1}^{n} |v_{i}(\mathbf{x}(t;\mathbf{x}_0),t)|^p \; dt \;,\quad p \in (0,1]
\label{Mp_function}
\end{equation}
For this LD one can mathematically prove in certain model problems that hyperbolic points and their stable and unstable manifolds are detected as singularities of the $M_p$ field, that is, points in which the function is non-differentiable. Therefore, this alternative definition provides a characterization for the concept of \textit{singular features}. Moreover, tori-like invariant structures are also related to time averages of $M_p$ by means of the Ergodic Partition Theory. An important aspect to highlight from the alternative definition given in Eq. (\ref{Mp_function}) is that it allows to decompose the phase space analysis by separating the integral into the different DoFs of the system under study, making it possible to isolate and assess their elliptic and hyperbolic dynamical contributions separately. This property has been shown to be relevant for the detection of unstable periodic orbits in Hamiltonian system with two DoF for the classical H\'{e}non-Heiles system\cite{demian2017}. Furthermore, this property has been used recently to address the Barbanis system with three DoF\cite{Naik2019b}, and also to mathematically prove that for the normal form of a Hamiltonian system with three DoFs with an rank-1 saddle, the NHIM and its stable and unstable manifolds are located at the singularities, which are also local minima, of the LD field. Hence, it is shown that if we decompose Eq. \eqref{Mp_function} into forward and backward integration as we also did for Eq. \eqref{M_classDef}, for a sufficiently large integration time $\tau$ we have:
\begin{equation}
\mathcal{W}^u(\mathbf{x}_{0},t_0) = \textrm{argmin } M_p^{(b)}(\mathbf{x}_{0},t_0,\tau) \quad,\quad \mathcal{W}^s(\mathbf{x}_{0},t_0) = \textrm{argmin } M_p^{(f)}(\mathbf{x}_{0},t_0,\tau)
\label{min_LD_manifolds}
\end{equation}
where $\mathcal{W}^u$ and $\mathcal{W}^s$ are, respectively, the unstable and stable manifolds calculated at time $t_0$ and $\textrm{argmin}$ denotes the phase space coordinates $\mathbf{x}_0$ that minimize the function $M_p$. In addition, the NHIM at time $t_0$ can be calculated as the intersection of the stable and unstable manifolds:
\begin{equation}
\mathcal{N}(\mathbf{x}_{0},t_0) = \mathcal{W}^u(\mathbf{x}_{0},t_0) \cap \mathcal{W}^s(\mathbf{x}_{0},t_0) = \textrm{argmin } M_p(\mathbf{x}_{0},t_0,\tau)
\label{min_NHIM_LD}
\end{equation}

At this stage, we would like to point out other relevant properties reated to the method of Lagrangian descriptors. First, the integration parameter $\tau$ that appears in the definition of both $M$ and $M_p$ plays a critical role in the identification of the underlying dynamical structures in phase space. For applications, the value of $\tau$ is chosen so that the dynamical history of particle trajectories covers all significant timescales for the problem under study. Therefore, there is no general rule on how to choose the appropriate value for $\tau$ in order to achieve the successful visualization of phase space structures. Consequently, all dynamical systems have to be investigated on a case by case basis by means of trial and error simulations. Observe that small values of $\tau$ would yield blurry phase space structures, which are barely recognizable, since this regime resembles the Eulerian (instantaneous) description of the flow. On the other hand, large $\tau$ values result in a richer and more complex geometrical description of phase space structures, because LDs resolve them in great detail, making the task of interpreting transport from the obained picture difficult. So there is always a compromise between these two situations. Another interesting feature of LDs is that its computational implementation is straightforward, even for high dimensions, and that the method can be parallelized easily to run on high performance computers, because it deals with initial conditions separately. 

But probably the most important capability of LDs in all its variants is that it provides us with a high-resolution methodology to explore and visualize high-dimensional phase space dynamics. The successful results obtained by applying this tool to uncover the phase space geometry are in part vindicated from its naive approach of emphasizing initial conditions of particle trajectories rather than focusing on their precise location and long-term evolution. As a result of this crucial point, it offers tremendous advantages for the analysis of high-dimensional phase space, where the evolution of ensembles of initial conditions may yield trajectories that become disperse and ``lost'' with respect to each other, making the interpretation of phase space structures problematic and challenging. The goal of identifying phase space structures for high-dimensional systems with LDs is achieved by analyzing the behavior of initial conditions over low-dimensional phase space slices \cite{demian2017,gg2018,Naik2019a,Naik2019b}. In fact, any low-dimensional surface can be selected as a probe, which can be sampled with arbitrary high resolution by defining an adequate grid of initial conditions. Therefore, no resolution will be lost in the analysis, as the trajectories corresponding to these initial conditions evolve in time, since phase space structure is encoded in the initial conditions of the trajectories themselves. 

\begin{figure}[!ht]
	\begin{center}		
		A) \includegraphics[scale=0.32]{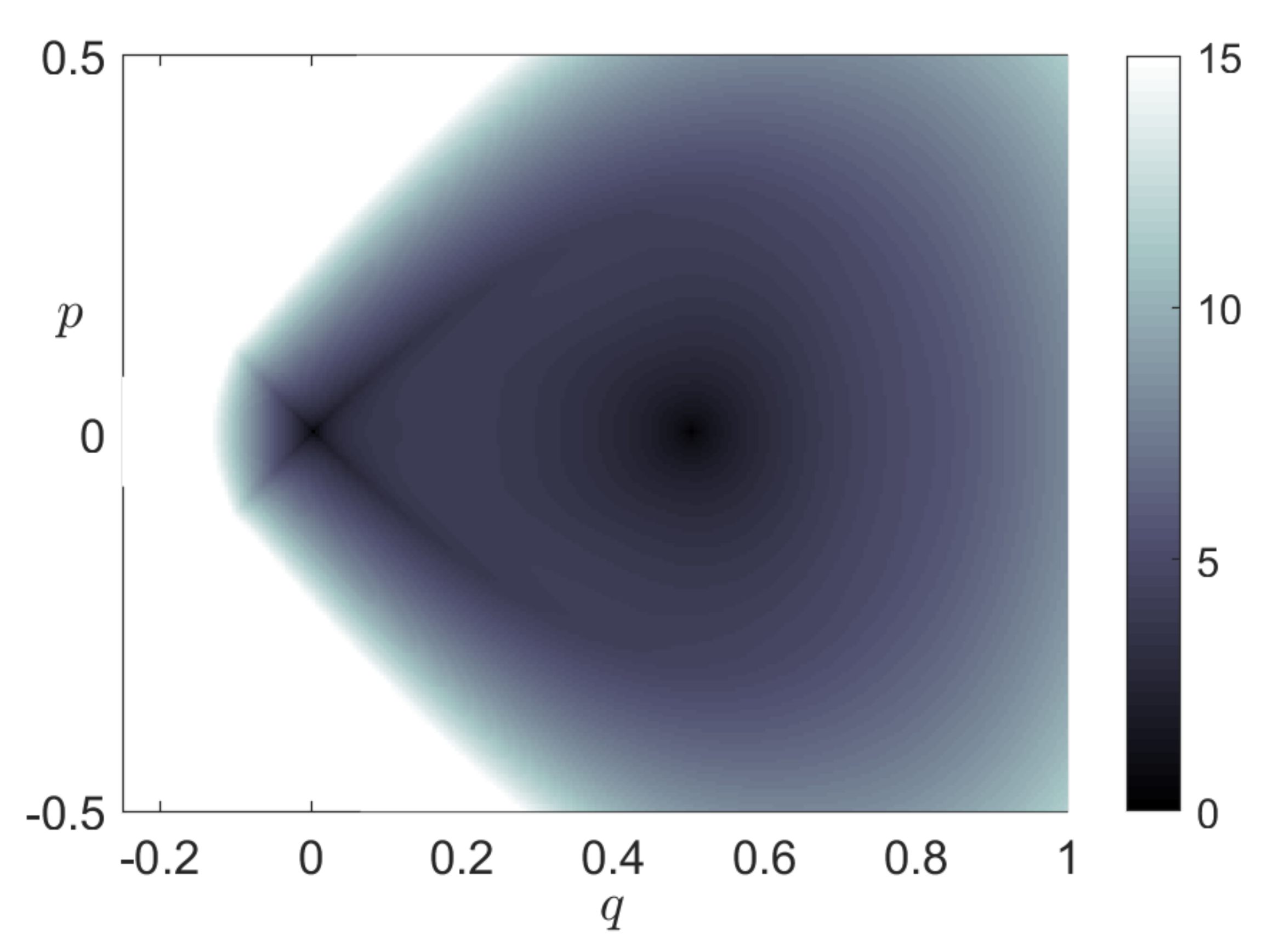}
		B) \includegraphics[scale=0.32]{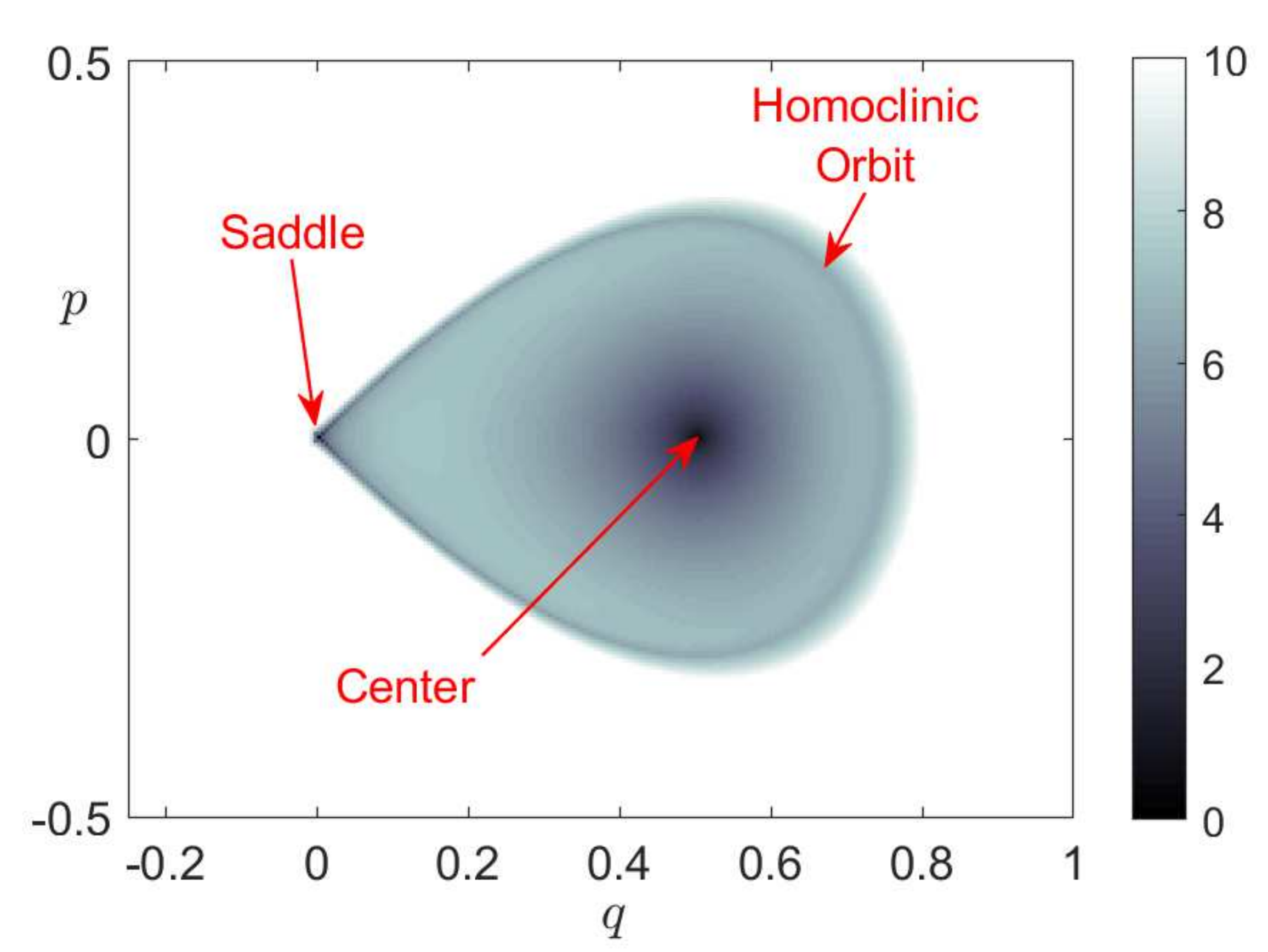}
	\end{center}
	\caption{Fixed-time $M_p$ function calculated with $p = 1/2$ for the 1 DoF Hamiltonian described by Eq. \eqref{hameq_1dof_gen} using the model parameters $\mu = 0.25$ and $\alpha = 2$. Panel A corresponds to $\tau = 3$ and $B$ is calculated for $\tau = 5$.}
	\label{ld_fixTime}
\end{figure}

In recent studies\cite{junginger2017chemical,Naik2019b} it has been highlighted that computing fixed-time LDs, that is, integrating all initial conditions chosen on a phase space surface for the same integration time $\tau$, could give rise to some issues. First, it can obscure the detection of the NHIM, which is a crucial step, for instance in Transition State Theory, in order to determine chemical reaction rates by analyzing flux across the dividing surface constructed from the NHIM. This difficulty appears as a result of bounded trajectories  recrossing the barrier region that surrounds the NHIM, due to the Poincar\'e recurrence theorem. Consequently, multiple minima and singularities occur in the LD plots, which makes the location of the true NHIM trajectories a challenging task. Another issue, which takes place in the saddle-node bifurcation problem that we study in this paper, is that some of the trajectories that escape the potential well of the PES can go to infinity in finite time. The trajectories that show this behavior will give NaN values in the LD scalar field, hiding some regions of the phase space, and therefore obscuring the detection of invariant manifolds. In order to illustrate this problem, we have calculated the fixed-time $M_p$ function given in Eq. \eqref{Mp_function} with $p = 1/2$ for the 1 DoF Hamiltonian described by Eq. \eqref{hameq_1dof_gen} using the model parameters $\mu = 0.25$ and $\alpha = 2$. The results are shown in Fig. \ref{ld_fixTime} for different values of the integration time. In Fig. \ref{ld_fixTime}A, which corresponds to the value $\tau = 3$, we can clearly see how LDs detect the hyperbolic fixed point at the origin and the elliptic point at $(1/2,0)$. Moreover, the stable and unstable manifolds that originate from the saddle point are barely visible, as well as the homoclinic orbit. This is a consequence of the integration time being small. However, flat regions on the left part of the phase space, corresponding to initial conditions that have escaped to infinity in finite time, start to appear as a result of NaN values in the LD. This effect is emphasized further in Fig. \ref{ld_fixTime}B, where we have calculated fixed-time LDs using $\tau = 5$. As we increase the integration time more trajectories escape to infinity and, consequently, a larger region of phase space disappears in the LD picture.

In order to circumvent these issues, we will apply in this work the approach that has been recently adopted in the literature \cite{junginger2017chemical,Naik2019b} known as variable time Lagrangian Descriptors. In this methodology, LDs are calculated for a given initial condition until the trajectory leaves a certain barrier (or saddle) region $\mathcal{R}$ defined in the phase space. Therefore, the total integration time in this strategy depends on the initial condition itself, that is $\tau(\mathbf{x}_0)$. In this formulation, the $p$-norm definition of LDs has the form:
\begin{equation}
M_p(\mathbf{x}_{0},t_0,\tau) = \int^{t_0 + \tau^{+}_{\mathbf{x}_0}}_{t_0 - \tau^{-}_{\mathbf{x}_0}} \sum_{i=1}^{n} |v_{i}(\mathbf{x}(t;\mathbf{x}_0),t)|^p \; dt \;,\quad p \in (0,1] \;.
\label{Mp_vt}
\end{equation}
and, for a fixed integration time $\tau_0$, the total integration time is defined as:
\begin{equation}
\tau^{\pm}_{\mathbf{x}_{0}} = \min \left\lbrace \tau_0 \, , \, t^{\pm} \big|_{\mathbf{x}(t^{\pm};\mathbf{x}_{0}) \notin \mathcal{R}} \right\rbrace
\end{equation}
where $t^{+}$ and $t^{-}$ are the times for which the trajecory leaves the barrier region $\mathcal{R}$ in forward and backward time, respectively. Since the origin is an rank-1 sddle for the Hamiltonian models that we are using to analyze saddle-node bifurcations, we choose for the barrier region:
\begin{equation}
\mathcal{R} = \left\lbrace \mathbf{x} = (q,x,p,p_x) \in \mathbb{R}^4 \; \big| \; |q| < 15 \; , \, |p| < 15 \right\rbrace
\end{equation}
To conclude, it is important to point out here that if the selected barrier region is large enough, the variable time LD definition given above in Eq. \eqref{Mp_vt} will approach the fixed-time LD definition in Eq. \eqref{Mp_vt}. Normally hyperbolic invariant manifolds and their stable and unstable manifolds will be captured by the phase space points for which the LD is non-differentiable and tori-like structures can be determined from time averages of LDs. Moreover, if the barrier region is very small, the NHIM and its invariant stable and unstable manifolds will appear as local maxima in the LD field, while for large barrier regions the local minimum behavior given in Eqs. \eqref{min_LD_manifolds} and \eqref{min_NHIM_LD} is recovered. Consequently, the variable integration time LD provides us with a suitable methodology to study the phase space geometrical structures that characterize the type of saddle-node problems that we discuss in this work, since it avoids the issue of trahectories escaping from the potential well of the PES to infinity in finite time. A detailed analysis on the theoretical background on the variable integration time Lagrangian Descriptors technique will be carried out in future work.

\section{Visualization of phase space structure for the quadratic normal form Hamiltonian}
\label{sec:app_viz}

\begin{figure}[!ht]
	\begin{center}		
		\subfigure{A) \includegraphics[scale=0.32]{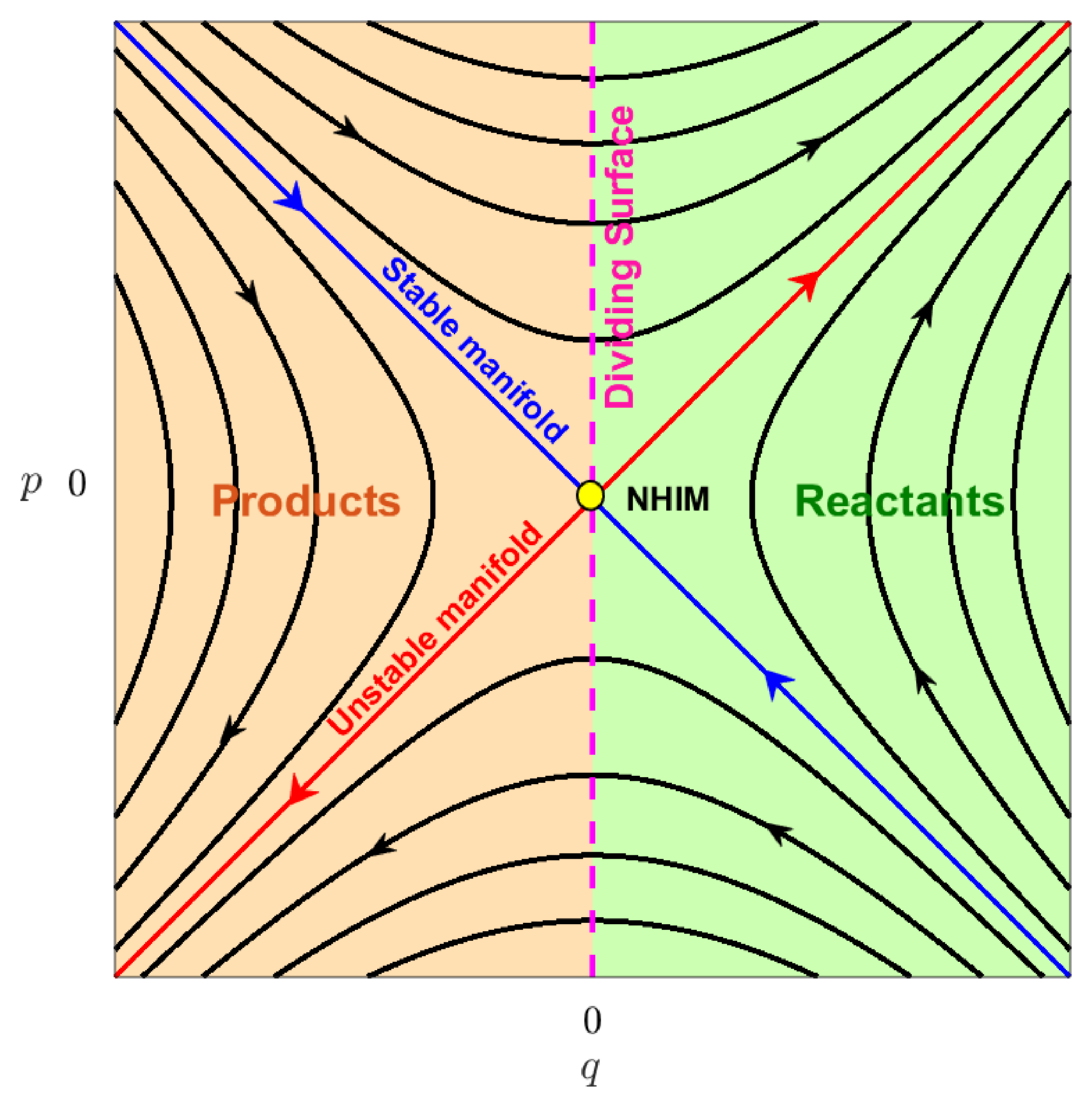}}
		\subfigure{B) \includegraphics[scale=0.31]{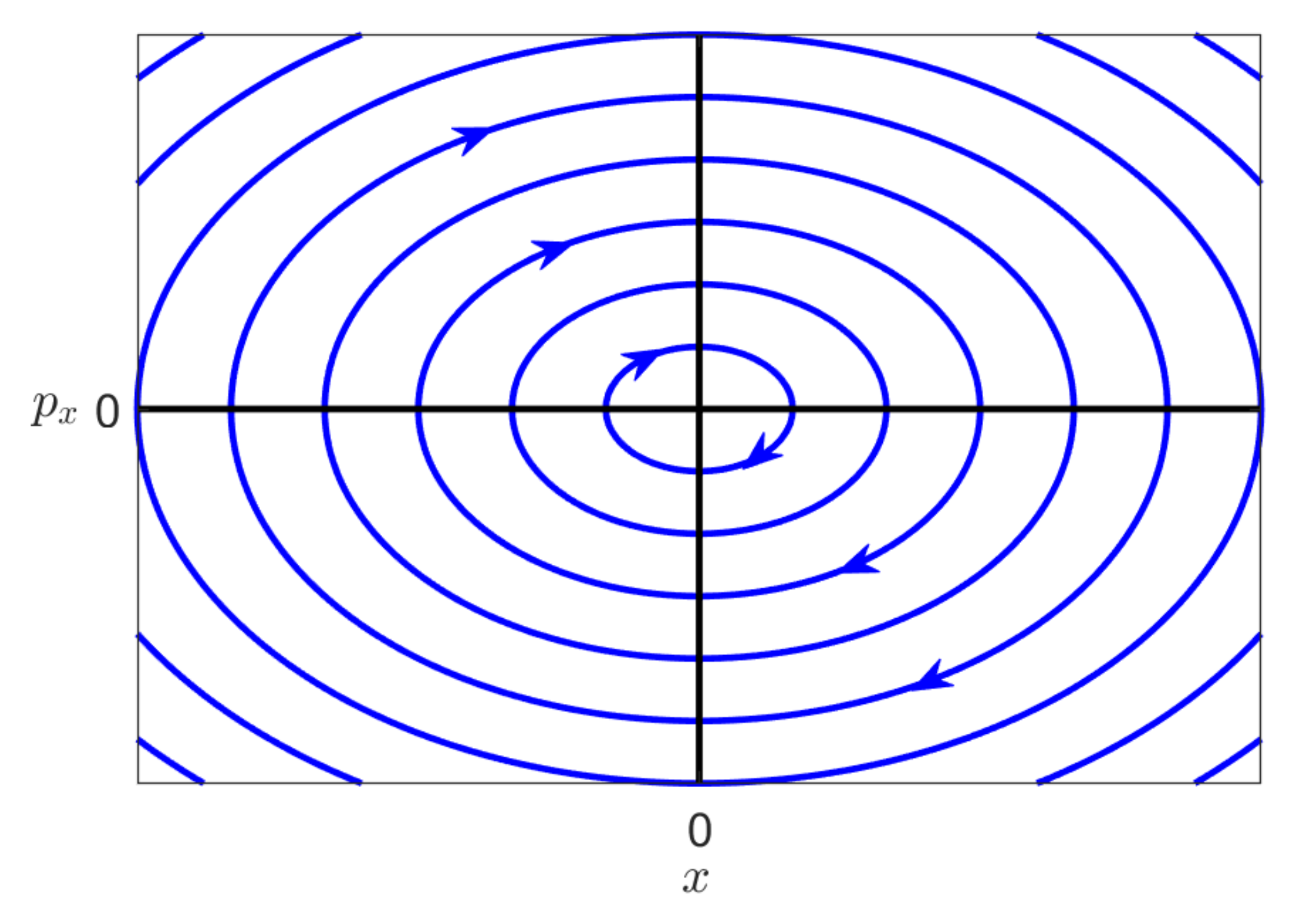}}
		\subfigure{C) \includegraphics[scale=0.33]{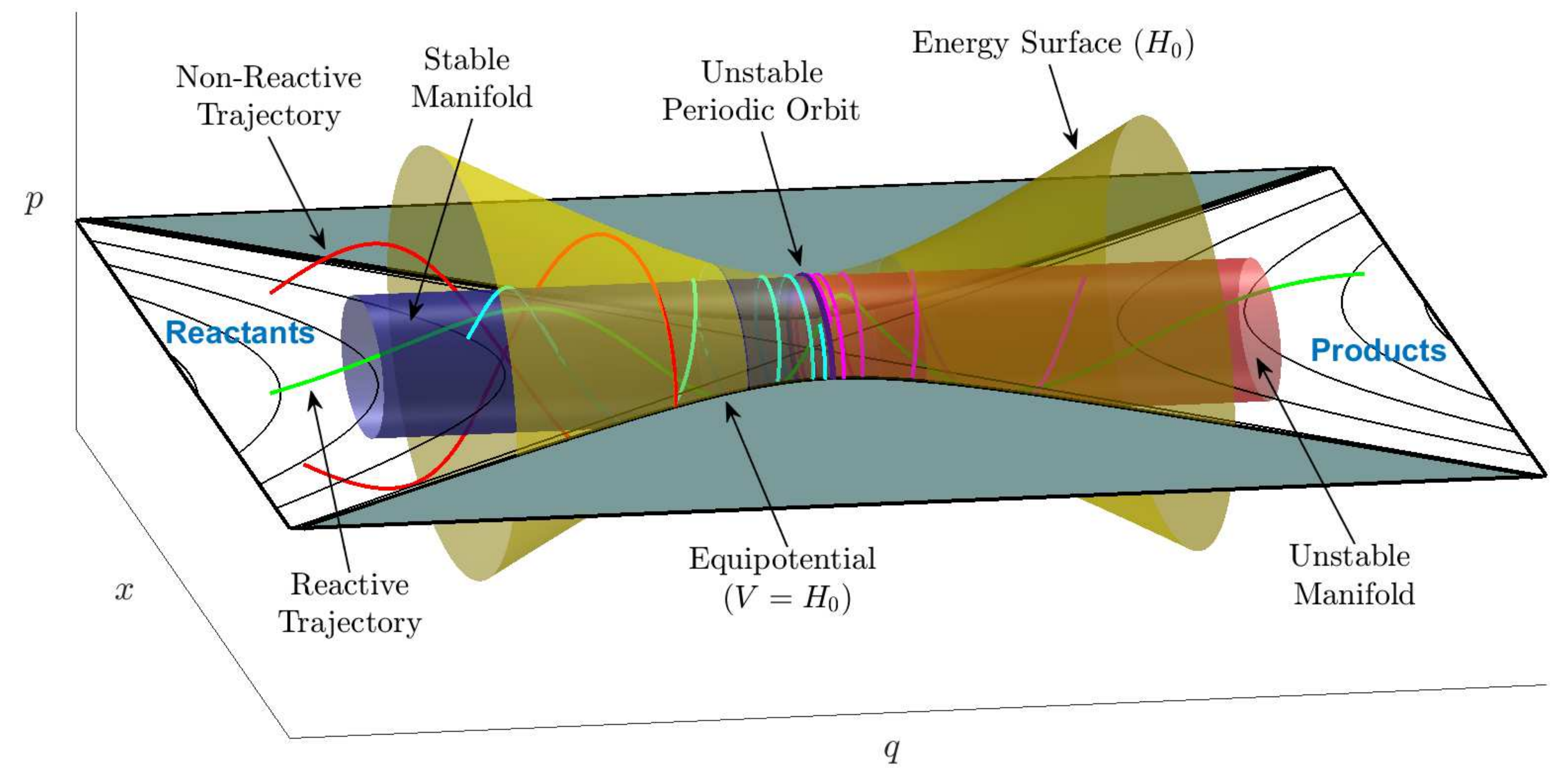}}	
	\end{center}
	\vspace{-3ex}
	\caption{Phase space structures in the neighborhood of the bottleneck for the uncoupled ($\varepsilon = 0$) Hamiltonian system \eqref{ham_2dof}. A) Linearized dynamics about the rank-1 saddle in the saddle space corresponding to the reactive DoF; B) Dynamics in the center space associated to the  harmonic oscillator DoF; C) Description of the phase space bottleneck region obtained for an energy of the system above that of the rank-1 saddle (the barrier of the PES). The stable and unstable manifolds of the NHIM act as conduits connecting reactants (well region) and products (escape to infinity) and characterize reaction dynamics.}
	\label{local_index1_ps}
\end{figure}

Consider the normal form for a quadratic Hamiltonian system with two DoF given by
\begin{equation}
H(q_1, q_1, p_1, p_2) = \underbrace{\frac{\lambda}{2} (p_1^2 - q_1^2)}_\text{$H_r$} + 
\underbrace{\frac{\omega}{2}(q_2^2 + p^2_2)}_\text{$H_b$} \; ,
\quad \lambda \, , \, \omega > 0 
\label{eqn:sqh_2dof}
\end{equation}
where Hamilton's equations are
\begin{equation}
\begin{cases}
\dot{q}_1 = \dfrac{\partial H}{\partial p_1} = \lambda \, p_1 \\[.3cm]
\dot{p}_1 = - \dfrac{\partial H}{\partial q_1} = \lambda \, q_1 \\[.3cm]
\dot{q}_2 = \dfrac{\partial H}{\partial p_2} = \omega \, p_2 \\[.3cm]
\dot{p}_2 = -\dfrac{\partial H}{\partial q_2} = - \omega \, q_2 
\end{cases}
\label{eqn:eom_nf_2dof}
\end{equation}
The equilibirum point is located at $(0,0,0,0)$ and has zero energy. It is trivial to check that the eigenvalues of the linearized system about the equilibrium point are $\pm \lambda$ and $\pm \omega \, i$, and hence the equilibrium point is of saddle$\times$center stability type, which is known as a rank-1 saddle. In this form, the Hamiltonian~\eqref{eqn:sqh_2dof} is decoupled into the ``reactive'' mode given by $H_r$ and the ``bath'' mode given by $H_b$. For this reason, it is known as a separable quadratic Hamiltonian (SQH). This representation allows us to address the phase space structures and discuss the distribution of the total energy of the system between the two modes in uncoupled coordinates. In this form, a chemical reaction is said to have occurred when the $q_1$ coordinate of a trajectory changes sign and thus, an isoenergetic, $H = H_0$ dividing surface (DS) can be defined by the $q_1 = 0$ hypersurface. The constant energy defines a three-dimensional energy surface in the four dimensional phase space given by
\begin{equation}
\frac{\lambda}{2} \left(p_1^2 - q_1^2\right) + \frac{\omega}{2}\left(p_2^2 + q^2_2\right) = H_r + H_b = H_0 > 0 \; , \quad H_r > 0 \; , \; H_b \geq 0
\end{equation}
The dividing surface, $q_1 = 0$, for a constant energy is 
\begin{equation}
\frac{\lambda}{2} p_1^2 + \frac{\omega}{2}\left(p_2^2 + q^2_2\right) = H_r + H_b = H_0 > 0 \; , \quad H_r > 0 \; , \; H_b \geq 0 
\label{eqn:ds_sqh_2dof}
\end{equation}
which is a two dimensional surface, and has the geometry $S^2$, that is, a 2-sphere on the three dimensional energy surface. Thus, it is codimension-1 and partitions the energy surface into reactant $p_1 - q_1 > 0$ and product $p_1 - q_1 < 0$ regions by the forward and backward ``reaction'' dividing surfaces as shown in Ref.~\cite{waalkens2004direct}, which are given by
\begin{equation}
\begin{aligned}
p_1 & = & \pm \sqrt{\frac{2}{\lambda}\left(H_0 - \frac{\omega}{2}\left(p_2^2 + q_2^2\right) \right)} \quad , \quad \text{forward/backward DS} 
\end{aligned}
\end{equation}
The forward and backward DS join at $p_1 = 0$ giving
\begin{equation}
\mathcal{N}(H_0) = \left\{ (q_1,q_2,p_1,p_2) \in \mathbb{R}^4 \; \vert \; q_1 = p_1 = 0 \; , \; \frac{\omega}{2}\left(p_2^2 + q_2^2 \right) = H_0 \geq 0 \right\} \quad , \quad \text{NHIM}
\label{eqn:sep_quad_ham2dof_nhim}
\end{equation}
which is of geometry $S^1$, that is a circle centered at the origin with radius $\sqrt{2H_0/\omega}$ in the $(q_2,p_2)$ plane. This is a {\em normally hyperbolic invariant manifold} (NHIM) associated with the rank-1 saddle and parametrized by total energy $H(q_1,q_2,p_1,p_2) = H_0$~\cite{wiggins2013normally}. {\em Invariance} follows from the vector field~\eqref{eqn:eom_nf_2dof}, since when we have that $q_1 = p_1 =0$, this gives $\dot{q}_1 = \dot{p}_1 =0$. Thus $q_1$ and $p_1$ always remain zero, and trajectories with these initial conditions remain on the NHIM, that is, $q_1 = p_1 =0$ is invariant. It is {\em normally hyperbolic} since the directions normal to the NHIM, that is, the $(q_1, p_1)$ plane, have linear saddle-like dynamics. For a two DoF system, this NHIM is more commonly referred to in the literature as an unstable periodic orbit and is shown in Fig.~\ref{local_index1_ps}.

In order to understand the relationship between the NHIM and the rank-1 saddle point, we note that for $H_r = H_b =0$ the NHIM reduces to the point $(q_1,q_2,p_1,p_2) = (0, 0, 0, 0)$, which is the rank-1 saddle point on the energy surface $H_0 = 0$. Moreover, recall that this rank-1 saddle equilibrium point is a configuration space concept and is located on the potential energy surface. Therefore, as the total energy of the system is increased from $0$, with the bath mode energy $H_b$ increasing from zero, the NHIM ``grows'' from the index-1 saddle point on the zero energy surface into an invariant 1-sphere. This shows how the ``influence'' of the rank-1 saddle point is carried to higher energy sufaces on which the saddle point does not exist. The stable and unstable manifolds of the NHIM~\eqref{eqn:sep_quad_ham2dof_nhim} are given by 
\begin{align}
\mathcal{W}^{\rm u} =& \big\{(q_1,q_2,p_1,p_2) \in \mathbb{R}^4 \; | \; q_1 = p_1 \; , \; \frac{\omega}{2}\left( p_2^2 + q_2^2 
\right) = H_b > 0  \big\} \label{eqn:quad_ham2dof_umani} \\[.1cm]
\mathcal{W}^{\rm s} =& \big\{(q_1,q_2,p_1,p_2) \in \mathbb{R}^4 \; | \; q_1 = - p_1 \; , \; \frac{\omega}{2}\left( p_2^2 + q_2^2 
\right) = H_b > 0  \big\} \label{eqn:quad_ham2dof_smani}
\end{align}
which are two-dimensional surfaces and have the geometry of $\mathbb{R} \times S^1$ for a fixed energy. Thus, the codimension-1 geometry of the manifolds partition the phase space into ``reactive'' and ``non-reactive'' trajectories as shown in Fig.~\ref{local_index1_ps}.

\section{Computation of NHIM and its invariant manifolds for the 2 DoF system}
\label{sec:appB}

In this aapendix we describe the steps followed in order to calculate the NHIM and its stable and unstable  invariant manifolds associated with the rank-1 saddle located at the origin for the Hamiltonian system with 2 DoF given by Eq. \eqref{ham_2dof}, which models the saddle-node bifurcation phenomena in phase space.

\textbf{Step 1: Select an excess energy above the critical value.} The Lyapunov subcenter theorem\cite{wiggins_normally_2014} tells us that when the energy of the Hamiltonian system is raised above that of the equilibrium point at the origin of saddle$\times$center (rank-1 saddle) stability, which is known as {\it critical energy}, a family of NHIM with geometry $S^{2N-3}$ bifurcates, where $N$ indicates the number of DoF. In our problem, $N = 2$ so the topology of the NHIM is $S^1$, that is, the NHIM is an unstable periodic orbit (UPO). So we pick a value for the total energy $H_0$ greater than the critical value $H_c = 0$, which gives an excess energy $\Delta H = H_0 - H_c = H_0$. This total energy is also the energy of trajectories on the invariant manifolds which partition the phase space into nonreactive and reactive trajectories at the same energy. The excess energy can be an arbitrary value, up to  the value  at which the NHIM  bifurcates. 

\textbf{Step 2: Obtain the NHIM at the selected excess energy.} As we have discussed, for the 2 DoF problem the NHIM is an UPO which touches the equipotential contour corresponding to the selected energy at the bottleneck region. However, due to the saddle-like dynamics in the transverse directions, any numerical error in the computation of the periodic orbit will get exponentially amplified with each time step, which will eventually destroy the periodic nature of the trajectory, since it will fail to come back to its starting point. 

In order to generate the desired unstable periodic orbit corresponding to a selected excess energy, we consider a procedure that starts with a small initial condition (``seed'') obtained from the linearized equations of motion near the rank-1 saddle, and uses {\it differential correction} and {\it numerical continuation}\cite{Koon2011,naik2017geometry,ross2018experimental} on that initial guess. The result is an unstable periodic orbit at target energy $H_0$ of period $T$ which will be close to $2\pi/\omega_{\varepsilon}$, where $\pm \omega_\varepsilon i$ is the pair of imaginary eigenvalues of the linearization about the rank-1 saddle point. In order to choose an initial guess for the search of the UPO we can use the linearized equations of motion about the rank-1 saddle equilibrium point $\mathbf{x}_1^e = \mathbf{0}$. The Jacobian matrix at the origin is given by Eq. \eqref{eqn:jacobian_2dof} and the general solution to the linearized dynamical system can be written as:
\begin{equation}
\mathbf{x}(t) = \mathbf{x}_1^e +  C_1 e^{\lambda_\varepsilon t} \mathbf{v}_+ + C_2 e^{-\lambda_\varepsilon t} \mathbf{v}_- + 2 {\rm Re}\left(\eta e^{i\omega_\varepsilon t} \mathbf{w}_+ \right)
\label{gensol_lineom_sn2dof}
\end{equation}
where $C_1,C_2 \in \mathbb{R}$ and $\eta = \eta_1 + \eta_2 \, i \in \mathbb{C}$. The real eigenvalues $\pm \lambda_\varepsilon$ with corresponding eigenvectors $\mathbf{v}_+, \mathbf{v}_-$ are described in Eqs.  \eqref{eigenvalues_2dof} and \eqref{saddle_eigenv} and the pair of complex eigenvalues $\pm \omega_\varepsilon i$ with corresponding eigenvectors $\mathbf{w}_+, \mathbf{w}_-$ are described in Eqs.  \eqref{eigenvalues_2dof} and \eqref{center_eigenv}. The idea is to use the complex eigenvalues and their corresponding eigenvectors to obtain a starting guess as an initial condition to initialize the method that searches for the UPO. To do so, we choose a small amplitude, $A_x \approx 2 \times 10^{-5}$, periodic orbit in the center manifold of the linearized system by selecting in Eq. \eqref{gensol_lineom_sn2dof} the values $\eta = - A_x/2$ (this eliminates the factor 2 in the formula), $C_1 = C_2 = 0$ and $t = 0$. Thus, the initial guess is:
\begin{equation}
\bar{\mathbf{x}}_{\rm 0,g} = (q_{0,g},x_{0,g},0,0) = \mathbf{x}^e_1 + 
2{\rm Re}(\eta w_+) = \left(-\dfrac{A_x \, \varepsilon}{\varepsilon - \left(\lambda_0^2 + \omega^2_\varepsilon\right)},-A_x,0,0\right)
\label{eqn:initial_guess_sn2dof}
\end{equation}
which has a period of $T_{0,g} = 2\pi / \omega_\varepsilon$.

\textbf{Step 3: Differential correction of the initial guess.} The differential correction procedure that we apply in this step to the initial guess will only yield a good approximation to the true UPO of the nonlinear Hamiltonian system whenever  $A_x \ll 1$, which corresponds to a much smaller excess energy $\Delta H$ than the one we originally selected in step 1. The reason for this is that the construction of the initial guess is based on the linear approximation near the rank-1 saddle equilibrium point. The convergence criterion that we use is based on the basic property that a periodic orbit returns to the starting point after a given period $T$, which is in fact the period of the PO. If $\bar{\mathbf{x}}_0 = \bar{\mathbf{x}}_{\rm po}(0)$ is a true initial condition on the PO $\mathbf{x}_{po}$ of period $T$, the convergence is checked using the condition:
\begin{equation}
\left\| \bar{\mathbf{x}}_{\rm po}(T) - 
\bar{\mathbf{x}}_{\rm po}(0) \right\| < \epsilon
\end{equation}
for some tolerance $\epsilon \ll 1$. In this approach, we hold the configuration coordinate $q$ constant, while applying correction to the $x$ configuration coordinate of the initial guess. While the $p_x$ momentum coordinate is used as a stopping criterion for the differential correction procedure, the $p$ momentum coordinate is used as a terminating event (crossing the $p = 0$ plane in phase space) for the integration. It is to be noted that this combination of coordinates is suitable for the structure of the initial guess at hand, so for other problems or other forms of the initial guess, would require some permutation of the phase space coordinates to achieve a stable implementation. 

Let us denote the flow map of a differential equation $\dot{\mathbf{x}} = \mathbf{f}(\mathbf{x})$ with initial condition $\mathbf{x}(t_0) = \mathbf{x}_0$ by $\phi(t;\mathbf{x}_0)$. Thus, the displacement of a reference trajectory $\bar{\mathbf{x}}(t)$ after a time $\delta t$ becomes:
\begin{align}
\delta \bar{\mathbf{x}}(t + \delta t) = \phi(t + \delta 
t;\bar{\mathbf{x}}_0 + \delta \bar{\mathbf{x}}_0) - 
\phi(t ;\bar{\mathbf{x}}_0) \;.
\end{align}
Thus, a Taylor expansion of the displacement after $t_1 + \delta t_1$ gives:
\begin{align}
\delta \bar{\mathbf{x}}(t_1 + \delta t_1) = 
\frac{\partial \phi(t_1;\bar{\mathbf{x}}_0)}{\partial 
	\mathbf{x}_0}\delta \bar{\mathbf{x}}_0 + \frac{\partial 
	\phi(t_1;\bar{\mathbf{x}}_0)}{\partial t_1}\delta t_1 + 
h.o.t
\end{align}
where the first term on the right hand side is the state transition matrix, $\mathbf{\Phi}(t_1,t_0)$, evaluated along the reference trajectory initialized at $t_0$. The state transition matrix along a trajectory is obtained from the numerical solution of the variational equations along with the Hamilton's equation of motion~\cite{Parker1989}. Suppose that we want to land at a point $\mathbf{x}_{\rm d}$ (this would be the starting initial condition for a periodic orbit), after an integration time interval $t_1$, and starting from the initial guess $\bar{\mathbf{x}}_{\rm 0, g}$, then we have:
\begin{align}
\bar{\mathbf{x}}(t_1) = \phi(t_1;\bar{\mathbf{x}}_{\rm 0, g}) 
= \bar{\mathbf{x}}_1 = \mathbf{x}_d - \delta 
\bar{\mathbf{x}}_1
\end{align}
where the error $\delta \bar{\mathbf{x}}_1$ is the applied first order correction obtained from the state transition matrix evaluated along the trajectory with initial condition $\bar{\mathbf{x}}_{\rm 0, g}$ and integrated for $t_1$ time units. For the rank-1 saddle equilibrium point under consideration, we initialize the guess as obtained in Eq. \eqref{eqn:initial_guess_sn2dof} with $\bar{\mathbf{x}}_{\rm 0, g}$. Using numerical integration, the initial condition is integrated until a $p = 0$ event occurs with a high specified tolerance (typically $10^{-14}$). This results in $\bar{\mathbf{x}}(t_1)$, which for the guess initial condition denotes the half-period location $t_1 = T_{0,g}/2$. Then, we evaluate the state transition matrix $\mathbf{\Phi}(t_1,0)$ for the trajectory obtained from the guess intial condition. Now, this is used to correct the initial value of $x_{0,g}$ while keeping $q_{0,g}$ constant and iterating until $p_x = 0$. Since the $x$ configuration coordinate is kept constant, the first order correction is given by:
\begin{align}
\delta p_{1} &= \Phi_{32} \, \delta x_0 + \dot{p}_{1} \, \delta t_1 + h.o.t \nonumber \\
\delta p_{x_1} &= \Phi_{42} \, \delta x_0 + \dot{p}_{x_1} \, \delta t_1 + h.o.t \nonumber
\end{align}
where $\Phi_{ij}$ is the $(i,j)^{th}$ entry of $\mathbf{\Phi}(t_1,0)$ and the acceleration terms come from the equations of motion evaluated at the crossing $t = t_1$ when $p_{1} = \delta p_{1} = 0$. Thus, we obtain the first order correction $\delta x_0$ as:
\begin{equation}
\delta x_0 \approx \left(\Phi_{42} - 
\Phi_{32}\frac{\dot{p}_{x_1}}{\dot{p}_{1}}\right)^{-1} \delta p_{x_1} \quad,\quad 
x_0 \rightarrow x_0 - \delta x_0 
\end{equation}
which is iterated until $|p_{x_1}| = |\delta p_{x_1}| < \epsilon$ for some tolerance $\epsilon$, since we want the final point of the periodic orbit to be of the form $\bar{\mathbf{x}}_{t_1} = (q_1,x_1,0,0)$. We remark that in all this process, differential correction assumes that the guess periodic orbit has a small error (for example in this system, of the order of $10^{-2}$) and can be corrected using first order form of the correction terms. If, however, large a corrective step is used, the half-orbit overshoots between successive steps leading to failure in converging to a closed orbit. Once the appropriate conditions are chosen, differential correction generates a family of periodic orbits and takes 2-3 iterations per unstable periodic orbit. 

\textbf{Step 4: Numerical continuation to the UPO at the selected excess energy.} The procedure described above yields an initial condition for an unstable periodic orbit from an initial guess. Since our initial guess came from the linearization near the rank-1 saddle equilibrium pointgiven by~\eqref{gensol_lineom_sn2dof}, we can use this procedure for small amplitudes of order $2 \times 10^{-5}$. We remark that this procedure is based on computations presented for this problem and will vary for a different nonlinear system. This small amplitude corresponds to small excess energy, typically of the order $10^{-5}$, and to obtain the unstable periodic orbit of arbitrarily large amplitude, we resort to {\it numerical continuation} for generating a family of periodic orbits that reach the selected excess energy. 

This procedure starts with the initial conditions for two nearby unstable periodic orbits of small amplitude to obtain an initial guess for the next periodic orbit. The initial guess obtained by a simple extrapolation can then be corrected using differential correction. To this end, we proceed as follows. Suppose we find two nearby small amplitude unstable periodic orbits with initial conditions $\bar{\mathbf{x}}_0^{(1)}$ and $\bar{\mathbf{x}}_0^{(2)}$, correct to within the tolerance $d_{\rm tol}$ obtained using the differential correction procedure described above. We can then generate a family of periodic orbits with increasing amplitudes around $\bar{\mathbf{x}}_{\rm eq}$ as:
\begin{align}
\Delta = \bar{\mathbf{x}}_0^{(2)} - \bar{\mathbf{x}}_0^{(1)} 
= (\Delta q_0, \Delta x_0, 0, 0)		
\end{align}
A linear extrapolation to an initial guess of slightly larger amplitude, 
$\bar{\mathbf{x}}_0^{(3)}$ is given by:
\begin{align}
\bar{\mathbf{x}}_{\rm 0,g}^{(3)} =&~\bar{\mathbf{x}}_0^{(2)} + \Delta = (q_0^{(2)} + \Delta q_0,x_0^{(2)} + \Delta x_0,0,0) = (q_0^{(3)},x_0^{(3)},0,0)
\end{align}
Thus, we can use differential correction on this guess initial condition to compute an accurate solution $\bar{\mathbf{x}}_0^{(3)}$ from the initial guess $\bar{\mathbf{x}}_{\rm 0,g}^{(3)}$ and repeat as an iterative step to generate a family of unstable periodic orbits. 

Next, to compute an unstable periodic orbit at the selected excess energy, we track the energy of each unstable periodic orbit in the family until we have two solutions, $\bar{\mathbf{x}}_0^{\rm (k)}$ and $\bar{\mathbf{x}}_0^{\rm (k+1)}$, whose energy brackets the selected excess energy $\Delta H$. Then, we can resort to combining a bisection type method with differential correction on the two periodic orbits until we converge to the desired periodic orbit to within a specified 
tolerance. Thus, the result is an unstable periodic orbit at the selected energy $H_0$ and is specified by the initial condition $\bar{\mathbf{x}}_{\rm po}(0)$ and time period $T_{\rm po}$. 

\textbf{Step 5: Computation of invariant manifolds of the NHIM.} We find the global 
approximation to the unstable and stable manifolds of the periodic orbit from the eigenvectors of the monodromy matrix. The local linear approximation of the unstable (or stable) manifolds (initial conditions displaced along the saddle space eigenvectors) is integrated using the full nonlinear equations of motion to produce the global approximation of the unstable (or stable) manifolds. This procedure is known as \textit{globalization of the manifolds} and we proceed as follows.

First, the state transition matrix $\Phi(t)$ along the periodic orbit with initial condition $X_0$ can be obtained by numerical integration of the variational equations 
along with the equations of motion from $t = 0$ to $t = T_{\rm po}$. This is gives the monodromy matrix $M = \Phi(T_{\rm po})$ and its eigenvalues are obtained. For Hamiltonian systems\cite{Meyer2009}, it is known that the eigenvalues of $M$ are of the form:
\begin{align}
\lambda_1 > 1, \qquad \lambda_2 = \frac{1}{\lambda_1}, \qquad \lambda_3 = 
\lambda_4 = 1
\end{align}
The eigenvector associated with eigenvalue $\lambda_1$ is in the unstable 
direction, the eigenvector associated with eigenvalue $\lambda_2$ is in the stable 
direction. Let $e^{s}(\bar{\mathbf{x}}_{\rm po}(0))$ denote the normalized stable eigenvector, and $e^{u}(\bar{\mathbf{x}}_{\rm po}(0))$ 
denote the normalized unstable eigenvector. We can compute the invariant manifolds by 
initializing along these eigenvectors as:
\begin{equation}
\mathbf{X}_0^{s/u}(\bar{\mathbf{x}}_{\rm po}(0)) = \bar{\mathbf{x}}_{\rm po}(0) + \epsilon e^{s/u}(\bar{\mathbf{x}}_{\rm po}(0))
\end{equation}
for the stable/unstable manifold of the desired periodic orbit. Here the small displacement from $\mathbf{X}_0$ is denoted by $\epsilon > 0$ and its magnitude is taken to be small enough for the validity of the linearization, yet not so small that the time of flight becomes too large due to asymptotic nature of the stable and unstable manifolds. It has been suggested\cite{Koon2011} that typical values of $\epsilon$ around $10^{-9}$ correspond to nondimensional position displacements of magnitude around $10^{-6}$.

By numerically integrating the unstable vector forwards in time, using both 
$\epsilon$ and $-\epsilon$, for the forward and backward branches respectively, we 
generate trajectories shadowing the two branches, $\mathcal{W}^u_{+}$ and $\mathcal{W}^u_{-}$. Similarly, by integrating the stable vector backwards in time, using both $\epsilon$ and $-\epsilon$, for forward and backward branches respectively, we generate trajectories shadowing the stable manifold, $\mathcal{W}^{s}_{+,-}$. 
For the manifold at $\mathbf{X}(t)$, one can simply use the state transition matrix to 
transport the eigenvectors from $\mathbf{X}_0$ to $\mathbf{X}(t)$
\begin{equation}
\mathbf{X}^{u/s}(\mathbf{X}(t)) = \Phi(t,0)\mathbf{X}^{u/s}(\mathbf{X}_0)
\end{equation}
It is to be noted that since the state transition matrix does not preserve the norm, and hence the resulting vector must be normalized. The globalized invariant manifolds associated with rank-1 saddles are known as Conley-McGehee tubes\cite{Marsden2006}. These tubes form the impenetrable phase space conduits (codimension-1 barriers) for the trajectories to react by crossing the $q = 0$ dividing surface.

In summary, the computation of invariant manifolds of the unstable periodic orbit associated with the rank-1 saddle begins with the linearized equations of motion. This is obtained after a coordinate transformation to the rank-1 saddle equilibrium point and a Taylor expansion of the equations of motion. Keeping the first order terms in this expansion, we obtain the eigenvalues and eigenvectors of the linearized system. The eigenvectors corresponding to the center subspace provide the starting guess for computing the unstable periodic orbits of small excess energy, $\Delta H << 1$. This iterative procedure performs a small correction to the starting guess for initial condition based on the terminal condition of the periodic orbit until a desired tolerance is satisfied. This procedure is known as differential correction and generates initial condition for an unstable periodic orbit at small excess energy. Next, a numerical continuation procedure is adopted to follow the small energy (small amplitude) periodic orbit out to high excess energy. Once the unstable periodic orbit is obtained, the globalization of its invariant manifolds is done using the initial conditions along the eigenvectors of the state transition matrix computed along the unstable periodic orbit. The result of the steps described above is shown in Fig.~\ref{fig:set2_manifolds_upo_energysurf}.

\end{document}